\documentclass{ptephy_v1}

\usepackage{graphics} 

\usepackage{dcolumn}
\usepackage{bm}
\usepackage{amsfonts}
\usepackage{latexsym}
\usepackage{amssymb}
\usepackage{verbatim}
\usepackage{amsthm}
\usepackage{amsmath}
\usepackage{mathrsfs}
\usepackage{wrapft}

\newcommand{\CF}{{{\mathbb C}}}

\newcommand{\RF}{{{\mathbb R}}}
\newcommand{\SF}{{{\mathbb S}}}

\newcommand{\ZF}{{{\mathbb Z}}}

\begin{document}


\numberwithin{equation}{section}

\title{
  Quantum noise and vacuum fluctuations \\
  in balanced homodyne detections \\
  through ideal multi-mode detectors
}

\author{
  Kouji Nakamura
  \footnote{E-mail address: kouji.nakamura@nao.ac.jp},
}
\address{%
  Gravitational-Wave Science Project,
  National Astronomical Observatory of Japan,\\
  2-21-1 Osawa, Mitaka, Tokyo 181-8588, Japan
}%

\date{\today}

\begin{abstract}
  The balanced homodyne detection as a readout scheme of
  gravitational-wave detectors is carefully examined from the quantum
  field theoretical point of view.
  The readout scheme in gravitational-wave detectors specifies the
  directly measured quantum operator in the detection.
  This specification is necessary when we apply the recently developed
  quantum measurement theory to gravitational-wave detections.
  We examine the two models of measurement.
  One is the model in which the directly measured quantum
  operator at the photodetector is Glauber's photon number operator,
  and the other is the model in which the power operator of the optical
  field is directly measured.
  These two are regarded as ideal models of photodetectors.
  We first show these two models yield the same expectation value of
  the measurement.
  Since it is consensus in the gravitational-wave community that
  vacuum fluctuations contribute to the noises in the detectors,
  we also clarify the contributions of vacuum fluctuations to the
  quantum noise spectral density without using the two-photon
  formulation which is used in the gravitational-wave community.
  We found that the conventional noise spectral density in the
  two-photon formulation includes vacuum fluctuations from the main
  interferometer but does not include those from the local
  oscillator.
  Although the contribution of vacuum fluctuations from the local
  oscillator theoretically yields the difference between the above two
  models in the noise spectral densities, this difference is
  negligible in realistic situations.
\end{abstract}

\maketitle


\section{Introduction}
\label{sec:Introduction}


Since gravitational waves are directly observed by the Laser
Interferometer Gravitational-wave
Observatory~\cite{LIGO-GW150914-2016} in 2015, the gravitational-wave
astronomy and the multi-messenger astronomy including
gravitational-wave are developing~\cite{LSC-homepage}.
To support these developments as more precise science, improvements of
the detector sensitivity are necessary.
For these improvements, it is important to continue the research and
development of the science of gravitational-wave detectors together
with the source sciences of gravitational waves.
Current gravitational-wave detectors already reached to the
fundamental noise that arises from quantum fluctuations of lasers in
the detector system and the reduction of these quantum noises is an
important topic in the science of gravitational-wave detectors.
Therefore, to improve the sensitivity of gravitational-wave detectors,
a rigorous quantum-theoretical description is required.


Recently, a mathematically rigorous quantum measurement theory is also
developed~\cite{Ozawa-2004}.
One of the motivations of this development was gravitational-wave
detections~\cite{Ozawa-1988}.
However, the actual application of this theory to the
gravitational-wave detectors requires its extension to the quantum
field theory, because the quantum noise in gravitational-wave
detectors is discussed through the quantum field theory of
lasers~\cite{H.J.Kimble-Y.Levin-A.B.Matsko-K.S.Thorne-S.P.Vyatchanin-2001}.
Furthermore, in the quantum measurement theory, we have to specify the
directly measured quantum operator.
In interferometric gravitational-wave detectors, we may regard that
the directly measured operator is specified at their ``{\it readout
  scheme}'' in the detectors.
The readout scheme in gravitational-wave detectors is the optical
system to specify the optical fields which are detected at the
photodetectors through the signal output from the ``{\it
  main-interferometer}'' and the reference field which is injected
from the ``{\it local oscillator}.''
The research on this readout scheme is important for the development
and application of the mathematical quantum measurement theory to
gravitational-wave detections.


Current gravitational-wave detectors use the DC readout scheme, in
which the output photons from the main interferometer are directly
measured.
On the other hand, ``{\it homodyne detections}'' are regarded as one
of candidates of the readout scheme in the near
future~\cite{S.Steinlechner-et-al-2015-T.Zhang-2017}.
In
Ref.~\cite{H.J.Kimble-Y.Levin-A.B.Matsko-K.S.Thorne-S.P.Vyatchanin-2001},
it is written that the output quadrature $\hat{b}_{\theta}$ defined by
\begin{eqnarray}
  \label{eq:DBHD_20180805_1.1}
  \hat{b}_{\theta}
  :=
  \cos\theta \hat{b}_{1} + \sin\theta \hat{b}_{2}
\end{eqnarray}
is measured by the ``{\it balanced homodyne detection}.''
Here, $\hat{b}_{1}$ and $\hat{b}_{2}$ are the amplitude and phase
quadrature in the two-photon
formulation~\cite{C.M.Caves-B.L.Schumaker-1985,B.L.Schumaker-C.M.Caves-1985},
which are defined by Eqs.~(\ref{eq:hatb1-hatb2-def}) in this paper, and
$\theta$ is called the homodyne angle.
The output operator $\hat{b}_{\theta}$ includes gravitational-wave
signal $h(\Omega)$ as
\begin{eqnarray}
  \label{eq:DBHD_20180805_1.2}
  \hat{b}_{\theta}
  &=&
      R(\Omega,\theta)\left(
      \hat{h}_{n}(\Omega,\theta) + h(\Omega)
      \right)
      ,
\end{eqnarray}
where $\hat{h}_{n}(\Omega)$ is the noise operator given by the linear
combination of the annihilation and creation operators of photons
injected to the main interferometer through the input-output relation
of the main interferometer.
However, it does not seem that there is clear description on the
actual measurement processes of the operator
(\ref{eq:DBHD_20180805_1.1}) or directly measured quantum operators in
these processes.


Following this motivation, in Refs.~\cite{K.Nakamura-M.-K.Fujimoto-double-balanced-letter,K.Nakamura-M.-K.Fujimoto-double-balanced-full},
we examined the case where the directly measured operators are the number operator:
\begin{eqnarray}
  \label{eq:each-mode-photon-number-def}
  \hat{n}(\omega) = \hat{a}^{\dagger}(\omega)\hat{a}(\omega)
\end{eqnarray}
for each mode frequency, where the annihilation [$\hat{a}(\omega)$]
and creation [$\hat{a}^{\dagger}(\omega)$] operators of the electric
field.
From the usual commutation relation
\begin{eqnarray}
  \label{eq:commutation-relation-a}
  \left[\hat{a}(\omega),\hat{a}^{\dagger}(\omega')\right]
  =
  2 \pi \delta(\omega-\omega'),
  \quad
  \left[\hat{a}(\omega),\hat{a}(\omega')\right]
  =
  \left[\hat{a}^{\dagger}(\omega),\hat{a}^{\dagger}(\omega')\right]
  =
  0
  ,
\end{eqnarray}
the eigenvalue of the operator $\hat{n}(\omega)$ becomes non-negative
countable number.
This countable number give rise to the notion of ``photon'' and we can
count this number, in principle.
Actually, there are many experiments in which detectors count the
photon number $\hat{n}(\omega)$ of the single mode.
As a result of the examination in
Refs.~\cite{K.Nakamura-M.-K.Fujimoto-double-balanced-letter,K.Nakamura-M.-K.Fujimoto-double-balanced-full},
we reached to a conclusion that we cannot measure the expectation
value of the operator (\ref{eq:DBHD_20180805_1.1}) by the balanced
homodyne detection~\cite{E.Shchukin-Th.Richter-W.Vogel-2005}.
However, the operator (\ref{eq:each-mode-photon-number-def}) is not
appropriate as a directly measured operator in gravitational-wave
detectors, because we take the time-sequence data detected at the
photodetector and discuss its Fourier spectrum.
This implies that the photodetection in gravitational-wave detectors
is essentially a multi-mode detection.


In this paper, we re-examine the measurement process of the balanced
homodyne detections for multi-mode detections.
As a directly measured operator in multi-mode detections, Glauber's
photon number operator
\begin{eqnarray}
  \label{eq:multi-mode-photon-number-propto}
  \hat{N}_{a}(t)
  \propto
  \hat{E}_{a}^{(-)}(t)\hat{E}_{a}^{(+)}(t)
\end{eqnarray}
is often used in many literatures.
Here $\hat{E}_{a}^{(+)}$ and $\hat{E}_{a}^{(-)}$ are the positive- and
negative-frequency parts of the electric field $\hat{E}_{a}(t)$ of the
detected optical field, respectively, which are introduced in
Sec.~\ref{sec:Basic_Notation} in this paper.
On the other hand, in the gravitational-wave community, it is often
said that the probability of the excitation of the photocurrent is
proportional to the power operator
\begin{eqnarray}
  \label{eq:multi-mode-photon-power-propto}
  \hat{P}_{a} \propto \frac{1}{4\pi} \left(\hat{E}_{a}(t)\right)^{2}
\end{eqnarray}
of the measured optical field.
It is also true that there are many literatures in which the
photo-detection is treated as a classical stochastic process, in which
the detection probability is proportional to the expectation value of
the power
operator~\cite{P.J.Winzer-1997,F.Quinlan-et-al.-2013a,F.Quinlan-et-al.-2013b}.


Historically speaking, in theoretical quantum measurement of optical
fields~\cite{R.J.Glauber-1963-130,R.J.Glauber-1963-131,P.L.Kelley-W.H.Kleiner-1964,B.R.Mollow-1968,R.J.Cook-1982a,R.J.Cook-1982b,H.P.Yuen-V.W.S.Chan-1983,R.S.Bondurant-J.H.Shapiro-1984,H.J.Kimble-L.Mandel-1984,B.Yurke-1985,R.S.Bondurant-1985,J.H.Shapiro-1985,S.L.Braunstein-D.D.Crouch-1991,Z.Y.Ou-H.J.Kimble-1995,M.J.Collett-R.Loudon-W.W.Gardiner-1987},
there was a controversy on which variable is the directly measured by
photodetectors for multi-frequency optical fields.
Some insisted that the direct observable of photodetectors is the
above photon number operator
(\ref{eq:multi-mode-photon-number-propto}) in the multi-mode optical
fields, some insisted that the direct observable of photodetectors is
the power of the optical field~\footnote{
  For example, discussions in
Refs.~\cite{R.J.Glauber-1963-130,R.J.Glauber-1963-131,P.L.Kelley-W.H.Kleiner-1964,H.J.Kimble-L.Mandel-1984,B.Yurke-1985}
  are as follows:
  From the microscopic point of view, the interaction between photon
  and charged particles is proportional to $\hat{p}\hat{A}$, where
  $\hat{p}$ is the momentum of the charged particles in the
  photodetector and $\hat{A}$ is the vector potential of the laser.
  This interaction is also given as
  $\hat{p}\hat{A}=\hat{p}(\hat{A}^{(+)}+\hat{A}^{(-)})$, where
  $A^{(+)}$ ($A^{(-)}$) is the positive- (negative-)  frequency part
  of the vector potential.
  In
  Refs.~\cite{R.J.Glauber-1963-130,R.J.Glauber-1963-131,P.L.Kelley-W.H.Kleiner-1964,H.J.Kimble-L.Mandel-1984},
  the term $\hat{p}\hat{A}^{(-)}$ was ignored at their starting point.
  In Ref.~\cite{H.J.Kimble-L.Mandel-1984}, with this ignorance, it was
  insisted that the macroscopic photo-current is proportional to the
  expectation value $\langle\hat{A}^{(-)}\hat{A}^{(+)}\rangle$ under
  the ``quasistationary field condition'' even in the wideband detection.
  This result supports Glauber's photon number operator
  (\ref{eq:multi-mode-photon-number-propto}) as the directly measured
  operator in photo-detections.
  However, in Ref.~\cite{B.Yurke-1985}, it was claimed that the term
  $\hat{p}\hat{A}^{(-)}$ gives a finite contribution to the
  macroscopic photo-current in the wideband photo-detections.


  In any case, the connection of the microscopic process and
  macroscopic measured photo-current will be necessary to reach the
  conclusion.
  However, the story is not so simple but the problem becomes quite
  complicated if we take into account the randomness of detected
  photo-currents~\cite{P.L.Kelley-W.H.Kleiner-1964,Z.Y.Ou-H.J.Kimble-1995}.
}
.
Although the author could not find a complete conclusion of this
controversy in literature, both of the operators
(\ref{eq:multi-mode-photon-number-propto}) and
(\ref{eq:multi-mode-photon-power-propto}) are used in recent many
literature.


Keeping in our mind this situation, in this paper, we examine the two
models of directly measured operators as two ideal cases.
One is the model in which the directly measured operator at
photodetectors is Glauber's photon number operator
(\ref{eq:multi-mode-photon-number-propto}), and the other is the model
in which the directly measured operator is the power operator
(\ref{eq:multi-mode-photon-power-propto}).
As the results of these examinations, we reached to the conclusion
that the expectation value of the operator $\hat{b}_{\theta}$ given by
Eq.~(\ref{eq:DBHD_20180805_1.1}) is measured by the balanced homodyne
detection contrary to the results in
Refs.~\cite{K.Nakamura-M.-K.Fujimoto-double-balanced-letter,K.Nakamura-M.-K.Fujimoto-double-balanced-full}.


In this paper, we also evaluate noise spectral density of these models.
In many literatures of the gravitational-wave detection, it is written
that the single sideband noise spectral density
$\bar{S}_{A}^{(s)}(\omega)$ for an arbitrary operator
$\hat{A}(\omega)$ with the vanishing expectation value in the
``{\it stationary}'' system is given by
\begin{eqnarray}
  \label{eq:Kimble-noise-spectral-density-single-side}
  \frac{1}{2} 2\pi \delta(\omega-\omega') \bar{S}_{A}^{(s)}(\omega)
  :=
  \frac{1}{2} \langle\mbox{in}|
  \hat{A}(\omega)\hat{A}^{\dagger}(\omega')
  +
  \hat{A}^{\dagger}(\omega')\hat{A}(\omega)
  |\mbox{in}\rangle
  .
\end{eqnarray}
The noise spectral density
(\ref{eq:Kimble-noise-spectral-density-single-side}) is introduced by
Kimble et al. in
Ref.~\cite{H.J.Kimble-Y.Levin-A.B.Matsko-K.S.Thorne-S.P.Vyatchanin-2001}
in the context of the two-photon
formulation~\cite{C.M.Caves-B.L.Schumaker-1985,B.L.Schumaker-C.M.Caves-1985}.
However, in this paper, we do not use the two-photon formulation,
though some final formulae are written in terms of the two-photon
formulation as much as possible.
We also examine the original meaning of Kimble's noise spectral density
(\ref{eq:Kimble-noise-spectral-density-single-side}) and derive
the deviations from this noise formula
(\ref{eq:Kimble-noise-spectral-density-single-side}) together with the
noise contributions due to the imperfection of the interferometer
configuration.
Although Collett et al.~\cite{M.J.Collett-R.Loudon-W.W.Gardiner-1987}
also discussed the quantum noises in the heterodyne and homodyne
detections, they used the normal-ordered generating function to
evaluate the noises in detections.
Therefore, it is not clear whether their arguments properly includes
vacuum fluctuations or not.
The contributions from the vacuum fluctuations are carefully discussed
in this paper.


The organization of this paper as follows.
In Sec.~\ref{sec:Preliminary}, we summarize the basic notation which
we use in this paper and introduce Glauber's photon number
operators and the power operator of the detected optical field in the
multi-mode detections.
In Sec.~\ref{sec:Homodyne_Detections_by_multi-mode_detectors}, we
examine the expectation values measured by the homodyne detections in
the two models with the above two different directly measured
operators.
In Sec.~\ref{sec:Noise-spectral-densities}, we show the general
arguments of the noise spectral density and introduce the definition
of the noise spectral density in this paper.
In Sec.~\ref{sec:Estimation_of_Quantum_noise}, we evaluate the quantum
noise in the homodyne detections.
The final section (Sec.~\ref{sec:Summary}) devoted to our summary and
discussions.


\section{Preliminary}
\label{sec:Preliminary}


In this section, we describe the notation of the electric fields,
quantum states, and quantum operators, which are used within this
paper.
In Sec.~\ref{sec:Basic_Notation}, we describe the basic
notations of quantum operators for the electric field of the laser in
gravitational-wave detectors.
In Sec.~\ref{sec:Multi-mode_number_and_power_operators}, we
introduce two quantum operators which are the candidates of the
directly measured operators at the multi-mode photodetectors.
One is Glauber's photon number operator and the other is the power
operator as mentioned in Sec.~\ref{sec:Introduction}.


\subsection{Basic Notation}
\label{sec:Basic_Notation}


In this paper, we denote the electric field of the lasers by
\begin{eqnarray}
  \hat{E}_{a}(t-z)
  &=&
      \int_{0}^{+\infty} \frac{d\omega}{2\pi}
      \sqrt{\frac{2\pi\hbar\omega}{{\cal A}c}} \left[
      \hat{a}(\omega) e^{-i\omega(t-z)}
      +
      \hat{a}^{\dagger}(\omega) e^{+i\omega(t-z)}
      \right]
      \nonumber\\
  &=&
      \hat{E}_{a}^{(+)}(t-z)
      +
      \hat{E}_{a}^{(-)}(t-z)
      ,
      \label{eq:electric-field-notation-total-electric-field}
\end{eqnarray}
where $\hat{E}_{a}^{(+)}(t-z)$ is the positive frequency part of
$\hat{E}_{a}(t-z)$:
\begin{eqnarray}
  \label{eq:electric-field-notation-total-electric-field-positive}
  \hat{E}_{a}^{(+)}(t-z)
  =
  \int_{0}^{+\infty} \frac{d\omega}{2\pi}
  \sqrt{\frac{2\pi\hbar\omega}{{\cal A}c}}
  \hat{a}(\omega) e^{-i\omega(t-z)}
\end{eqnarray}
and $\hat{E}_{a}^{(-)}(t-z)$ is the adjoint operator of
$\hat{E}_{a}^{(+)}(t-z)$ as
\begin{eqnarray}
  \label{eq:electric-field-notation-total-electric-field-negative}
  \hat{E}_{a}^{(-)}(t-z)
  =
  \left[
  \hat{E}_{a}^{(+)}(t-z)
  \right]^{\dagger}
  .
\end{eqnarray}
The ${\cal A}$ in the factor of
Eq.~(\ref{eq:electric-field-notation-total-electric-field}) is the
sectional area of the laser beam.
The operators $\hat{a}(\omega)$ and $\hat{a}^{\dagger}(\omega)$ in
Eq.~(\ref{eq:electric-field-notation-total-electric-field}) are
annihilation and creation operators of the photon, respectively, and
these satisfy the commutation relations (\ref{eq:commutation-relation-a}).
We explicitly denote the quadrature $\hat{a}(\omega)$ of the electric
field in the subscript of the electric field
(\ref{eq:electric-field-notation-total-electric-field}) itself.
We may write
Eq.~(\ref{eq:electric-field-notation-total-electric-field}) so that
\begin{eqnarray}
  \label{eq:electric-field-notation-total-electric-field-using-step-func.}
  \hat{E}_{a}(t-z)
  =
  \int_{-\infty}^{+\infty} \frac{d\omega}{2\pi}
  \sqrt{\frac{2\pi\hbar\omega}{{\cal A}c}} \left[
  \Theta(\omega)
  \hat{a}(\omega)
  +
  \Theta(-\omega)
  \hat{a}^{\dagger}(-\omega)
  \right]
  e^{-i\omega(t-z)}
  ,
\end{eqnarray}
where $\Theta(\omega)$ is the Heaviside step function
\begin{eqnarray}
  \label{eq:Heavisite-step-function-def}
  \Theta(\omega)
  =
  \left\{
  \begin{array}{lcccl}
    1 & \quad & \mbox{for} & \quad & \omega>0; \\
    0 & \quad & \mbox{for} & \quad & \omega<0.
  \end{array}
  \right.
\end{eqnarray}


From the commutation relations (\ref{eq:commutation-relation-a}), we
can derive the commutation relation between the positive- and
negative-frequency part $\hat{E}_{a}^{(\pm)}(t)$ of the electric field as
\begin{eqnarray}
  &&
     \label{eq:commutation-relation-E+E-}
     \left[
     \hat{E}_{a}^{(+)}(t), \hat{E}_{a}^{(-)}(t')
     \right]
     =
     \frac{2\pi\hbar}{{\cal A}c}
     \int_{0}^{+\infty} \frac{d\omega}{2\pi} \omega
     e^{-i\omega(t-t')}
     =:
     \frac{2\pi\hbar}{{\cal A}c} \Delta_{a}(t-t').
  \\
     \label{eq:commutation-relation-E-E-E+E+}
  &&
     \left[
     \hat{E}_{a}^{(+)}(t), \hat{E}_{a}^{(+)}(t')
     \right]
     =
     \left[
     \hat{E}_{a}^{(-)}(t), \hat{E}_{a}^{(-)}(t')
     \right]
     =
     0
     .
\end{eqnarray}
The subscription ``$a$'' of the function $\Delta_{a}(t-t')$ indicates
that this is the vacuum fluctuations originated from the electric
field $\hat{E}_{a}$ with the quadrature $\hat{a}(\omega)$.


We note that the function $\Delta_{a}(t-t')$ has the ultra-violet
divergence.
This divergence is clearly seen from the $\omega$-integration in the
definition (\ref{eq:commutation-relation-E+E-}) of the function
$\Delta_{a}(t-t')$ and is famous one which comes from the infinite sum
of the vacuum fluctuations.
However, in the actual measurement of the time sequence of the
variables, the time in a measurement is discrete with a finite time
bin.
Namely, the time in a measurement has the minimal time interval.
This time interval is adjusted in experiments so that the time
scale of the interest is sufficiently resolved and it gives the
maximum frequency $\omega_{\max}$ which becomes natural ultra-violet
cut off of the frequency in the obtained data.
Incidentally, in the actual measurement of the time sequence of the
variables, the whole measurement time is also finite and this
finiteness of the whole measurement time gives the minimum frequency
$\omega_{min}$ which corresponds to a natural infra-red cut off in
frequency.
Therefore, we may regard that the integration range over $\omega$ in
the definition of the function $\Delta_{a}(t-t')$ in
Eq.~(\ref{eq:commutation-relation-E+E-}) as
$[\omega_{\min},\omega_{\max}]$ instead of $[0,+\infty]$.
For this reason, throughout this paper, we do not regard the
divergence in the definition of the function $\Delta_{a}(t-t')$ as a
serious one.


Finally, we introduce the vacuum state $|0\rangle_{a}$ and the
coherent state $|\alpha\rangle_{a}$ of the electric field whose
complex amplitude is $\alpha(\omega)$.
As usual, we introduce the vacuum state $|0\rangle_{a}$ through the
operation of annihilation operator $\hat{a}(\omega)$ as
\begin{eqnarray}
  \label{eq:vacuum_state_a_def}
  \hat{a}(\omega)|0\rangle_{a} = 0.
\end{eqnarray}
We also introduce the coherent state $|\alpha\rangle_{a}$
associate with the annihilation operator $\hat{a}(\omega)$ as an eigen
state of the operator $\hat{a}(\omega)$ as
\begin{eqnarray}
  \label{eq:coherent-state-def-a}
  \hat{a}(\omega)|\alpha\rangle_{a} = \alpha(\omega) |\alpha\rangle_{a}.
\end{eqnarray}
Here, we note that the dimension of the complex amplitude
$\alpha(\omega)$ is Hz$^{-1/2}$.
Incidentally, we also note this coherent state is theoretically
produced by the operation of the displacement operator
$\hat{D}[\alpha]$ from the vacuum state $|0\rangle_{a}$ defined by
Eq.~(\ref{eq:vacuum_state_a_def}) as follows:
\begin{eqnarray}
  \label{eq:coherent-state-and-vacuum-state}
  |\alpha\rangle_{a}
  =
  \hat{D}[\alpha] |0\rangle_{a}
  =:
  \exp\left[
  \int_{0}^{\infty} \frac{d\omega}{2\pi} \left(
  \alpha(\omega) \hat{a}^{\dagger}(\omega)
  -
  \alpha^{*}(\omega) \hat{a}(\omega)
  \right)
  \right]
  |0\rangle_{a}
  .
\end{eqnarray}
Here, we note that the subscriptions ``$a$'' in the states
$|0\rangle_{a}$ and $|\alpha\rangle_{a}$ indicates that these states
are associated with the electric field operator $\hat{E}_{a}(t)$ with the
quadrature $\hat{a}(\omega)$.


In the time domain operators $\hat{E}_{a}^{(\pm)}(t)$, the definition
(\ref{eq:vacuum_state_a_def}) of the vacuum state is given by
\begin{eqnarray}
  \label{eq:vacuum_state_a_def-time-domain}
  \hat{E}_{a}^{(+)}(t)|0\rangle_{a} = 0.
\end{eqnarray}
On the other hand, the definition (\ref{eq:coherent-state-def-a}) of
the coherent state is also given by
\begin{eqnarray}
  \label{eq:coherent-state-def-a-time-domain}
  \hat{E}_{a}^{(+)}(t)|\alpha\rangle_{a} =
  \sqrt{\frac{2\pi\hbar}{{\cal A}c}} \alpha(t) |\alpha\rangle_{a},
\end{eqnarray}
where
\begin{eqnarray}
  \label{eq:alphat-alphaomega-relation}
  \alpha(t)
  :=
  \int_{0}^{\infty} \frac{d\omega}{2\pi}
  \sqrt{|\omega|}
  \alpha(\omega)
  e^{-i\omega t}
  .
\end{eqnarray}
These properties
(\ref{eq:vacuum_state_a_def-time-domain})--(\ref{eq:alphat-alphaomega-relation})
are often used in arguments of the homodyne detections below.


\subsection{Multi-mode number and power operators}
\label{sec:Multi-mode_number_and_power_operators}


It is often said that the operator
(\ref{eq:multi-mode-photon-number-propto}) is regarded as the ``photon
number'' in the multi-mode photon system from Glauber's pioneer
papers~\cite{R.J.Glauber-1963-130,R.J.Glauber-1963-131}.
One of reasons for regarding the operator
(\ref{eq:multi-mode-photon-number-propto}) as the photon number is the
fact that the superposition of the electric field operator is possible
within the field equation, i.e., the Maxwell equations, while the
superposition of the operator $\hat{n}(\omega)$ defined by
Eq.~(\ref{eq:each-mode-photon-number-def}) is not possible within the
field equations.
Therefore, it is said that the natural extension of the photon number
to the multi-mode case is the operator given by
Eq.~(\ref{eq:multi-mode-photon-number-propto}).
Furthermore, Glauber~\cite{R.J.Glauber-1963-130} discussed this
operator through the state transition.
The matrix element for the transition from the initial state
$|i\rangle$ to a final state $|f\rangle$ in which one photon has been
absorbed is given by $\langle f|\hat{E}^{(+)}(r,t)|i\rangle$.
The probability per unit time that a photon be absorbed by an ideal
detector is at point ``$r$'' at time ``$t$'' is proportional to
\begin{eqnarray}
  \label{eq:sum_f_fE+rti2-Glauber}
  \sum_{f} \left|\langle f|\hat{E}^{(+)}(r,t)|i\rangle\right|^{2}
  =
  \sum_{f} \langle i|\hat{E}^{(-)}(r,t)|f\rangle \langle f|\hat{E}^{(+)}(r,t)|i\rangle
  =
  \langle i|\hat{E}^{(-)}(r,t)\hat{E}^{(+)}(r,t)|i\rangle.
\end{eqnarray}
The vacuum condition (\ref{eq:vacuum_state_a_def-time-domain}) implies
that the rate at which photons are detected in the empty, or vacuum,
state vanishes~\cite{R.J.Glauber-1963-130}.


Although the above Glauber's argument seems to be plausible, the
contributions of the vacuum fluctuations are not clear.
On the other hand, it is consensus of the gravitational-wave community
that the vacuum fluctuations are contribute to the noise in
gravitational-wave detectors.
For this reason, in this paper, we examine two models of the
photodetection as two ideal cases.
One is the model in which the directly measured operator at the
photodetectors is the multi-mode photon number defined by
\begin{eqnarray}
  \label{eq:multi-mode-photon-number-def}
  \hat{N}_{a}(t)
  :=
  \frac{\kappa_{n}c}{2\pi\hbar} {\cal A}
  \hat{E}_{a}^{(-}(t)\hat{E}_{a}^{(+)}(t)
  .
\end{eqnarray}
The other is the model in which the photocurrents is proportional to
the power operator $\hat{P}_{a}$ of optical field, which is defined by
\begin{eqnarray}
  \label{eq:multi-mode-power-def}
  \hat{P}_{a}(t)
  :=
  \frac{\kappa_{p}c}{4\pi\hbar} {\cal A}
  \left(\hat{E}_{a}(t)\right)^{2}
  .
\end{eqnarray}
Here, the coefficients $\kappa_{n}$ and $\kappa_{p}$ is a
phenomenological constant whose dimension is [time].
These coefficients include so called ``quantum efficiency''.
However, these are not important within our discussion of this
paper.


\section{Homodyne Detections by Multi-mode Detectors}
\label{sec:Homodyne_Detections_by_multi-mode_detectors}


In this section, we show the quantum field theoretical description of
the homodyne detection motivated by the quantum measurement theory.
In quantum measurement theories, we have to specify the directly
measured operator to describe the measurement process.
As mentioned above, we consider the two measurements.
In Sec.~\ref{sec:Homodyne_Detections_by_photon-number-counting}, we
discuss the measurement process in which the directly measured quantum
operator is Glauber's photon number (\ref{eq:multi-mode-photon-number-def}).
In Sec.~\ref{sec:Homodyne_Detections_by_power-counting}, we discuss
the measurement process in which the directly measured quantum
operator is the power operator (\ref{eq:multi-mode-power-def})
of the optical field.
In this section, we concentrate only on the output expectation value
of the homodyne detection.


\subsection{Homodyne detections by photon-number counting detectors}
\label{sec:Homodyne_Detections_by_photon-number-counting}


Now, we describe the homodyne detections in the case where the
directly measured quantum operator is Glauber's photon
number~(\ref{eq:multi-mode-photon-number-def}).
We also describe the details of the homodyne detections, here.
We start our arguments from the simple homodyne detection in
Sec.~\ref{sec:simple_Homodyne_Detection}.
This simple homodyne detection is extended to the balanced homodyne
detection, which described in
Sec.~\ref{sec:Balanced_Homodyne_Detection}.


\subsubsection{Simple Homodyne Detection}
\label{sec:simple_Homodyne_Detection}

\begin{figure}[ht]
  \centering
  \includegraphics[width=0.7\textwidth]{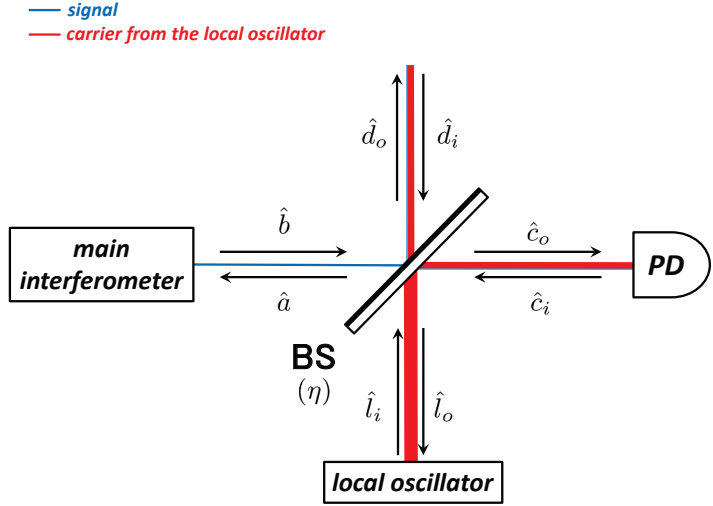}
  \caption{
    Configuration of the interferometer for the simple homodyne
    detection.
    The notations of the quadratures $\hat{a}$, $\hat{b}$,
    $\hat{c}_{o}$, $\hat{c}_{i}$, $\hat{d}_{o}$, $\hat{d}_{i}$,
    $\hat{l}_{o}$, and $\hat{l}_{i}$ are also given in this figure.
  }
  \label{fig:Simple_homodyne_detection}
\end{figure}

Here, we review the simple homodyne detection depicted in
Fig.~\ref{fig:Simple_homodyne_detection}.
In this paper, we want to evaluate the signal in the electric field
$\hat{E}_{b}(t)$.
The electric field from the local oscillator is the coherent state
(\ref{eq:coherent-state-def-a-time-domain}) with the complex amplitude
$\gamma(\omega)$.
The output signal field $\hat{E}_{b}(t)$ and the additional optical
field from the local oscillator is mixed through the beam splitter
with the transmissivity $\eta$.
In the ideal case, this transmissivity is $\eta=1/2$.
However, in this paper, we dare to denote this transmissivity of the
beam splitter for the homodyne detection by $\eta$ as a simple model
of the imperfection of the interferometer.


As introduced in Sec.~\ref{sec:Basic_Notation}, the output electric
field $\hat{E}_{b}$ from the main interferometer is given by
\begin{eqnarray}
  \label{eq:DBHD_20180805_2.7}
  \hat{E}_{b}(t)
  =
  \hat{E}_{b}^{(+)}(t)
  +
  \hat{E}_{b}^{(-)}(t)
\end{eqnarray}
and the electric field from the local oscillator is given by
\begin{eqnarray}
  \label{eq:DBHD_20180805_2.8}
  \hat{E}_{l_{i}}(t)
  =
  \hat{E}_{l_{i}}^{(+)}(t)
  +
  \hat{E}_{l_{i}}^{(-)}(t)
  .
\end{eqnarray}
Furthermore, the electric field to be detected through the
photodetector is given by
\begin{eqnarray}
  \label{eq:DBHD_20180805_2.9}
  \hat{E}_{c_{o}}(t)
  =
  \hat{E}_{c_{o}}^{(+)}(t)
  +
  \hat{E}_{c_{o}}^{(-)}(t)
  .
\end{eqnarray}


At the beam splitter, the signal electric field $\hat{E}_{b}(t)$ and
the electric field $\hat{E}_{l_{i}}(t)$ are mixed and the beam
splitter output the electric field $\hat{E}_{c_{o}}(t)$ to the one of
the ports.
These fields are related through the relation
\begin{eqnarray}
  \label{eq:DBHD_20180805_2.10}
  \hat{E}_{c_{o}}(t) = \sqrt{\eta} \hat{E}_{b}(t) + \sqrt{1-\eta} \hat{E}_{l_{i}}(t).
\end{eqnarray}
Next, we assign the states for the independent electric field
described in Fig.~\ref{fig:Simple_homodyne_detection}.
As mentioned above, the electric field $\hat{E}_{l_{i}}$ from the
local oscillator is in the coherent state $|\gamma\rangle_{l_{i}}$
(\ref{eq:coherent-state-def-a-time-domain}) with the complex amplitude
$\gamma(\omega)$ in the frequency domain, or equivalently,
\begin{eqnarray}
  \label{eq:gammat-gammaomega-relation}
  \gamma(t)
  :=
  \int_{0}^{\infty} \frac{d\omega}{2\pi}
  \sqrt{|\omega|}
  \gamma(\omega)
  e^{-i\omega t}
  .
\end{eqnarray}
in the time domain as Eq.~(\ref{eq:alphat-alphaomega-relation}).
In addition to this state, the electric fields $\hat{E}_{d_{i}}(t)$
and $\hat{E}_{c_{i}}(t)$ are in their vacua, respectively.
The junction condition for the electric fields at the beam splitter is
given by
\begin{eqnarray}
  \label{eq:DBHD_20180805_2.12}
  \hat{E}_{a}(t)
  =
  \sqrt{\eta} \hat{E}_{c_{i}}(t) - \sqrt{1-\eta} \hat{E}_{d_{i}}(t)
  .
\end{eqnarray}
Due to the relation (\ref{eq:DBHD_20180805_2.12}), the state
associated with the quadrature $\hat{a}$ is described by the vacuum
states for the quadratures $\hat{d}_{i}$ and $\hat{c}_{i}$.


Usually, the state associated with the quadrature $\hat{b}$ depends on
the state of the input field $\hat{E}_{a}$ into the main
interferometer and the other optical fields which inject to the main
interferometer~\cite{H.J.Kimble-Y.Levin-A.B.Matsko-K.S.Thorne-S.P.Vyatchanin-2001}.
Furthermore, in this paper, we consider the situation where the output
electric field $\hat{E}_{b}$ includes the information of classical
forces as in Eq.~(\ref{eq:DBHD_20180805_1.2}) and this information are
measured through the expectation value of the operator $\hat{E}_{b}$.


To evaluate the expectation value of the signal field $\hat{E}_{b}$,
we have to specify the state of the total system.
Here, we assume that this state of the total system is given by
\begin{eqnarray}
  \label{eq:DBHD_20180805_2.14}
  |\Psi\rangle
  =
  |\gamma\rangle_{i_{i}}\otimes|0\rangle_{c_{i}}\otimes|0\rangle_{d_{i}}\otimes|\psi\rangle_{main},
\end{eqnarray}
where the state $|\psi\rangle_{main}$ is the state for the electric
fields associated with the main interferometer, which is independent
of the state $|\gamma\rangle_{l_{i}}$, $|0\rangle_{c_{i}}$, and
$|0\rangle_{d_{i}}$.
As noted above, the signal field $\hat{E}_{b}$ may depends on the
input field $\hat{E}_{a}$, and this input field $\hat{E}_{a}$ is
related to the quadratures $\hat{E}_{c_{i}}$ and  $\hat{E}_{d_{i}}$
through Eq.~(\ref{eq:DBHD_20180805_2.12}).
Therefore, strictly speaking, the expectation value of the signal
field $\hat{E}_{b}$ means that
\begin{eqnarray}
  \label{eq:DBHD_20180805_2.15}
  \langle\hat{E}_{b}(t)\rangle
  =
  \langle\psi|_{main}\otimes\langle 0|_{d_{i}}\otimes\langle 0|_{c_{i}}
  \hat{E}_{b}(t)
  |0\rangle_{c_{i}}\otimes|0\rangle_{d_{i}}\otimes|\psi\rangle_{main},
\end{eqnarray}
but we denote simply
\begin{eqnarray}
  \label{eq:DBHD_20180805_2.16}
  \langle\hat{E}_{b}\rangle
  :=
  \langle\Psi|\hat{E}_{b}|\Psi\rangle.
\end{eqnarray}


In this section, we specify the directly measured operator of the
photodetector is the multi-mode photon number operator defined by
Eq.~(\ref{eq:multi-mode-photon-number-def}) associated with the
optical field which enters the photodetector.
In the case of the simple homodyne detection depicted in
Fig.~\ref{fig:Simple_homodyne_detection}, the direct observable which
measured at the photodetector (P.D. in
Fig.~\ref{fig:Simple_homodyne_detection}) is
\begin{eqnarray}
  \label{eq:simple-homodyne-direct-observable-number}
  \hat{N}_{c_{o}}(t)
  :=
  \frac{\kappa_{n}c}{2\pi\hbar} {\cal A}
  \hat{E}_{c_{o}}^{(-)}(t)\hat{E}_{c_{o}}^{(+)}(t)
  .
\end{eqnarray}
Through the photodetector, we measure the time sequence of the
operator $\hat{N}_{c_{o}}(t)$ and we can perform the Fourier
transformation of this time sequence as
\begin{eqnarray}
  \label{eq:simple-homodyne-direct-observable-number-Fourier}
  \hat{{\cal N}}_{c_{o}}(\omega)
  :=
  \int_{-\infty}^{+\infty} dt \hat{N}_{c_{o}}(t) e^{+i\omega t}
  .
\end{eqnarray}


Substituting Eq.~(\ref{eq:DBHD_20180805_2.10}) into
Eq.~(\ref{eq:simple-homodyne-direct-observable-number}), we obtain
\begin{eqnarray}
  \hat{N}_{c_{o}}(t)
  &=&
      \eta \hat{N}_{b}(t)
      +
      \sqrt{\eta(1-\eta)} \kappa_{n} \frac{{\cal A}c}{2\pi\hbar}
      \left(
      \hat{E}_{l_{i}}^{(-)}(t) \hat{E}_{b}^{(+)}(t)
      +
      \hat{E}_{b}^{(-)}(t) \hat{E}_{l_{i}}^{(+)}(t)
      \right)
      \nonumber\\
  &&
     +
     (1-\eta) \hat{N}_{l_{i}}(t)
     .
     \label{eq:simple-homodyne-direct-observable-number-b-li}
\end{eqnarray}
The expectation value of the P.D. output field
$\langle\hat{N}_{c_{o}}(t)\rangle$ under the state
(\ref{eq:DBHD_20180805_2.14}) is given by
\begin{eqnarray}
  \label{eq:simple-homodyne-direct-observable-number-b-li-exp}
  \left\langle\hat{N}_{c_{o}}(t)\right\rangle
  &=&
      \eta \left\langle\hat{N}_{b}(t)\right\rangle
      +
      \sqrt{\eta(1-\eta)} \kappa_{n} \frac{{\cal A}c}{2\pi\hbar}
      \left\langle
      \gamma^{*}(t) \hat{E}_{b}^{(+)}(t)
      +
      \gamma(t) \hat{E}_{b}^{(-)}(t)
      \right\rangle
      \nonumber\\
  &&
     +
     (1-\eta) \kappa_{n} |\gamma(t)|^{2}
     .
\end{eqnarray}


The second term in
Eq.~(\ref{eq:simple-homodyne-direct-observable-number-b-li-exp}) is
the linear combination of the electric field $\hat{E}_{b}$ from the
main interferometer.
Here, we consider the case of the monochromatic local oscillator case,
in which the complex amplitude $\gamma(\omega)$ of the coherent state
from the local oscillator is given by
\begin{eqnarray}
  \label{eq:monochromatic_gamma_omega_is_delta}
  \gamma(\omega)
  =
  2\pi \gamma \delta(\omega-\omega_{0})
  , \quad \gamma\in\CF, \quad \omega_{0}>0
  .
\end{eqnarray}
In this case, the Fourier transformation of this second term in
Eq.~(\ref{eq:simple-homodyne-direct-observable-number-b-li-exp}) is
given by
\begin{eqnarray}
  &&
     \int_{-\infty}^{+\infty} dt e^{+i\omega t}
     \sqrt{\eta(1-\eta)} \kappa_{n} \frac{{\cal A}c}{2\pi\hbar}
     \left\langle
     \gamma^{*}(t) \hat{E}_{b}^{(+)}(t)
     +
     \gamma(t) \hat{E}_{b}^{(-)}(t)
     \right\rangle
     \nonumber\\
  &=&
      \sqrt{\eta(1-\eta)} \kappa_{n}
      \left\langle
      \Theta(\omega_{0}+\omega) \gamma^{*}
      \sqrt{\omega_{0}(\omega_{0}+\omega)} \hat{b}(\omega_{0}+\omega)
      \right.
      \nonumber\\
  && \quad\quad\quad\quad\quad\quad\quad
     \left.
     +
     \Theta(\omega_{0}-\omega) \gamma
     \sqrt{\omega_{0}(\omega_{0}-\omega)} \hat{b}^{\dagger}(\omega_{0}-\omega)
     \right\rangle
     .
      \label{eq:middle-simple-homodyne-direct-observable-number-b-li-exp-tmp1}
\end{eqnarray}
Here, we denote the complex function $\gamma$ as
\begin{eqnarray}
  \label{eq:gamma_is_absgamma_with_phase}
  \gamma =: |\gamma| e^{+i\theta},
\end{eqnarray}
and consider the situation where $\omega_{0}\gg\omega>0$.
In this case,
Eq.~(\ref{eq:middle-simple-homodyne-direct-observable-number-b-li-exp-tmp1})
is given by
\begin{eqnarray}
  &&
     \int_{-\infty}^{+\infty} dt e^{+i\omega t}
     \sqrt{\eta(1-\eta)} \kappa_{n} \frac{{\cal A}c}{2\pi\hbar}
     \left\langle
     \gamma^{*}(t) \hat{E}_{b}^{(+)}(t)
     +
     \gamma(t) \hat{E}_{b}^{(-)}(t)
     \right\rangle
     \nonumber\\
  &\sim&
         \sqrt{\eta(1-\eta)} \kappa_{n}
         \omega_{0}
         |\gamma|
         \left\langle
         e^{-i\theta}
         \hat{b}(\omega_{0}+\omega)
         +
         e^{+i\theta}
         \hat{b}^{\dagger}(\omega_{0}-\omega)
         \right\rangle
         .
         \label{eq:middle-simple-homodyne-direct-observable-number-b-li-exp-tmp2}
\end{eqnarray}
We note that, at this moment, $\omega_{0}$ is just the central
frequency of the coherent state from the local oscillator and have
nothing to do with the central frequency of the signal field
$\hat{E}_{b}(t)$ from the main interferometer.
Therefore,
Eq.~(\ref{eq:middle-simple-homodyne-direct-observable-number-b-li-exp-tmp2})
is still valid even in the case ``{\it heterodyne detection}'' in which
the central frequency of the coherent state from the local oscillator
may not coincide with the central frequency of the signal field
$\hat{E}_{b}(t)$.


Now, we choose $\omega_{0}$ so that this frequency coincides with the
central frequency of the signal field $\hat{E}_{b}(t)$.
This choice is the ``{\it homodyne detection}''.
In this choice, we may identify the quadratures
$\hat{b}(\omega_{0}+\omega)$ and $\hat{b}(\omega_{0}-\omega)$ with the
upper- and lower-sideband quadratures $\hat{b}_{+}(\omega)$ and
$\hat{b}_{-}(\omega)$ in the two-photon formulation~\cite{C.M.Caves-B.L.Schumaker-1985,B.L.Schumaker-C.M.Caves-1985},
respectively.
Therefore, we may introduce the amplitude quadrature
$\hat{b}_{1}(\omega)$ and the phase quadrature $\hat{b}_{2}(\omega)$
by
\begin{eqnarray}
  \label{eq:hatb1-hatb2-def}
  \hat{b}_{1}
  :=
  \frac{1}{\sqrt{2}}\left(\hat{b}_{+}+\hat{b}_{-}^{\dagger}\right)
  , \quad
  \hat{b}_{2}
  :=
  \frac{1}{\sqrt{2}i}\left(\hat{b}_{+}-\hat{b}_{-}^{\dagger}\right)
  .
\end{eqnarray}
In terms of these quadratures $\hat{b}_{1,2}(\omega)$
Eq.~(\ref{eq:middle-simple-homodyne-direct-observable-number-b-li-exp-tmp2})
is given by
\begin{eqnarray}
  &&
     \int_{-\infty}^{+\infty} dt e^{+i\omega t}
     \sqrt{\eta(1-\eta)} \kappa_{n} \frac{{\cal A}c}{2\pi\hbar}
     \left\langle
     \gamma^{*}(t) \hat{E}_{b}^{(+)}(t)
     +
     \gamma(t) \hat{E}_{b}^{(-)}(t)
     \right\rangle
     \nonumber\\
  &\sim&
         \sqrt{2\eta(1-\eta)} \kappa_{n}
         \omega_{0}
         |\gamma|
         \left\langle
         \hat{b}_{\theta}(\omega)
         \right\rangle
         ,
         \label{eq:middle-simple-homodyne-direct-observable-number-b-li-exp-final}
\end{eqnarray}
where the operator $\hat{b}_{\theta}(\omega)$ is defined by
Eq.~(\ref{eq:DBHD_20180805_1.1}) through the definitions
(\ref{eq:hatb1-hatb2-def}) of the amplitude, and the phase quadratures
$\hat{b}_{1}(\omega)$ and $\hat{b}_{2}(\omega)$.
Thus, the middle term in
Eq.~(\ref{eq:simple-homodyne-direct-observable-number-b-li-exp})
yields the expectation value of the operator
$\hat{b}_{\theta}(\omega)$ in the two-photon
formulation~\cite{H.J.Kimble-Y.Levin-A.B.Matsko-K.S.Thorne-S.P.Vyatchanin-2001}.


On the other hand, the Fourier transformation of the final term in
Eq.~(\ref{eq:simple-homodyne-direct-observable-number-b-li-exp}) in
the monochromatic local oscillator case
(\ref{eq:monochromatic_gamma_omega_is_delta}), we obtain
\begin{eqnarray}
  (1-\eta)
  \kappa_{n}
  \int_{-\infty}^{+\infty} dt e^{+i\omega t}
  |\gamma(t)|^{2}
  =
  (1-\eta) \kappa_{n}
  2 \pi \omega_{0} |\gamma|^{2} \delta(\omega)
  .
  \label{eq:final-simple-homodyne-direct-observable-number-b-li-exp-final}
\end{eqnarray}
Then, the Fourier transformation of the expectation value
(\ref{eq:simple-homodyne-direct-observable-number-b-li-exp}) is given by
\begin{eqnarray}
  \label{eq:simple-homodyne-direct-observable-number-b-li-exp-Fourier}
  \left\langle\hat{{\cal N}}_{c_{o}}(\omega)\right\rangle
  \sim
  \eta \left\langle\hat{N}_{b}(\omega)\right\rangle
  +
  \sqrt{\eta(1-\eta)} \kappa_{n} \omega_{0} |\gamma|
  \left\langle
  \hat{b}_{\theta}(\omega)
  \right\rangle
  +
  (1-\eta) \kappa_{n} 2 \pi \omega_{0} |\gamma|^{2} \delta(\omega)
  .
\end{eqnarray}
Since the final term in
Eq.~(\ref{eq:simple-homodyne-direct-observable-number-b-li-exp-Fourier})
is classically predictable, we can eliminate this term from the data,
or we may ignore the data at the frequency $\omega=0$.
In any case, the final term is not important, though this term is the
first dominant term in
Eq.~(\ref{eq:simple-homodyne-direct-observable-number-b-li-exp-Fourier})
in the case where $|\gamma|$ is sufficiently large.
In the same case, i.e., the case where $|\gamma|$ is sufficiently
large, the first term in
Eq.~(\ref{eq:simple-homodyne-direct-observable-number-b-li-exp-Fourier})
is negligible.
Then, we can measure the middle term in
Eq.~(\ref{eq:simple-homodyne-direct-observable-number-b-li-exp-Fourier}).
Thus, we may say that the expectation value
$\langle\hat{b}_{\theta}(\omega)\rangle$ can be measured through the
simple homodyne detection, though the first term in
Eq.~(\ref{eq:simple-homodyne-direct-observable-number-b-li-exp-Fourier})
becomes a noise in the expectation value.


\subsubsection{Balanced Homodyne Detection}
\label{sec:Balanced_Homodyne_Detection}

\begin{figure}[ht]
  \centering
  \includegraphics[width=0.7\textwidth]{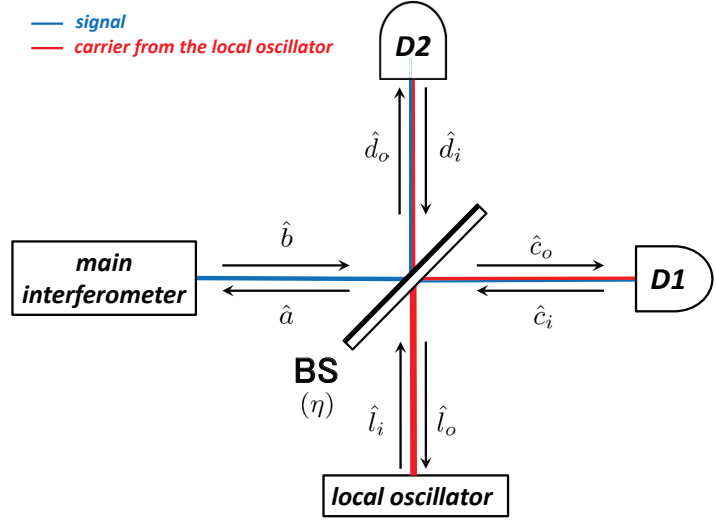}
  \caption{
    Configuration of the interferometer for the balanced homodyne
    detection.
    This notation of the quadratures are same as those in
    Fig.~\ref{fig:Simple_homodyne_detection}.
  }
  \label{fig:Balanced_homodyne_detection}
\end{figure}

In the above simple homodyne detection, the term
$\eta\langle\hat{{\cal N}}_{b}(\omega)\rangle$ in
Eq.~(\ref{eq:simple-homodyne-direct-observable-number-b-li-exp-Fourier})
becomes a noise in expectation value
$\langle\hat{{\cal N}}_{c_{o}}(\omega)\rangle$ if we want
to measure the expectation value
$\langle\hat{b}_{\theta}(\omega)\rangle$.
This noise can be eliminated through the balanced homodyne detection
depicted in Fig.~\ref{fig:Balanced_homodyne_detection}.


As depicted in Fig.~\ref{fig:Balanced_homodyne_detection}, in the
balanced homodyne detection, we detect the optical field from another
port of the beam splitter (BS) in addition to the interferometer setup
of the simple homodyne detection depicted in
Fig.~\ref{fig:Simple_homodyne_detection}.
We denote the photodetectors D1 and D2 as in
Fig.~\ref{fig:Balanced_homodyne_detection}.


At D2, we detect the operator
\begin{eqnarray}
  \label{eq:balanced-homodyne-direct-observable-number-D2}
  \hat{N}_{d_{o}}(t)
  :=
  \frac{\kappa_{n}c}{2\pi\hbar} {\cal A}
  \hat{E}_{d_{o}}^{(-)}(t) \hat{E}_{d_{o}}^{(+)}(t)
  , \quad
  \hat{{\cal N}}(\omega)
  :=
  \int_{-\infty}^{+\infty} dt \hat{N}_{d_{o}}(t) e^{+i\omega t}
  .
\end{eqnarray}
At the BS, the electric fields from the main interferometer and the
local oscillator are mixed as
\begin{eqnarray}
  \label{eq:hatEdo-hatEb-hatEli-beamsplitter-relation-BHD}
  \hat{E}_{d_{o}}(t)
  =
  - \sqrt{1-\eta} \hat{E}_{l_{i}}(t)
  + \sqrt{\eta} \hat{E}_{b}(t)
  .
\end{eqnarray}
and the detected operator $\hat{N}_{d_{o}}(t)$ is given by
\begin{eqnarray}
  \hat{N}_{d_{o}}(t)
  &=&
      (1-\eta) \hat{N}_{b}(t)
      -
      \sqrt{\eta(1-\eta)} \kappa_{n} \frac{{\cal A}c}{2\pi\hbar}
      \left(
      \hat{E}_{l_{i}}^{(-)}(t) \hat{E}_{b}^{(+)}(t)
      +
      \hat{E}_{b}^{(-)}(t) \hat{E}_{l_{i}}^{(+)}(t)
      \right)
      \nonumber\\
  &&
     +
     \eta \hat{N}_{l_{i}}(t)
     .
     \label{eq:D2-balanced-homodyne-direct-observable-number-b-li}
\end{eqnarray}
The expectation value of this operator under the state
(\ref{eq:DBHD_20180805_2.14}) is given by
\begin{eqnarray}
  \left\langle\hat{N}_{d_{o}}(t)\right\rangle
  &=&
      (1-\eta) \left\langle\hat{N}_{b}(t)\right\rangle
      -
      \sqrt{\eta(1-\eta)} \kappa_{n} \frac{{\cal A}c}{2\pi\hbar}
      \left\langle
      \gamma^{*}(t) \hat{E}_{b}^{(+)}(t)
      +
      \gamma(t) \hat{E}_{b}^{(-)}(t)
      \right\rangle
      \nonumber\\
  &&
     +
     \eta |\gamma(t)|^{2}
     .
     \label{eq:D2-balanced-homodyne-direct-observable-number-b-li-exp}
\end{eqnarray}


From the expectation values
(\ref{eq:simple-homodyne-direct-observable-number-b-li-exp}) and
(\ref{eq:D2-balanced-homodyne-direct-observable-number-b-li-exp}), we
can eliminate the term $\langle\hat{N}_{b}(t)\rangle$ which
was a noise term in the expectation value
(\ref{eq:simple-homodyne-direct-observable-number-b-li-exp}) of the
simple homodyne detection.
This elimination is accomplished by the combination
\begin{eqnarray}
  \frac{1}{\kappa_{n}\sqrt{\eta(1-\eta)}}
  \left\langle
  (1-\eta) \hat{N}_{c_{o}}(t)
  -
  \eta \hat{N}_{d_{o}}(t)
  \right\rangle
  &=&
      \frac{{\cal A}c}{2\pi\hbar}
      \left\langle
      \gamma^{*}(t) \hat{E}_{b}^{(+)}(t)
      +
      \gamma(t) \hat{E}_{b}^{(-)}(t)
      \right\rangle
      \nonumber\\
  &&
     +
     \frac{1-2\eta}{\sqrt{\eta(1-\eta)}} |\gamma(t)|^{2}
     .
     \label{eq:balanced-homodyne-direct-observable-number-balanced-exp}
\end{eqnarray}
From this expectation value, we define the signal operator
$\hat{s}_{N}(t)$ as
\begin{eqnarray}
  \hat{s}_{N}(t)
  &:=&
       \frac{1}{\kappa_{n}\sqrt{\eta(1-\eta)}}
       \left[
       (1-\eta) \hat{N}_{c_{o}}(t)
       - \eta \hat{N}_{d_{o}}(t)
       \right]
       -
       \frac{1 -2 \eta}{\sqrt{\eta(1-\eta)}}
       |\gamma(t)|^{2}
       \label{eq:BHD-signal-operator-number-time-domain-def}
\end{eqnarray}
so that
\begin{eqnarray}
  \langle \hat{s}_{N}(t) \rangle
  &=&
      \sqrt{\frac{{\cal A}c}{2\pi\hbar}}
      \left[
      \gamma^{*}(t)
      \left\langle \hat{E}_{b}^{(+)}(t) \right\rangle
      +
      \gamma(t)
      \left\langle \hat{E}_{b}^{(-)}(t) \right\rangle
      \right]
      .
      \label{eq:sNt-exp-valu}
\end{eqnarray}
We note that $\hat{s}_{N}(t)$ is self-adjoint, i.e.,
$\hat{s}_{N}^{\dagger}(t)=\hat{s}_{N}(t)$.
In
Eq.~(\ref{eq:middle-simple-homodyne-direct-observable-number-b-li-exp-final}),
we have already shown that the Fourier transformation of the
expectation value (\ref{eq:sNt-exp-valu}) is proportional to the
expectation value $\langle\hat{b}_{\theta}(\omega)\rangle$.
Actually, in the case where the local oscillator is monochromatic with
the central frequency $\omega_{0}$ and the situation
$\omega_{0}\gg\omega>0$, we conclude that
\begin{eqnarray}
  \frac{1}{\sqrt{2}\omega_{0}|\gamma|}
  \left\langle\hat{s}_{N}(\omega)\right\rangle
  \sim
  \left\langle
  \hat{b}_{\theta}(\omega)
  \right\rangle
  .
  \label{eq:hatsnomega-expvalue-delta-gamma-sideband-picture-result}
\end{eqnarray}
From the view point of quantum measurement theories, we may regard
that the operator $\hat{s}_{N}(t)$ is the signal operator to be
measured in the measurement process of the balanced homodyne detection
through the measurements of Glauber's photon number
$\hat{N}_{c_{o}}(t)$ and $ \hat{N}_{d_{o}}(t)$.


\subsection{Balanced Homodyne Detections by power counting detectors}
\label{sec:Homodyne_Detections_by_power-counting}


In this section, we re-examine the arguments on the balanced homodyne
detection in
Sec.~\ref{sec:Homodyne_Detections_by_photon-number-counting} under the
premise that the direct observable is the power operator
(\ref{eq:multi-mode-power-def}).


At the beam splitter, the signal field $\hat{E}_{b}(t)$ and the field
$\hat{E}_{l_{i}}(t)$ from the local oscillator is mixed through the
conditions (\ref{eq:DBHD_20180805_2.10}) and
(\ref{eq:hatEdo-hatEb-hatEli-beamsplitter-relation-BHD}).
Through these conditions and the definition
(\ref{eq:multi-mode-power-def}) of the operator $\tilde{P}_{a}(t)$,
the power operators $\hat{P}_{c_{l}}(t)$ at the D1 and
$\hat{P}_{d_{o}}(t)$ at D2 are given by
\begin{eqnarray}
  \hat{P}_{c_{o}}(t)
  &=&
      \eta \hat{P}_{b}(t) + (1-\eta) P_{l_{i}}(t)
      \nonumber\\
  &&
     +
     \sqrt{\eta(1-\eta)} \kappa_{p} \frac{{\cal A}c}{2\pi\hbar} \left(
     \hat{E}_{b}(t) \hat{E}_{l_{i}}(t) + \hat{E}_{l_{i}}(t) \hat{E}_{b}(t)
     \right)
     ,
     \label{eq:balanced-homodyne-direct-observable-power-D1}
  \\
  \hat{P}_{d_{o}}(t)
  &=&
      (1-\eta) \hat{P}_{b}(t) + \eta \hat{P}_{l_{i}}(t)
      \nonumber\\
  &&
      -
      \sqrt{\eta(1-\eta)} \kappa_{p} \frac{{\cal A}c}{2\pi\hbar} \left(
      \hat{E}_{b}(t) \hat{E}_{l_{i}}(t) + \hat{E}_{l_{i}}(t) \hat{E}_{b}(t)
      \right)
      .
      \label{eq:balanced-homodyne-direct-observable-power-D2}
\end{eqnarray}
Here, we note that the expectation value of $\hat{P}_{l_{i}}(t)$ is
given by
\begin{eqnarray}
  \left\langle
  \hat{P}_{l_{i}}(t)
  \right\rangle
  &:=&
       \frac{\kappa_{p}c}{4\pi\hbar} {\cal A}
       \left\langle
       \left(\hat{E}_{l_{i}}(t)\right)^{2}
       \right\rangle
       =
       \kappa_{p} \left[
       (\gamma(t)+\gamma^{*}(t))^{2} + \Delta_{l_{i}}(0)
       \right]
       .
       \label{eq:exp-Plit-pwer}
\end{eqnarray}
Here, $\Delta_{l_{i}}(0)$ is the vacuum fluctuations from the local
oscillator.
Inspecting the arguments in
Sec.~\ref{sec:Balanced_Homodyne_Detection}, we define the signal
operator $\hat{s}_{P}(t)$ by
\begin{eqnarray}
  \hat{s}_{P}(t)
  &:=&
       \frac{1}{2\kappa_{p}\sqrt{\eta(1-\eta)}} \left[
       (1-\eta) \hat{P}_{c_{o}}(t)
       -
       \eta \hat{P}_{d_{o}}(t)
       \right]
       -
       \frac{1-2\eta}{2\kappa_{p}\sqrt{\eta(1-\eta)}}
       \left\langle
       \hat{P}_{l_{i}}(t)
       \right\rangle
       \label{eq:hatsPt-def}
  \\
  &=&
      \frac{{\cal A}c}{4\pi\hbar} \left[
      \hat{E}_{l_{i}}(t) \hat{E}_{b}(t) + \hat{E}_{b}(t) \hat{E}_{l_{i}}(t)
      \right]
      \nonumber\\
  &&
     +
     \frac{1-2\eta}{2\sqrt{\eta(1-\eta)}} \left[
     \frac{{\cal A}c}{2\pi\hbar} \left(\hat{E}_{l_{i}}(t)\right)^{2}
     -
     \left\{
     (\gamma(t) + \gamma^{*}(t))^{2} + \Delta_{l_{i}}(0)
     \right\}
     \right]
     \label{eq:hatsPt-field-expression}
     .
\end{eqnarray}


Now, we evaluate the expectation value of the signal operator
$\hat{s}_{P}(t)$.
Here, we assume the commutation relation
\begin{eqnarray}
  \label{eq:commutation-Eb-Eli}
  \left[\hat{E}_{b}(t), \hat{E}_{l_{i}}(t)\right] = 0.
\end{eqnarray}
This assumption is justified in the Appendix~\ref{sec:Commutation_relation_Eb_Eli}.
Under the assumption (\ref{eq:commutation-Eb-Eli}), the expectation
value of the signal operator $\hat{s}_{P}(t)$ is given by
\begin{eqnarray}
  \label{eq:hatsPt-exp-valu}
  \left\langle\hat{s}_{P}(t)\right\rangle
  =
  \sqrt{\frac{{\cal A}c}{2\pi\hbar}}
  \left(\gamma(t)+\gamma^{*}(t)\right)
  \left\langle\hat{E}_{b}(t)\right\rangle
  .
\end{eqnarray}
When the monochromatic local oscillator with the central frequency
$\omega_{0}$ and the frequency $\omega$ of interest satisfy the
condition $\omega_{0}\gg\omega>0$, the Fourier transformation of the
expectation value of the operator $\hat{s}_{P}(t)$ is given by
\begin{eqnarray}
  \label{eq:hatsPt-exp-valu-Fourier-def}
  \left\langle\hat{s}_{P}(\omega)\right\rangle
  &:=&
       \int_{-\infty}^{+\infty} dt e^{+i\omega t}
       \left\langle\hat{s}_{P}(t)\right\rangle
  \\
  \label{eq:hatsPt-exp-valu-Fourier-monochro-local-omega0gtromegagrt0}
  &\sim&
         \sqrt{2} \omega_{0} |\gamma|
         \left\langle
         \hat{b}_{\theta}(\omega)
         \right\rangle
         .
\end{eqnarray}
Here, we note that the factor $1/2$ in the definition
(\ref{eq:hatsPt-def}) is chosen so that the expectation value
(\ref{eq:hatsPt-exp-valu-Fourier-monochro-local-omega0gtromegagrt0})
coincides with
Eq.~(\ref{eq:hatsnomega-expvalue-delta-gamma-sideband-picture-result}).
From the view point of quantum measurement theories, we may regard
that the operator $\hat{s}_{P}(t)$ is the signal operator to be
measured in the measurement process of the balanced homodyne detection
through the measurements of the power operators $\hat{P}_{c_{o}}(t)$
and $\hat{P}_{d_{o}}(t)$.


\section{Noise Spectral Densities}
\label{sec:Noise-spectral-densities}


In many literature of the gravitational-wave detection, it is written
that the single sideband noise spectral density
$\bar{S}_{A}^{(s)}(\omega)$ for an arbitrary operator
$\hat{A}(\omega)$ in the frequency domain with the condition
$\langle\hat{A}(\omega)\rangle=0$ in the ``{\it stationary}'' system
is given by Eq.~(\ref{eq:Kimble-noise-spectral-density-single-side})
in the context of the two-photon formulation.
In the two-photon
formulation~\cite{C.M.Caves-B.L.Schumaker-1985,B.L.Schumaker-C.M.Caves-1985},
we consider the sideband fluctuations in the frequency
$\omega_{0}\pm\omega$ with the central frequency $\omega_{0}$ of the
optical field.
The ``single sideband'' means that the noise spectral density is
evaluated only in the frequency range $\omega>0$ of the positive
sideband $\omega_{0}+\omega$ or the range $\omega>0$ of the negative
sideband $\omega_{0}-\omega$.
This is due to implicit assumption of the symmetry of the data around
the central frequency $\omega_{0}$.
The frequencies $\omega$ and $\omega'$ in
Eq.~(\ref{eq:Kimble-noise-spectral-density-single-side}) is the
sideband frequencies in the two-photon formulation.


The noise spectral density in which we take into account of the both
sideband $\omega_{0}\pm\omega$ with $\omega\gtrless 0$ is called the
double sideband noise spectral density.
The double sideband noise spectral density
$\bar{S}_{A}^{(d)}(\omega)$ for an arbitrary operator
$\hat{A}(\omega)$ in the frequency domain with the condition
$\langle\hat{A}(\omega)\rangle=0$ is also given by
\begin{eqnarray}
  \label{eq:Kimble-noise-spectral-density-double-side}
  2\pi \delta(\omega-\omega') \bar{S}_{A}^{(d)}(\omega)
  :=
  \frac{1}{2} \langle\mbox{in}|
  \hat{A}(\omega)\hat{A}^{\dagger}(\omega')
  +
  \hat{A}^{\dagger}(\omega')\hat{A}(\omega)
  |\mbox{in}\rangle
  .
\end{eqnarray}
Furthermore, we consider the (double sideband) ``{\it correlation
  spectral density}'' of the observables $\hat{A}(\omega)$ and
$\hat{B}(\omega)$ with the condition
$\langle\hat{A}(\omega)\rangle=\langle\hat{B}(\omega)\rangle=0$ is
given by
\begin{eqnarray}
  \label{eq:Kimble-noise-correlation-spectral-density-double-side}
  2\pi \delta(\omega-\omega') \bar{S}_{AB}^{(d)}(\omega)
  :=
  \frac{1}{2} \langle\mbox{in}|
  \hat{A}(\omega)\hat{B}^{\dagger}(\omega')
  +
  \hat{B}^{\dagger}(\omega')\hat{A}(\omega)
  |\mbox{in}\rangle
  .
\end{eqnarray}


Here, we examine the meaning of the noise spectral density
(\ref{eq:Kimble-noise-spectral-density-double-side}) and noise
correlation spectral density
(\ref{eq:Kimble-noise-correlation-spectral-density-double-side}).
In this paper, we do not explicitly introduce the central frequency
$\omega_{0}$ nor distinguish the upper- or the lower-sideband which
are basis of the two-photon formulation.
We consider the noise-spectral density in the frequency range
$\omega\in[0,+\infty]$.
For this reason, we first consider the double-sideband noise spectral
density
(\ref{eq:Kimble-noise-correlation-spectral-density-double-side}) in
which we take into account both of the upper- and lower-sideband
frequency range.
We also consider the single-sideband noise spectral density
(\ref{eq:Kimble-noise-spectral-density-single-side}), if necessary.


To examine the meaning of the spectral densities
(\ref{eq:Kimble-noise-correlation-spectral-density-double-side}), we
consider the time-domain expression of this formulae for the
correlation spectral density through the Fourier transformation of the
above formulae. Performing the double inverse Fourier transformations
associated with $\omega$ and $\omega'$ in noise correlation spectral
density
(\ref{eq:Kimble-noise-correlation-spectral-density-double-side}) and
introduce the time-domain variables $\hat{A}(t)$ and $\hat{B}(t)$ as
\begin{eqnarray}
  \label{eq:time-domain-A}
  \hat{A}(t)
  &=&
      \int_{-\infty}^{+\infty} \frac{d\omega}{2\pi}
      \left(
      \Theta(\omega) \hat{A}(\omega)
      +
      \Theta(-\omega) \hat{A}^{\dagger}(-\omega)
      \right)
      e^{-i\omega t}
      ,
  \\
  \label{eq:time-domain-B}
  \hat{B}(t')
  &=&
      \int_{-\infty}^{+\infty} \frac{d\omega'}{2\pi}
      \left(
      \Theta(\omega') \hat{B}(\omega')
      +
      \Theta(-\omega') \hat{B}^{\dagger}(-\omega')
      \right)
      e^{-i\omega' t'}
      ,
\end{eqnarray}
Eq.~(\ref{eq:Kimble-noise-correlation-spectral-density-double-side})
yields
\begin{eqnarray}
  \label{eq:Kimble-noise-correlation-function-time-domain}
  \bar{C}_{AB}(t-t')
  =
  \frac{1}{2} \langle\mbox{in}|
  \hat{A}(t)\hat{B}^{\dagger}(t')
  +
  \hat{B}^{\dagger}(t')\hat{A}(t)
  |\mbox{in}\rangle
  ,
\end{eqnarray}
where
\begin{eqnarray}
  \bar{C}_{AB}(t'-t)
  :=
  \int_{-\infty}^{+\infty} \frac{d\omega}{2\pi}
  \bar{S}_{AB}(\omega)
  e^{+i\omega(t'-t)}
  .
\end{eqnarray}
Equivalently, we may represent
Eq.~(\ref{eq:Kimble-noise-correlation-function-time-domain}) and
\begin{eqnarray}
  \label{eq:Kimble-noise-correlation-function-time-domain-tau}
  \bar{C}_{AB}(\tau)
  =
  \frac{1}{2} \langle\mbox{in}|
  \hat{A}(t)\hat{B}^{\dagger}(t+\tau)
  +
  \hat{B}^{\dagger}(t+\tau)\hat{A}(t)
  |\mbox{in}\rangle
  .
\end{eqnarray}


We see that the left-hand side of
Eq.~(\ref{eq:Kimble-noise-correlation-function-time-domain-tau})
depends only on $\tau$, while the right-hand side may depend both
on $t$ and $\tau$.
This dependence implies that the ``{\it stationarity}'' of the
system.
Here, the ``stationarity'' of the correlation means that the
correlation function does not depend on the absolute value of $t$
but depends only on the time-difference $\tau=t'-t$.
If we take into account of the non-stationary cases, the correlation
function may depends on $t$ as
\begin{eqnarray}
  \label{eq:noise-correlation-function-time-domain-t-tau}
  \bar{C}_{AB}(t,\tau)
  =
  \frac{1}{2} \langle\mbox{in}|
  \hat{A}(t)\hat{B}^{\dagger}(t+\tau)
  +
  \hat{B}^{\dagger}(t+\tau)\hat{A}(t)
  |\mbox{in}\rangle
  .
\end{eqnarray}
This is general form of the correlation function.
From the general correlation function
(\ref{eq:noise-correlation-function-time-domain-t-tau}), we can
obtain the correlation function for the stationary noise by the
time-average as
\begin{eqnarray}
  C_{({\rm av})AB}(\tau)
  &:=&
       \lim_{T\rightarrow\infty} \frac{1}{T}
       \int_{-T/2}^{T/2} dt C_{AB}(t,\tau)
       \nonumber\\
  &=&
      \lim_{T\rightarrow\infty} \frac{1}{T}
      \int_{-T/2}^{T/2} dt
      \frac{1}{2}
      \langle\mbox{in}|
      \hat{A}(t+\tau) \hat{B}^{\dagger}(t)
      +
      \hat{B}^{\dagger}(t) \hat{A}(t+\tau)
      |\mbox{in}\rangle
      .
      \label{eq:noise-correlation-function-time-domain-t-tau-time-average}
\end{eqnarray}
We use this expression $C_{({\rm av})AB}(\tau)$ defined by
(\ref{eq:noise-correlation-function-time-domain-t-tau-time-average})
of the correlation function for the stationary noise, instead of
$\bar{C}_{AB}(\tau)$ given by
Eq.~(\ref{eq:Kimble-noise-correlation-function-time-domain-tau}).
When $\hat{A}(t)=\hat{B}(t)$, the auto-correlation function
$C_{({\rm av})A}(\tau)$ for ``stationary noise'' is given by
\begin{eqnarray}
  \label{eq:auto-correlation-function-time-domain-t-tau-time-average}
  C_{({\rm av})A}(\tau)
  :=
  \lim_{T\rightarrow\infty} \frac{1}{T} \int_{-T/2}^{T/2} dt
  \frac{1}{2} \langle\mbox{in}|
  \hat{A}(t+\tau) \hat{A}^{\dagger}(t)
  +
  \hat{A}^{\dagger}(t) \hat{A}(t+\tau)
  |\mbox{in}\rangle
  .
\end{eqnarray}
If the operator $\hat{A}(t)$ is self-adjoint, i.e.,
$\hat{A}^{\dagger}(t)=A(t)$, the auto-correlation function is given by
\begin{eqnarray}
  \label{eq:auto-correlation-time-domain-tau-time-average-SA-A}
  C_{({\rm av})A}(\tau)
  :=
  \lim_{T\rightarrow\infty} \frac{1}{T} \int_{-T/2}^{T/2} dt
  \frac{1}{2} \langle\mbox{in}|
  \hat{A}(t+\tau) \hat{A}(t)
  +
  \hat{A}(t) \hat{A}(t+\tau)
  |\mbox{in}\rangle
  .
\end{eqnarray}
Here, we note that $[\hat{A}(t),\hat{A}(t+\tau)]\neq 0$, in
general.


In the above argument, we only consider the operator $\hat{A}(t)$
whose expectation value vanishes $\langle\hat{A}(t)\rangle = 0$.
When the operator $A(t)$ has a non-trivial expectation value
$\langle\hat{A}(t)\rangle$ $:=$ $\langle\mbox{in}|\hat{A}(t)|\mbox{in}\rangle$
under the state $|\mbox{in}\rangle$, we consider the noise operator
$\hat{A}_{n}(t)$ for the operator $\hat{A}(t)$, which is defined by
\begin{eqnarray}
  \label{eq:noise-operator-def-An}
  \hat{A}(t) =: \hat{A}_{n}(t) + \langle\hat{A}(t)\rangle.
\end{eqnarray}
Furthermore, we can also evaluate the noise correlation function by
\begin{eqnarray}
  C_{({\rm av})A_{n}}(\tau)
  &:=&
       \lim_{T\rightarrow\infty} \frac{1}{T} \int_{-T/2}^{T/2} dt
       \frac{1}{2} \langle\mbox{in}|
       \hat{A}_{n}(t+\tau) \hat{A}_{n}(t)
       +
       \hat{A}_{n}(t) \hat{A}_{n}(t+\tau)
       |\mbox{in}\rangle
       \nonumber\\
  &=:&
       C_{({\rm av})A}(\tau)
       -
       C_{({\rm av},{\rm cl})A}(\tau)
       ,
       \label{eq:auto-correlation-time-domain-tau-time-average-SA-A-nonise}
\end{eqnarray}
where $C_{({\rm av},{\rm cl})A}(\tau)$ is the classical correlation
function defined by
\begin{eqnarray}
  \label{eq:classical-correlation-function-def}
  C_{({\rm av},{\rm cl})A}(\tau)
  :=
  \lim_{T\rightarrow\infty} \frac{1}{T} \int_{-T/2}^{T/2} dt
  \langle\hat{A}(t+\tau)\rangle\langle\hat{A}(t)\rangle
  .
\end{eqnarray}


Thus, the quantum noise correlation function for the operator $A(t)$
is given by
Eq.~(\ref{eq:auto-correlation-time-domain-tau-time-average-SA-A-nonise}),
where $C_{({\rm av})A}(\tau)$ and $C_{({\rm av},{\rm cl})A}(\tau)$ are
defined by
Eq.~(\ref{eq:auto-correlation-time-domain-tau-time-average-SA-A-nonise})
and (\ref{eq:classical-correlation-function-def}), respectively.
The noise spectral density $S_{A_{n}}(\omega)$ is given by the Fourier
transformation of $C_{({\rm av})A_{n}}(\tau)$ as
\begin{eqnarray}
  \label{eq:noise-spectral-density-An-def}
  S_{A_{n}}(\omega)
  :=
  \int_{-\infty}^{+\infty} d\tau
  C_{({\rm av})A_{n}}(\tau)
  e^{+i\omega\tau}
  .
\end{eqnarray}
This is the generalization of the noise spectral density of
Eq.~(\ref{eq:Kimble-noise-spectral-density-double-side}).
In this paper, we evaluate the quantum noise through the noise
spectral density (\ref{eq:noise-spectral-density-An-def}) instead of Eq.~(\ref{eq:Kimble-noise-spectral-density-double-side}).


\section{Estimation of Quantum Noise}
\label{sec:Estimation_of_Quantum_noise}


Following the discussion on the noise spectral density, we evaluate
the quantum noise in homodyne detections.
As discussed above, the noise spectral density
(\ref{eq:noise-spectral-density-An-def}) is not based on the
conventional two-photon formulation and the evaluated noise spectral
density here is beyond the two-photon formulation.
However, we express our results in terms of the two-photon formulation
as much as possible.


As the first case, we evaluate the noise spectral density in the case
that the directly measured quantum operator is Glauber's photon
number (\ref{eq:multi-mode-photon-number-def}) in
Sec.~\ref{sec:Quantum_Noise_in_Balanced_Homodyne_Detections_By_PhotonNumberDetectors}.
In this case, the measured signal operator in the balanced homodyne
detection is given by
Eq.~(\ref{eq:BHD-signal-operator-number-time-domain-def}).
Furthermore, we carefully examine the contributions of the vacuum
fluctuations to the noise spectral density in this case.


As the second case, we evaluate the noise spectral density in the case
where the directly measured quantum operator is the power operator
(\ref{eq:multi-mode-power-def}) in Sec.~\ref{sec:Quantum_Noise_in_Balanced_Homodyne_Detections_By_PowerCountingDetectors}.
In this case, the measured signal operator in the balanced homodyne
detection is defined by Eq.~(\ref{eq:hatsPt-def}).
Although we do not carry out the careful examination of the
contribution of vacuum fluctuations to the noise spectral density in
this case, it will be trivial from the considerations in Sec.~\ref{sec:Quantum_Noise_in_Balanced_Homodyne_Detections_By_PhotonNumberDetectors}.


\subsection{Quantum Noise in Balanced Homodyne Detections by Photon-Number Detectors}
\label{sec:Quantum_Noise_in_Balanced_Homodyne_Detections_By_PhotonNumberDetectors}


Here, we evaluate the noise spectral density for the measurement
operator $\hat{s}_{N}(t)$ defined by
Eq.~(\ref{eq:BHD-signal-operator-number-time-domain-def}).
In terms of the electric fields, the operator $\hat{s}_{N}(t)$ is
given by
\begin{eqnarray}
  \hat{s}_{N}(t)
  &=&
      \frac{{\cal A}c}{2\pi\hbar} \left[
      \hat{E}_{l_{i}}^{(-)}(t) \hat{E}_{b}^{(+)}(t)
      +
      \hat{E}_{b}^{(-)}(t) \hat{E}_{l_{i}}^{(+)}(t)
      \right.
      \nonumber\\
  && \quad\quad\quad
     \left.
     +
     \frac{1-2\eta}{\sqrt{\eta(1-\eta)}} \left(
     \hat{E}_{l_{i}}^{(-)} \hat{E}_{l_{i}}^{(+)}
     -
     \frac{2\pi\hbar}{{\cal A}c} |\gamma(t)|^{2}
     \right)
     \right]
     .
     \label{eq:hatsNt-electric-field}
\end{eqnarray}
To carry out the evaluation of the noise spectral density, it is
convenient to introduce the states $|\hat{s}_{N}(t)\rangle$ and
$\langle\hat{s}_{N}(t)|$, which are defined by
\begin{eqnarray}
  |\hat{s}_{N}(t)\rangle
  &:=&
       \hat{s}_{N}(t)|\Psi\rangle
       \nonumber\\
  &=&
      \sqrt{\frac{{\cal A}c}{2\pi\hbar}}
      \left[
      \sqrt{\frac{{\cal A}c}{2\pi\hbar}} \hat{E}_{l_{i}}^{(-)}(t) \hat{E}_{b}^{(+)}(t)
      + \hat{E}_{b}^{(-)}(t) \gamma(t)
      \right.
      \nonumber\\
  && \quad\quad\quad\quad
     \left.
     + \frac{1-2\eta}{\sqrt{\eta(1-\eta)}} \gamma(t) \left[
     \hat{E}_{l_{i}}^{(-)}(t)
     - \sqrt{\frac{2\pi\hbar}{{\cal A}c}} \gamma^{*}(t)
     \right]
     \right]
     |\Psi\rangle
     ,
     \label{eq:sNt-state-ket}
  \\
  \langle\hat{s}_{N}(t)|
  &:=&
       \langle\Psi|
       \hat{s}_{N}(t)
       \nonumber\\
  &=&
      \sqrt{\frac{{\cal A}c}{2\pi\hbar}}
      \langle\Psi|
      \left[
      \gamma^{*}(t) \hat{E}_{b}^{(+)}(t)
      + \hat{E}_{b}^{(-)}(t) \sqrt{\frac{{\cal A}c}{2\pi\hbar}} \hat{E}_{l_{i}}^{(+)}(t)
      \right.
      \nonumber\\
  && \quad\quad\quad\quad\quad
     \left.
     + \frac{1-2\eta}{\sqrt{\eta(1-\eta)}} \gamma^{*}(t) \left[
     \hat{E}_{l_{i}}^{(+)}(t)
     - \sqrt{\frac{2\pi\hbar}{{\cal A} c}} \gamma(t)
     \right]
     \right]
      .
      \label{eq:sNt-state-bra}
\end{eqnarray}
We also use the states $|\hat{s}_{N}(t+\tau)\rangle$ and
$\langle\hat{s}_{N}(t+\tau)|$, which are given by the replacement
$t\rightarrow t+\tau$ in Eqs.~(\ref{eq:sNt-state-ket}) and
(\ref{eq:sNt-state-bra}), respectively.


We evaluate the noise spectral density $S_{s_{Nn}}(\omega)$ of the
noise operator
\begin{eqnarray}
  \label{eq:sNt-noise-operator}
  \hat{s}_{Nn}(t)
  :=
  \hat{s}_{N}(t)
  -
  \langle\hat{s}_{N}(t)\rangle
  ,
\end{eqnarray}
step by step.
The aim of these steps is to clarify the contributions of vacuum
fluctuations to the noise spectral density.
First, we evaluate the normal-ordered noise spectral density
$S_{S_{Nn}}^{({\rm normal})}(\omega)$, in which all vacuum
fluctuations are neglected in
Sec.~\ref{sec:Normal_ordered_noise_spectral_density};
Second, we evaluate the contribution from the vacuum fluctuations of
the signal field $\hat{E}_{b}(t)$ in
Sec.~\ref{sec:Vacuum_fluctuations_from_main};
Finally, we consider the contribution from the vacuum fluctuations
from the optical field $\hat{E}_{l_{i}}(t)$ from the local
oscillator in
Sec.~\ref{sec:Vacuum_fluctuations_from_local_oscillator}.


\subsubsection{Normal ordered noise spectral density}
\label{sec:Normal_ordered_noise_spectral_density}


To evaluate the normal-ordered noise spectral density
$S_{s_{Nn}}^{({\rm normal})}(\omega)$, we consider the normal-ordered
correlation function
\begin{eqnarray}
  \label{eq:normal-ordered-correlation-function-number-def}
  C_{({\rm av})s_{N}}^{({\rm normal})}(\tau)
  :=
  \lim_{T\rightarrow\infty} \frac{1}{T} \int_{-T/2}^{T/2} dt
  \frac{1}{2}
  \left\langle
  :
  \hat{s}_{N}(t+\tau)\hat{s}_{N}(t)
  +
  \hat{s}_{N}(t)\hat{s}_{N}(t+\tau)
  :
  \right\rangle
  .
\end{eqnarray}
As shown in Sec.~\ref{sec:Balanced_Homodyne_Detection}, the
expectation value of the operator $\hat{s}_{N}$ is given by
Eq.~(\ref{eq:sNt-exp-valu}).
This can also be verified through the states (\ref{eq:sNt-state-ket})
or (\ref{eq:sNt-state-bra}).
Furthermore, from the states (\ref{eq:sNt-state-ket}) and
(\ref{eq:sNt-state-bra}) with the appropriate replacement
$t\rightarrow t+\tau$, we obtain
\begin{eqnarray}
  && \!\!\!\!\!\!\!\!\!\!\!\!\!\!\!\!\!\!\!\!\!\!\!\!\!\!\!\!
     \left\langle
     :
     \hat{s}_{N}(t+\tau) \hat{s}_{N}(t)
     +
     \hat{s}_{N}(t) \hat{s}_{N}(t+\tau)
     :
     \right\rangle
     \nonumber\\
  &=&
      2 \frac{{\cal A}c}{2\pi\hbar} \left\langle\Psi\right|
      \left[
      \gamma^{*}(t+\tau)
      \gamma^{*}(t)
      \hat{E}_{b}^{(+)}(t+\tau)
      \hat{E}_{b}^{(+)}(t)
      \right.
      \nonumber\\
  && \quad\quad\quad\quad\quad
     \left.
     +
     \gamma^{*}(t)
     \gamma(t+\tau)
     \hat{E}_{b}^{(-)}(t+\tau)
     \hat{E}_{b}^{(+)}(t)
     \right.
     \nonumber\\
  && \quad\quad\quad\quad\quad
     \left.
     +
     \gamma^{*}(t+\tau)
     \gamma(t)
     \hat{E}_{b}^{(-)}(t)
     \hat{E}_{b}^{(+)}(t+\tau)
     \right.
     \nonumber\\
  && \quad\quad\quad\quad\quad
     \left.
     +
     \gamma(t+\tau)
     \gamma(t)
     \hat{E}_{b}^{(-)}(t+\tau)
     \hat{E}_{b}^{(-)}(t)
     \right]
     \left|\Psi\right\rangle
     .
     \label{eq:normal-ordered-hatsNt+tauhatsNt+hatsNthatsNt+tau-exp}
\end{eqnarray}
By the subtraction  of the classical part and using
\begin{eqnarray}
  \label{e:noisehatEbn-def}
  \hat{E}_{b_{n}}^{(\pm)}(t)
  :=
  \hat{E}_{b}^{(\pm)}(t)
  -
  \langle\hat{E}_{b}^{(\pm)}(t)\rangle
  ,
\end{eqnarray}
we obtain
\begin{eqnarray}
  && \!\!\!\!\!\!\!\!\!\!\!\!\!\!\!\!\!\!\!\!\!\!\!\!\!\!\!\!
     \frac{1}{2}
     \left\langle
     :
     \hat{s}_{N}(t+\tau) \hat{s}_{N}(t)
     +
     \hat{s}_{N}(t) \hat{s}_{N}(t+\tau)
     :
     \right\rangle
     -
     \langle\hat{s}_{N}(t+\tau)\rangle
     \langle\hat{s}_{N}(t)\rangle
     \nonumber\\
  &=&
      2 \frac{{\cal A}c}{2\pi\hbar} \left\langle\Psi\right|
      \left[
      \gamma^{*}(t+\tau)
      \gamma^{*}(t)
      \hat{E}_{bn}^{(+)}(t+\tau)
      \hat{E}_{bn}^{(+)}(t)
      \right.
      \nonumber\\
  && \quad\quad\quad\quad\quad
     \left.
      +
      \gamma^{*}(t)
      \gamma(t+\tau)
      \hat{E}_{bn}^{(-)}(t+\tau)
      \hat{E}_{bn}^{(+)}(t)
      \right.
      \nonumber\\
  && \quad\quad\quad\quad\quad
     \left.
     +
     \gamma^{*}(t+\tau)
     \gamma(t)
     \hat{E}_{bn}^{(-)}(t)
     \hat{E}_{bn}^{(+)}(t+\tau)
      \right.
      \nonumber\\
  && \quad\quad\quad\quad\quad
     \left.
     +
     \gamma(t+\tau)
     \gamma(t)
     \hat{E}_{bn}^{(-)}(t+\tau)
     \hat{E}_{bn}^{(-)}(t)
     \right]
     \left|\Psi\right\rangle
     .
     \label{eq:normal-ordered-hatsNt+tauhatsNt+hatsNthatsNt+tau-noise-exp}
\end{eqnarray}
Then, the normal-ordered noise correlation function
$C_{({\rm av})s_{Nn}}^{({\rm normal})}(\tau)$ is given by
\begin{eqnarray}
  C_{({\rm av})s_{Nn}}^{({\rm normal})}(\tau)
  &=&
      \frac{{\cal A}c}{2\pi\hbar} \lim_{T\rightarrow\infty}
      \frac{1}{T} \int_{-T/2}^{T/2} dt \left[
      \gamma^{*}(t+\tau)
      \gamma^{*}(t)
      \left\langle
      \hat{E}_{bn}^{(+)}(t+\tau)
      \hat{E}_{bn}^{(+)}(t)
      \right\rangle
      \right.
      \nonumber\\
  && \quad\quad\quad\quad\quad\quad\quad\quad\quad\quad
     \left.
     +
     \gamma^{*}(t)
     \gamma(t+\tau)
     \left\langle
     \hat{E}_{bn}^{(-)}(t+\tau)
     \hat{E}_{bn}^{(+)}(t)
     \right\rangle
     \right.
     \nonumber\\
  && \quad\quad\quad\quad\quad\quad\quad\quad\quad\quad
     \left.
     +
     \gamma^{*}(t+\tau)
     \gamma(t)
     \left\langle
     \hat{E}_{bn}^{(-)}(t)
     \hat{E}_{bn}^{(+)}(t+\tau)
     \right\rangle
     \right.
     \nonumber\\
  && \quad\quad\quad\quad\quad\quad\quad\quad\quad\quad
     \left.
     +
     \gamma(t+\tau)
     \gamma(t)
     \left\langle
     \hat{E}_{bn}^{(-)}(t+\tau)
     \hat{E}_{bn}^{(-)}(t)
     \right\rangle
     \right]
     .
     \label{eq:normal-ordered-noise-correlation-function}
\end{eqnarray}
The noise spectral density $S_{s_{Nn}}^{({\rm normal})}(\omega)$ is
the Fourier transformation of this noise correlation function
$C_{({\rm av})s_{Nn}}^{({\rm normal})}(\tau)$:
\begin{eqnarray}
  \label{eq:normal-ordered-noise-spectral-density-def}
  S_{s_{Nn}}^{({\rm normal})}(\omega)
  :=
  \int_{-\infty}^{+\infty} d\tau C_{({\rm av})s_{Nn}}^{({\rm normal})}(\tau)
  e^{+i\omega\tau}.
\end{eqnarray}
This noise spectral density
(\ref{eq:normal-ordered-noise-spectral-density-def}) is one of the
targets of this section.


Here, we consider the monochromatic local oscillator case, where
$\gamma(t)$ is given by
Eq.~(\ref{eq:monochromatic_gamma_omega_is_delta}) and
\begin{eqnarray}
  \label{eq:gammat-monochromatic}
  \gamma(t)
  :=
  \int_{0}^{+\infty} \frac{d\omega}{2\pi} \sqrt{\omega} \gamma(\omega)
  e^{-i\omega t}
  =
  \sqrt{\omega_{0}} |\gamma| e^{+i\theta} e^{-i\omega_{0}t}
\end{eqnarray}
with Eq.~(\ref{eq:gamma_is_absgamma_with_phase}).
Substituting (\ref{eq:gammat-monochromatic}) into
Eq.~(\ref{eq:normal-ordered-noise-correlation-function}) and
(\ref{eq:normal-ordered-noise-spectral-density-def}), we obtain
\begin{eqnarray}
  S_{s_{Nn}}^{({\rm normal})}(\omega)
  &=&
      \frac{{\cal A}c}{2\pi\hbar} \omega_{0} |\gamma|^{2}
      \int_{-\infty}^{+\infty} d\tau e^{+i\omega\tau}
      \nonumber\\
  &&
     \times
     \lim_{T\rightarrow\infty} \frac{1}{T} \int_{-T/2}^{T/2} dt \left[
     e^{-2i\theta} e^{+i\omega_{0}(2t+\tau)}
     \left\langle
     \hat{E}_{bn}^{(+)}(t+\tau) \hat{E}_{bn}^{(+)}(t)
     \right\rangle
     \right.
     \nonumber\\
  && \quad\quad\quad\quad\quad\quad\quad\quad
     \left.
      +
      e^{-i\omega_{0}\tau}
      \left\langle
      \hat{E}_{bn}^{(-)}(t+\tau) \hat{E}_{bn}^{(+)}(t)
      \right\rangle
      \right.
      \nonumber\\
  && \quad\quad\quad\quad\quad\quad\quad\quad
     \left.
      +
      e^{+i\omega_{0}\tau}
      \left\langle
      \hat{E}_{bn}^{(-)}(t) \hat{E}_{bn}^{(+)}(t+\tau)
      \right\rangle
      \right.
      \nonumber\\
  && \quad\quad\quad\quad\quad\quad\quad\quad
     \left.
      +
      e^{+2i\theta} e^{-i\omega_{0}(2t+\tau)}
      \left\langle
      \hat{E}_{bn}^{(-)}(t+\tau) \hat{E}_{bn}^{(-)}(t)
      \right\rangle
      \right]
     .
     \label{eq:normal-ordered-noise-spectral-density-monochro}
\end{eqnarray}


Here, we introduce the the Fourier transformed expression of the
field operator $\hat{E}_{bn}^{(\pm)}(t)$ like
Eq.~(\ref{eq:electric-field-notation-total-electric-field-positive}),
\begin{eqnarray}
  \label{eq:signal-electric-field-noise-positive}
  \hat{b}_{n}(\omega)
  &:=&
       \hat{b}(\omega) - \langle\hat{b}(\omega)\rangle
       ,
       \nonumber\\
  \hat{E}_{bn}^{(+)}(t)
  &:=&
       \int_{0}^{+\infty} \frac{d\omega}{2\pi}
       \sqrt{\frac{2\pi\hbar\omega}{{\cal A}c}}
       \hat{b}_{n}(\omega) e^{-i\omega t}
       ,
       \quad
       \hat{E}_{bn}^{(-)}(t)
       :=
       \left[\hat{E}_{bn}^{(+)}(t)\right]^{\dagger}
       .
\end{eqnarray}
Substituting these operators into
Eq.~(\ref{eq:normal-ordered-noise-spectral-density-monochro}), we
obtain
\begin{eqnarray}
  S_{s_{Nn}}^{({\rm normal})}(\omega)
  &=&
      \omega_{0} |\gamma|^{2} \left(
      {\cal I}_{1}(\omega)
      +
      {\cal I}_{2}(\omega)
      +
      {\cal I}_{3}(\omega)
      +
      {\cal I}_{4}(\omega)
      \right)
      ,
     \label{eq:normal-ordered-noise-spectral-density-monochro-calIi}
\end{eqnarray}
where we defined ${\cal I}_{i}(\omega)$ by
\begin{eqnarray}
  \label{eq:normal-ordered-noise-spectral-density-monochro-calI1}
  {\cal I}_{1}(\omega)
  &:=&
       e^{-2i\theta}
       \int_{0}^{+\infty} \frac{d\omega_{2}}{2\pi}
       \sqrt{(\omega+\omega_{0})\omega_{2}}
       \left\langle
       \hat{b}_{n}(\omega+\omega_{0})
       \hat{b}_{n}(\omega_{2})
       \right\rangle
       \nonumber\\
  && \quad\quad\quad\quad
     \times
     \lim_{T\rightarrow\infty} \frac{1}{T} \int_{-T/2}^{T/2} dt
     e^{+i(\omega_{0}-\omega-\omega_{2}) t}
     ,
  \\
  \label{eq:normal-ordered-noise-spectral-density-monochro-calI2}
  {\cal I}_{2}(\omega)
  &:=&
       \int_{0}^{+\infty} \frac{d\omega_{2}}{2\pi}
       \sqrt{(\omega_{0}-\omega)\omega_{2}}
       \left\langle
       \hat{b}_{n}^{\dagger}(\omega_{0}-\omega)
       \hat{b}_{n}(\omega_{2})
       \right\rangle
       \nonumber\\
  && \quad\quad\quad\quad
     \times
     \lim_{T\rightarrow\infty} \frac{1}{T} \int_{-T/2}^{T/2} dt
     e^{+i(\omega_{0}-\omega-\omega_{2})t}
      ,
  \\
  \label{eq:normal-ordered-noise-spectral-density-monochro-calI3}
  {\cal I}_{3}(\omega)
  &:=&
      \int_{0}^{+\infty} \frac{d\omega_{1}}{2\pi}
      \int_{0}^{+\infty} \frac{d\omega_{2}}{2\pi}
      \sqrt{(\omega_{0}+\omega)\omega_{2}}
      \left\langle
      \hat{b}_{n}^{\dagger}(\omega_{2})
      \hat{b}_{n}(\omega_{0}+\omega)
      \right\rangle
      \nonumber\\
  && \quad\quad\quad\quad
     \times
     \lim_{T\rightarrow\infty} \frac{1}{T} \int_{-T/2}^{T/2} dt
     e^{-i(\omega_{0}+\omega-\omega_{2})t}
     ,
  \\
  \label{eq:normal-ordered-noise-spectral-density-monochro-calI4}
  {\cal I}_{4}(\omega)
  &:=&
      e^{+2i\theta}
      \int_{0}^{+\infty} \frac{d\omega_{2}}{2\pi}
      \sqrt{(\omega_{0}-\omega)\omega_{2}}
      \left\langle
      \hat{b}_{n}^{\dagger}(\omega_{0}-\omega)
      \hat{b}_{n}^{\dagger}(\omega_{2})
      \right\rangle
      \nonumber\\
  && \quad\quad\quad\quad
     \times
     \lim_{T\rightarrow\infty} \frac{1}{T} \int_{-T/2}^{T/2} dt
     e^{-i(\omega_{0}+\omega-\omega_{2})t}
     .
\end{eqnarray}
Here, we used the situation where $\omega_{0}\gg\omega>0$ in the
derivation of
Eqs.~(\ref{eq:normal-ordered-noise-spectral-density-monochro-calI1})--(\ref{eq:normal-ordered-noise-spectral-density-monochro-calI4}).
The properties of the function
\begin{eqnarray}
  \label{eq:lim1overTintdte-iat-def-main}
  f(a) := \lim_{T\rightarrow+\infty} \frac{1}{T} \int_{-T/2}^{T/2} dt
  e^{-iat}
  ,
  \quad a\in\RF
\end{eqnarray}
in the factor of
Eqs.~(\ref{eq:normal-ordered-noise-spectral-density-monochro-calI1})--(\ref{eq:normal-ordered-noise-spectral-density-monochro-calI4})
is summarized in
Appendix~\ref{sec:Properties_of_a_time-averaged_function}.


In Appendix~\ref{sec:calI1omega-in-Michelson}, we showed the explicit
form (\ref{eq:calI1omega-Michelson-explicit-result}) of
${\cal I}_{1}(\omega)$ of the Michelson interferometer as an example.
Equation (\ref{eq:calI1omega-Michelson-explicit-result}) is a finite
result and this result implies that the expectation value
$\langle\hat{b}_{n}(\omega+\omega_{0})\hat{b}_{n}(\omega_{2})\rangle$
includes the $\delta$-function $2 \pi \delta(\omega_{2}-(\omega_{0}-\omega))$.
Due to this $\delta$-function, ${\cal I}_{1}(\omega)$ has a finite
value even after the averaging process of the integration by $t$.
The expectation value
$\langle\hat{b}_{n}(\omega+\omega_{0})\hat{b}_{n}(\omega_{2})\rangle$
may have more $\delta$-functions whose support is different from the
point $\omega_{2}=\omega_{0}-\omega$.
However, even in this case, such $\delta$ function does not contribute
to the result ${\cal I}_{1}(\omega)$ due to the property of the
function (\ref{eq:lim1overTintdte-iat-def-main}) as explained in Appendix~\ref{sec:Properties_of_a_time-averaged_function}.
This situation is also true in the case of ${\cal I}_{2}(\omega)$,
${\cal I}_{3}(\omega)$, and ${\cal I}_{4}(\omega)$ given by
Eqs.~(\ref{eq:normal-ordered-noise-spectral-density-monochro-calI2})--(\ref{eq:normal-ordered-noise-spectral-density-monochro-calI4}),
respectively.
These finite results of ${\cal I}_{i}(\omega)$ ($i=1,2,3,4$) also
imply that if we force to omit the integration by $\omega_{2}$ and to
specify $\omega_{2}$ so that the exponent in the average function
vanishes in
Eqs.~(\ref{eq:normal-ordered-noise-spectral-density-monochro-calI1})--(\ref{eq:normal-ordered-noise-spectral-density-monochro-calI4}),
respectively, the resulting expression of ${\cal I}_{i}(\omega)$
includes $2\pi\delta(0)$.
This corresponds to the imposition of the stationarity to the noise
spectral density.
Thus, we conclude that we can obtain the correct results if we regard
the expression of ${\cal I}_{1}(\omega)$, for example, as
\begin{eqnarray}
  \label{eq:calI1omega-general-delta0-expression-2}
  2 \pi \delta(\omega-\omega') {\cal I}_{1}(\omega)
  =
  e^{-2i\theta}
  \sqrt{(\omega_{0}+\omega)(\omega_{0}-\omega')}
  \left\langle\hat{b}_{n}(\omega_{0}+\omega)\hat{b}_{n}(\omega_{0}-\omega')\right\rangle
  .
\end{eqnarray}
In the case where $\omega_{0}\gg\omega>0$, we conclude
\begin{eqnarray}
  \label{eq:calI1omega-general-delta0-expression-4}
  2 \pi \delta(\omega-\omega') {\cal I}_{1}(\omega)
  \sim
  \omega_{0}
  e^{-2i\theta}
  \left\langle
  \hat{b}_{n}(\omega_{0}+\omega)
  \hat{b}_{n}(\omega_{0}-\omega')
  \right\rangle
  .
\end{eqnarray}
Similarly, we obtain
\begin{eqnarray}
  \label{eq:calI2omega-general-delta0-expression-4}
  2 \pi \delta(\omega-\omega') {\cal I}_{2}(\omega)
  &\sim&
         \omega_{0}
         \left\langle
         \hat{b}_{n}^{\dagger}(\omega_{0}-\omega)
         \hat{b}_{n}(\omega_{0}-\omega)
         \right\rangle
         ,
  \\
  \label{eq:calI3omega-general-delta0-expression-4}
  2 \pi \delta(\omega-\omega') {\cal I}_{3}(\omega)
  &\sim&
         \omega_{0}
         \left\langle
         \hat{b}_{n}^{\dagger}(\omega_{0}+\omega)
         \hat{b}_{n}(\omega_{0}+\omega)
         \right\rangle
         ,
  \\
  \label{eq:calI4omega-general-delta0-expression-4}
  2 \pi \delta(\omega-\omega') {\cal I}_{4}(\omega)
  &\sim&
         \omega_{0}
         e^{+2i\theta}
         \left\langle
         \hat{b}_{n}^{\dagger}(\omega_{0}-\omega)
         \hat{b}_{n}^{\dagger}(\omega_{0}+\omega)
         \right\rangle
         .
\end{eqnarray}


Thus, from
Eqs.~(\ref{eq:calI1omega-general-delta0-expression-4})--(\ref{eq:calI4omega-general-delta0-expression-4}),
the normal-ordered noise spectral density in the situation where
$\omega_{0}\gg\omega>0$ is given by
\begin{eqnarray}
  \label{eq:2pideltaSNsnormal-expression}
  2 \pi \delta(\omega-\omega')
  S_{s_{Nn}}^{({\rm normal})}(\omega)
  &=&
      2 \pi \delta(\omega-\omega')
      \omega_{0} |\gamma|^{2}
      \left(
      {\cal I}_{1}(\omega)
      +
      {\cal I}_{2}(\omega)
      +
      {\cal I}_{3}(\omega)
      +
      {\cal I}_{4}(\omega)
      \right)
      \nonumber\\
  &\sim&
         \omega_{0}^{2} |\gamma|^{2}
         \left\langle
         e^{-2i\theta} \hat{b}_{n}(\omega_{0}+\omega) \hat{b}_{n}(\omega_{0}-\omega')
         \right.
         \nonumber\\
  && \quad\quad\quad\quad
     \left.
     +
     \hat{b}_{n}^{\dagger}(\omega_{0}-\omega) \hat{b}_{n}(\omega_{0}-\omega')
     \right.
     \nonumber\\
  && \quad\quad\quad\quad
     \left.
     +
     \hat{b}_{n}^{\dagger}(\omega_{0}+\omega) \hat{b}_{n}(\omega_{0}+\omega')
     \right.
     \nonumber\\
  && \quad\quad\quad\quad
     \left.
     +
     e^{+2i\theta} \hat{b}_{n}^{\dagger}(\omega_{0}+\omega) \hat{b}_{n}^{\dagger}(\omega_{0}-\omega')
     \right\rangle
     .
\end{eqnarray}
Note that $\omega_{0}$ is the central frequency of the optical field
from the local oscillator.
This frequency $\omega_{0}$ may not coincide with the central
frequency of the signal field $\hat{E}_{b}(t)$.
In this sense, the above noise spectral density includes the
``heterodyne detection.''


Here, we regard that the central frequency $\omega_{0}$ of the optical
field from the local oscillator coincides with the central frequency
from the main interferometer.
This is the ``homodyne detection.''
In this case, we can use the sideband picture
$\hat{b}_{\pm}(\omega):=\hat{b}(\omega_{0}\pm\omega)$ and the above
noise spectral density is given by
\begin{eqnarray}
  \label{eq:2pideltaSNsnormal-expression-homodyne}
  2 \pi \delta(\omega-\omega')
  S_{s_{Nn}}^{({\rm normal})}(\omega)
  &\sim&
         \omega_{0}^{2} |\gamma|^{2}
         \left\langle
         e^{-2i\theta} \hat{b}_{n+}(\omega) \hat{b}_{n-}(\omega')
         +
         \hat{b}_{n-}^{\dagger}(\omega) \hat{b}_{n-}(\omega')
         \right.
         \nonumber\\
  && \quad\quad\quad\quad
     \left.
         +
         \hat{b}_{n+}^{\dagger}(\omega) \hat{b}_{n+}(\omega')
         +
         e^{+2i\theta} \hat{b}_{n+}^{\dagger}(\omega) \hat{b}_{n-}^{\dagger}(\omega')
         \right\rangle
         .
\end{eqnarray}
Through the amplitude and phase quadratures $\hat{b}_{1 n}$,
$\hat{b}_{2 n}$, and
$\hat{b}_{\theta n}:=\cos\theta\hat{b}_{1 n}+\sin\theta\hat{b}_{2 n}$ is
given by
\begin{eqnarray}
  \label{eq:2pideltaSNsnormal-expression-homodyne-btheta}
  2 \pi \delta(\omega-\omega')
  S_{s_{Nn}}^{({\rm normal})}(\omega)
  &\sim&
         \omega_{0}^{2} |\gamma|^{2}
         \left[
         \left\langle
         \hat{b}_{\theta n}^{\dagger}(\omega')
         \hat{b}_{\theta n}(\omega)
         +
         \hat{b}_{\theta n}(\omega)
         \hat{b}_{\theta n}^{\dagger}(\omega')
         \right\rangle
         \right.
         \nonumber\\
  && \quad\quad\quad\quad
     \left.
         -
         2 \pi \delta(\omega-\omega')
         \right]
         .
\end{eqnarray}
Here, we note that
$\left[\hat{b}_{\theta n}(\omega),\hat{b}_{\theta n}(\omega')\right]=0$.
Since we only consider the positive frequency $\omega$, the first term
in the right-hand side of
Eq.~(\ref{eq:2pideltaSNsnormal-expression-homodyne-btheta}) is
identical to the Kimble single-sideband noise spectral density
$S_{b_{\theta}}^{(s)}(\omega)$ introduced in
Ref.~\cite{H.J.Kimble-Y.Levin-A.B.Matsko-K.S.Thorne-S.P.Vyatchanin-2001}
as
\begin{eqnarray}
  \label{eq:2pideltaSNsnormal-expression-homodyne-btheta-vs-Kimble}
  S_{s_{Nn}}^{({\rm normal})}(\omega)
  &\sim&
         \omega_{0}^{2} |\gamma|^{2}
         \left[
         \bar{S}_{b_{\theta}}^{(s)}(\omega)
         -
         1
         \right]
         .
\end{eqnarray}


\subsubsection{Including vacuum fluctuations from the main interferometer}
\label{sec:Vacuum_fluctuations_from_main}


Here, we take into account of the vacuum fluctuations from the signal
field $\hat{E}_{b}(t)$.
To clarify the contribution of the vacuum fluctuations from the signal
field $\hat{E}_{b}(t)$, we ignore the vacuum fluctuations of the local
oscillator $\hat{E}_{l_{i}}(t)$, but take into account of the vacuum
fluctuations from the signal field $\hat{E}_{b}(t)$.


From the definition (\ref{eq:hatsNt-electric-field}) of the signal
operator $\hat{s}_{N}(t)$ and its expectation value
(\ref{eq:hatsPt-exp-valu}), we defined the noise operator
$\hat{s}_{Nn}(t)$ (\ref{eq:sNt-noise-operator}) and considered states
$|\hat{s}_{N}(t)\rangle$ and $\langle\hat{s}_{N}(t)|$ as
Eqs.~(\ref{eq:sNt-state-ket}) and (\ref{eq:sNt-state-bra}),
respectively.
From Eqs.~(\ref{eq:sNt-state-ket}), (\ref{eq:sNt-state-bra}), and
the replacement $t\rightarrow t+\tau$, we also derived
$|\hat{s}_{N}(t+\tau)\rangle$ and $\langle\hat{s}_{N}(t+\tau)|$.
From these states, here, we evaluate the inner products
$\langle\hat{s}_{Nn}(t)|\hat{s}_{Nn}(t+\tau)\rangle$ and
$\langle\hat{s}_{Nn}(t+\tau)|\hat{s}_{Nn}(t)\rangle$ under the
premises
\begin{eqnarray}
  &&
     \label{eq:commutation-relation-E+E-b-nonvanish}
     \left[
     \hat{E}_{b}^{(+)}(t), \hat{E}_{b}^{(-)}(t')
     \right]
     =:
     \frac{2\pi\hbar}{{\cal A}c} \Delta_{b}(t-t')
     \neq
     0
     ,
  \\
  &&
     \label{eq:commutation-relation-E+E-li-vanish}
     \left[
     \hat{E}_{l_{i}}^{(+)}(t), \hat{E}_{l_{i}}^{(-)}(t')
     \right]
     =:
     \frac{2\pi\hbar}{{\cal A}c} \Delta_{l_{i}}(t-t')
     =
     0
     .
\end{eqnarray}
Of course, this premise is not consistent within the quantum field
theory of electromagnetic fields.
However, we dare to use these premise
(\ref{eq:commutation-relation-E+E-b-nonvanish}) and
(\ref{eq:commutation-relation-E+E-li-vanish}) to clarify from which
field $\hat{E}_{b}$ or $\hat{E}_{li}$ the vacuum fluctuations
contribute to the noise spectral density.
Furthermore, we denote the inner products of the states
$\langle\hat{s}_{Nn}(t)|$, $|\hat{s}_{Nn}(t+\tau)\rangle$,
$\langle\hat{s}_{Nn}(t+\tau)|$, and $|\hat{s}_{Nn}(t)\rangle$ under the premise
(\ref{eq:commutation-relation-E+E-b-nonvanish}) and
(\ref{eq:commutation-relation-E+E-li-vanish}) by
\begin{eqnarray}
  &&
     \frac{1}{2}
     \langle
     \hat{s}_{Nn}(t)\hat{s}_{Nn}(t+\tau)
     +
     \hat{s}_{Nn}(t+\tau)\hat{s}_{Nn}(t)
     \rangle^{({\rm normal}+{\rm sig.vac.})}
     \nonumber\\
  &:=&
       \frac{1}{2} \left(
       \langle
       \hat{s}_{Nn}(t)
       |
       \hat{s}_{Nn}(t+\tau)
       \rangle^{({\rm normal}+{\rm sig.vac.})}
       \right.
       \nonumber\\
  && \quad\quad
     \left.
     +
     \langle
     \hat{s}_{Nn}(t+\tau)
     |
     \hat{s}_{Nn}(t)
     \rangle^{({\rm normal}+{\rm sig.vac.})}
     \right)
     .
     \label{eq:half-hatsNnthatsNnt+tau+hatsNnt+tauhatNnt-exp}
\end{eqnarray}
The straightforward calculations yields
\begin{eqnarray}
  &&
     \frac{1}{2}
     \langle
     \hat{s}_{Nn}(t)\hat{s}_{Nn}(t+\tau)
     +
     \hat{s}_{Nn}(t+\tau)\hat{s}_{Nn}(t)
     \rangle^{({\rm normal}+{\rm sig.vac.})}
     \nonumber\\
  &=&
      \frac{1}{2} \left(
      \langle
      :
      \hat{s}_{Nn}(t)
      \hat{s}_{Nn}(t+\tau)
      +
      \hat{s}_{Nn}(t+\tau)
      \hat{s}_{Nn}(t)
      :
      \rangle
      \right)
      \nonumber\\
  &&
      +
      \frac{1}{2} \left(
      \gamma^{*}(t+\tau) \gamma(t) \Delta_{b}(\tau)
      +
      \gamma^{*}(t) \gamma(t+\tau) \Delta_{b}(-\tau)
      \right)
      .
     \label{eq:normal+sig.va.-noise-correlation}
\end{eqnarray}
Then, we obtain the correlation functions and its average version as
\begin{eqnarray}
  \label{eq:sNn-correlation-normal-sig.vac.-is-normal+sig.vac.}
  C_{({\rm av})s_{Nn}}^{({\rm normal}+{\rm sig.vac.})}(\tau)
  =
  C_{({\rm av})s_{N}}^{({\rm normal})}(\tau)
  +
  C_{({\rm av})s_{N}}^{({\rm sig.vac.})}(\tau)
  ,
\end{eqnarray}
where
\begin{eqnarray}
  \label{eq:sNn-correlation-sig.vac.-def}
  C_{({\rm av})s_{N}}^{({\rm sig.vac.})}(\tau)
  :=
  \lim_{T\rightarrow+\infty} \frac{1}{T}
  \int_{-T/2}^{T/2} dt \frac{1}{2} \left(
  \gamma^{*}(t+\tau) \gamma(t) \Delta_{b}(\tau)
  +
  \gamma^{*}(t) \gamma(t+\tau) \Delta_{b}(-\tau)
  \right)
  .
\end{eqnarray}
In the monochromatic local oscillator case,
$\gamma(t)=\sqrt{\omega_{0}}e^{-i\omega_{0}t}=\sqrt{\omega_{0}}|\gamma|
e^{+i\theta} e^{-i\omega_{0}t}$, we obtain
\begin{eqnarray}
  \label{eq:sNn-correlation-sig.vac.-monochro-local}
  C_{({\rm av})s_{Nn}}^{({\rm sig.vac.})}(\tau)
  =
  \frac{1}{2} \omega_{0} |\gamma|^{2} \left(
  e^{+i\omega_{0}\tau} \Delta_{b}(\tau)
  +
  e^{-i\omega_{0}\tau} \Delta_{b}(-\tau)
  \right)
  .
\end{eqnarray}
Using the explicit expression (\ref{eq:commutation-relation-E+E-}) of
the $\Delta_{b}(\tau)$ and the situation where
$\omega_{0}\gg\omega>0$, the Fourier transformation of
Eq.~(\ref{eq:sNn-correlation-sig.vac.-monochro-local}) is given by
\begin{eqnarray}
  \label{eq:sNn-noise-spectral-sig.vac.-monochro-local}
  S_{s_{Nn}}^{({\rm sig.vac.})}(\omega)
  =
  \omega_{0}^{2} |\gamma|^{2}
  .
\end{eqnarray}
Together with the previous result
(\ref{eq:2pideltaSNsnormal-expression-homodyne-btheta-vs-Kimble}), we
obtain
\begin{eqnarray}
  \label{eq:2pideltaSNsnormal+sig.vac.-expression-homodyne-btheta-vs-Kimble}
  S_{s_{Nn}}^{({\rm normal}+{\rm sig.vac.})}(\omega)
  &\sim&
         \omega_{0}^{2} |\gamma|^{2}
         \left[
         \bar{S}_{b_{\theta}}^{(s)}(\omega)
         -
         1
         \right]
         +
         \omega_{0}^{2} |\gamma|^{2}
         =
         \omega_{0}^{2} |\gamma|^{2}
         \bar{S}_{b_{\theta}}^{(s)}(\omega)
         .
\end{eqnarray}


\subsubsection{Including vacuum fluctuations from the local oscillator}
\label{sec:Vacuum_fluctuations_from_local_oscillator}


Here, we take into account of the vacuum fluctuations of the field
$\hat{E}_{l_{i}}(t)$ from the local oscillator in addition to the
previous results.
We evaluate the inner products
$\langle\hat{s}_{Nn}(t)|\hat{s}_{Nn}(t+\tau)\rangle$ and
$\langle\hat{s}_{Nn}(t+\tau)|\hat{s}_{Nn}(t)\rangle$ under the premise
\begin{eqnarray}
  \label{eq:Deltalineq0-Deltabneq0}
  \left[
  \hat{E}_{l_{i}}^{(+)}(t), \hat{E}_{l_{i}}^{(-)}(t')
  \right]
  =
  \frac{2\pi\hbar}{{\cal A}c} \Delta_{l_{i}}(t-t')
  \neq
  0
  ,
  \quad
  \left[
  \hat{E}_{b_{n}}^{(+)}(t), \hat{E}_{b_{n}}^{(-)}(t')
  \right]
  =
  \frac{2\pi\hbar}{{\cal A}c} \Delta_{b}(t-t')
  \neq
  0
  .
\end{eqnarray}
This is the complete consideration which is taking into account of the
vacuum fluctuations of all optical fields.
We denote these inner products through this evaluation as
$\langle\hat{s}_{Nn}(t)|\hat{s}_{Nn}(t+\tau)\rangle^{({\rm
    normal}+{\rm sig.vac.}+{\rm loc.vac.})}$ and
$\langle\hat{s}_{Nn}(t+\tau)|\hat{s}_{Nn}(t)\rangle^{({\rm
    normal}+{\rm sig.vac.}+{\rm loc.vac.})}$.
From the states given by Eqs.~(\ref{eq:sNt-state-ket}),
(\ref{eq:sNt-state-bra}), and the definition
(\ref{eq:sNt-noise-operator}) of the noise operator $\hat{s}_{Nn}(t)$,
we can include the vacuum fluctuations from the local oscillator field as
\begin{eqnarray}
  &&
     \frac{1}{2}
     \left\langle
     \hat{s}_{Nn}(t) \hat{s}_{Nn}(t+\tau)
     +
     \hat{s}_{Nn}(t+\tau) \hat{s}_{Nn}(t)
     \right\rangle
     \nonumber\\
  &=&
      \frac{1}{2} \left(
      \langle\hat{s}_{Nn}(t)|\hat{s}_{Nn}(t+\tau)\rangle^{({\rm
      normal}+{\rm sig.vac.}+{\rm loc.vac.})}
      \right.
      \nonumber\\
  && \quad\quad
     \left.
     +
     \langle\hat{s}_{Nn}(t+\tau)|\hat{s}_{Nn}(t)\rangle^{({\rm
     normal}+{\rm sig.vac.}+{\rm loc.vac.})}
     \right)
     \nonumber\\
  &=&
      \frac{1}{2} \left\langle
      \hat{s}_{Nn}(t)\hat{s}_{Nn}(t+\tau)
      +
      \hat{s}_{Nn}(t+\tau)\hat{s}_{Nn}(t)
      \right\rangle^{({\rm normal}+{\rm sig.vac.})}
      \nonumber\\
  &&
      +
      \frac{1}{2} \frac{{\cal A}c}{2\pi\hbar} \left\{
      \left\langle
      \hat{E}_{b}^{(-)}(t) \hat{E}_{b}^{(+)}(t+\tau)
      \right\rangle
      \Delta_{l_{i}}(-\tau)
      +
      \left\langle
      \hat{E}_{b}^{(-)}(t+\tau) \hat{E}_{b}^{(+)}(t)
      \right\rangle
      \Delta_{l_{i}}(\tau)
      \right\}
      \nonumber\\
  &&
      +
      \frac{1}{2} \frac{1-2\eta}{\sqrt{\eta(1-\eta)}}
      \sqrt{\frac{{\cal A}c}{2\pi\hbar}} \Delta_{l_{i}}(\tau) \left\{
      \gamma^{*}(t+\tau) \left\langle\hat{E}_{b}^{(+)}(t)\right\rangle
      +
      \gamma(t) \left\langle\hat{E}_{b}^{(-)}(t+\tau)\right\rangle
      \right\}
      \nonumber\\
  &&
      +
      \frac{1}{2} \frac{1-2\eta}{\sqrt{\eta(1-\eta)}}
      \sqrt{\frac{{\cal A}c}{2\pi\hbar}} \Delta_{l_{i}}(-\tau) \left\{
      \gamma(t+\tau) \left\langle\hat{E}_{b}^{(-)}(t)\right\rangle
      +
      \gamma^{*}(t) \left\langle\hat{E}_{b}^{(+)}(t+\tau)\right\rangle
      \right\}
      \nonumber\\
  &&
      +
      \frac{1}{2} \frac{(1-2\eta)^{2}}{\eta(1-\eta)}
      \left\{
      \gamma^{*}(t) \gamma(t+\tau) \Delta_{l_{i}}(-\tau)
      \gamma^{*}(t+\tau) \gamma(t) \Delta_{l_{i}}(\tau)
      \right\}
      .
     \label{eq:total-sNn-correlation-normal-sig.vac.-loc.vac.}
\end{eqnarray}
The lines from the second to the last in
Eq.~(\ref{eq:total-sNn-correlation-normal-sig.vac.-loc.vac.}) are all
the vacuum fluctuations contributions from the local oscillator.
Then we denote the averaged correlation function
\begin{eqnarray}
  \label{eq:total-averaged-correlation-function}
  C_{({\rm av})s_{Nn}}(\tau)
  =
  C_{({\rm av})s_{Nn}}^{({\rm normal}+{\rm sig.vac.}+{\rm loc.vac.})}(\tau)
  =
  C_{({\rm av})s_{Nn}}^{({\rm normal}+{\rm sig.vac.})}(\tau)
  +
  C_{({\rm av})s_{Nn}}^{({\rm loc.vac.})}(\tau)
  ,
\end{eqnarray}
where we defined
\begin{eqnarray}
  C_{({\rm av})s_{Nn}}^{({\rm loc.vac.})}(\tau)
  &:=&
       {\cal J}_{1}(\tau)
       +
       {\cal J}_{2}(\tau)
       +
       {\cal J}_{3}(\tau)
       +
       {\cal J}_{4}(\tau)
       +
       {\cal J}_{5}(\tau)
       ,
       \label{eq:calJiomega-defs}
  \\
  {\cal J}_{1}(\tau)
  &:=&
       \frac{{\cal A}c}{4\pi\hbar}
       \lim_{T\rightarrow+\infty} \frac{1}{T} \int_{-T/2}^{T/2} dt
       \left\{
       \left\langle
       \hat{E}_{bn}^{(-)}(t) \hat{E}_{bn}^{(+)}(t+\tau)
       \right\rangle
       \Delta_{l_{i}}(-\tau)
       \right.
       \nonumber\\
  && \quad\quad\quad\quad\quad\quad\quad\quad\quad\quad\quad
     \left.
     +
     \left\langle
     \hat{E}_{bn}^{(-)}(t+\tau) \hat{E}_{bn}^{(+)}(t)
     \right\rangle
     \Delta_{l_{i}}(\tau)
     \right\}
     ,
     \label{eq:calJ1tau-def}
  \\
  {\cal J}_{2}(\tau)
  &:=&
       \frac{{\cal A}c}{4\pi\hbar}
       \lim_{T\rightarrow+\infty} \frac{1}{T} \int_{-T/2}^{T/2} dt
       \left\{
       \left\langle
       \hat{E}_{b}^{(-)}(t)
       \right\rangle
       \left\langle
       \hat{E}_{b}^{(+)}(t+\tau)
       \right\rangle
       \Delta_{l_{i}}(-\tau)
       \right.
       \nonumber\\
  && \quad\quad\quad\quad\quad\quad\quad\quad\quad\quad\quad
     \left.
     +
     \left\langle
     \hat{E}_{b}^{(-)}(t+\tau)
     \right\rangle
     \left\langle
     \hat{E}_{b}^{(+)}(t)
     \right\rangle
     \Delta_{l_{i}}(\tau)
     \right\}
     ,
     \label{eq:calJ2tau-def}
  \\
  {\cal J}_{3}(\tau)
  &:=&
       \frac{1-2\eta}{2\sqrt{\eta(1-\eta)}}
       \sqrt{\frac{{\cal A}c}{2\pi\hbar}}
       \lim_{T\rightarrow+\infty} \frac{1}{T} \int_{-T/2}^{T/2} dt
       \Delta_{l_{i}}(\tau)
       \nonumber\\
  && \quad\quad\quad\quad\quad
     \times
     \left\{
     \gamma^{*}(t+\tau) \left\langle\hat{E}_{b}^{(+)}(t)\right\rangle
     +
     \gamma(t) \left\langle\hat{E}_{b}^{(-)}(t+\tau)\right\rangle
     \right\}
     ,
     \label{eq:calJ3tau-def}
  \\
  {\cal J}_{4}(\tau)
  &:=&
       \frac{1-2\eta}{2\sqrt{\eta(1-\eta)}}
       \sqrt{\frac{{\cal A}c}{2\pi\hbar}}
       \lim_{T\rightarrow+\infty} \frac{1}{T} \int_{-T/2}^{T/2} dt
       \Delta_{l_{i}}(-\tau)
       \nonumber\\
  && \quad\quad\quad\quad\quad
     \times
     \left\{
     \gamma(t+\tau) \left\langle\hat{E}_{b}^{(-)}(t)\right\rangle
     +
     \gamma^{*}(t) \left\langle\hat{E}_{b}^{(+)}(t+\tau)\right\rangle
     \right\}
     ,
     \label{eq:calJ4tau-def}
  \\
  {\cal J}_{5}(\tau)
  &:=&
       \frac{(1-2\eta)^{2}}{2\eta(1-\eta)}
       \lim_{T\rightarrow+\infty} \frac{1}{T} \int_{-T/2}^{T/2} dt
       \nonumber\\
  && \quad\quad\quad\quad\quad
     \times
     \left\{
     \gamma^{*}(t) \gamma(t+\tau) \Delta_{l_{i}}(-\tau)
     \gamma^{*}(t+\tau) \gamma(t) \Delta_{l_{i}}(\tau)
     \right\}
     .
     \label{eq:calJ5tau-def}
\end{eqnarray}
We evaluate ${\cal J}_{i}(\tau)$ ($i=1,2,3,4,5$) and their Fourier
transformation ${\cal J}_{i}(\omega)$, separately.


First, we evaluate ${\cal J}_{1}(\omega)$.
Substituting Eqs.~(\ref{eq:signal-electric-field-noise-positive}) and
the definition (\ref{eq:commutation-relation-E+E-}) of the vacuum
fluctuations $\Delta_{l_{i}}(\tau)$, we obtain
\begin{eqnarray}
  {\cal J}_{1}(\omega)
  &:=&
       \frac{1}{2} \int_{-\infty}^{+\infty} d\tau e^{+i\omega\tau}
       \lim_{T\rightarrow+\infty} \frac{1}{T} \int_{-T/2}^{T/2} dt
       \nonumber\\
  && \quad\quad
     \times
     \frac{{\cal A}c}{2\pi\hbar} \left[
     \left\langle
     \hat{E}_{bn}^{(-)}(t)
     \hat{E}_{bn}^{(+)}(t+\tau)
     \right\rangle
     \Delta_{l_{i}}(-\tau)
     +
     \left\langle
     \hat{E}_{bn}^{(-)}(t+\tau)
     \hat{E}_{bn}^{(+)}(t)
     \right\rangle
     \Delta_{l_{i}}(\tau)
     \right]
     \nonumber\\
  &=&
      \frac{1}{2}
      \int_{0}^{+\infty} \frac{d\omega_{1}}{2\pi}
      \int_{0}^{+\infty} \frac{d\omega_{2}}{2\pi}
      \sqrt{\omega_{1}\omega_{2}}
      \left\langle
      \hat{b}_{n}^{\dagger}(\omega_{1})
      \hat{b}_{n}(\omega_{2})
      \right\rangle
      \nonumber\\
  &&  \quad
     \times
      \left[
      \Theta(\omega_{2}-\omega) (\omega_{2}-\omega)
      +
      \Theta(\omega_{2}+\omega) (\omega_{1}+\omega)
      \right]
      \nonumber\\
  &&  \quad
     \times
      \lim_{T\rightarrow+\infty} \frac{1}{T}
      \int_{-T/2}^{T/2}dt e^{+i(\omega_{1}-\omega_{2})t}
     .
  \label{eq:calJ1omega-tmp}
\end{eqnarray}
Here, we use the introduce the noise-spectral density
$\SF_{b_{n}}(\omega)$ by
\begin{eqnarray}
  \label{eq:calS-bn-omega-def}
  \frac{1}{2} 2 \pi \delta(\omega_{1}-\omega_{2})
  \SF_{b_{n}}(\omega_{1})
  :=
  \frac{1}{2} \left\langle
  \hat{b}_{n}(\omega_{1}) \hat{b}_{n}^{\dagger}(\omega_{2})
  +
  \hat{b}_{n}^{\dagger}(\omega_{2}) \hat{b}_{n}(\omega_{1})
  \right\rangle
  .
\end{eqnarray}
This definition of $\SF_{b_{n}}(\omega)$ has the same form of
the Kimble single-sideband noise spectral density
(\ref{eq:Kimble-noise-spectral-density-single-side}).
However, we have to emphasize that the noise-spectral density
$\SF_{b_{n}}(\omega)$ have nothing to do with the two-photon
formulation nor upper- and lower-sideband with the central frequency
$\omega_{0}$, while the Kimble single-sideband noise spectral density
(\ref{eq:Kimble-noise-spectral-density-single-side}) defined within
the sideband picture of the optical fluctuations in the two-photon
formulation.
The frequencies $\omega_{1}$ and $\omega_{2}$ in
Eq.~(\ref{eq:calS-bn-omega-def}) is not sideband frequencies, but the
frequency $\omega$ in the definition
(\ref{eq:electric-field-notation-total-electric-field-positive}) of
the mode function of the electric field.
Through the noise spectral density $\SF_{b_{n}}(\omega)$ defined by
Eq.~(\ref{eq:calS-bn-omega-def}) and $\omega>0$, we obtain
\begin{eqnarray}
  {\cal J}_{1}(\omega)
  &=&
      \frac{1}{2}
      \int_{0}^{+\infty} \frac{d\omega_{1}}{2\pi}
      (\omega_{1})^{2}
      \left(
      \SF_{b_{n}}(\omega_{1}) - 1
      \right)
      \nonumber\\
  &&
     -
     \frac{1}{4}
     \int_{0}^{\omega} \frac{d\omega_{1}}{2\pi}
     \omega_{1}(\omega_{1}-\omega)
     \left(
     \SF_{b_{n}}(\omega_{1}) - 1
     \right)
     .
     \label{eq:calJ1omega-result}
\end{eqnarray}


Second, we evaluate the Fourier transformation ${\cal J}_{2}(\omega)$
of ${\cal J}_{2}(\tau)$ defined by Eq.~(\ref{eq:calJ2tau-def}).
Here, we define the expectation value of the output quadrature
$\hat{b}(\omega)$ as
\begin{eqnarray}
  \label{eq:output-quadrature-expectation-value-main}
  \left\langle\hat{b}(\omega)\right\rangle
  =:
  \alpha(\omega) + \beta 2 \pi \delta(\omega-\omega_{0})
  ,
\end{eqnarray}
where $\omega_{0}$ is the central frequency of the signal field
$\hat{E}_{b}(t)$ and we assume that $\alpha(\omega)$ and $\beta$ are
finite.
From this expectation value
(\ref{eq:output-quadrature-expectation-value-main}), we can evaluate
the expectation value of the electric field $\hat{E}_{b}^{(\pm)}$.
The vacuum fluctuations $\Delta_{l_{i}}(\pm\tau)$ from the local
oscillator is also given by Eq.~(\ref{eq:Deltalineq0-Deltabneq0}) and
(\ref{eq:commutation-relation-E+E-}).
Through the properties of the averaged function
(\ref{eq:lim1overTintdte-iat-def-main}), we obtain
the Fourier transformation ${\cal J}_{2}(\omega)$ as
\begin{eqnarray}
  \label{eq:calJ2omega-result}
  {\cal J}_{2}(\omega) = (\omega_{0})^{2} |\beta|^{2}.
\end{eqnarray}


Third, we evaluate the Fourier transformation ${\cal J}_{3}(\omega)$
of ${\cal J}_{3}(\tau)$ defined by Eq.~(\ref{eq:calJ3tau-def}).
The expectation value of the electric field $\hat{E}^{(\pm)}_{b}$ are
evaluated from the expectation value
(\ref{eq:output-quadrature-expectation-value-main}) and the vacuum
fluctuations $\Delta_{l_{i}}(\tau)$ from the local oscillator is also
given by Eq.~(\ref{eq:Deltalineq0-Deltabneq0}) and
(\ref{eq:commutation-relation-E+E-}) as in the case of
${\cal J}_{2}(\omega)$.
Furthermore, we consider the monochromatic local oscillator case
(\ref{eq:gammat-monochromatic}).
From these, we obtain the Fourier transformation
${\cal J}_{3}(\omega)$ for the case $\omega>0$ as
\begin{eqnarray}
  \label{eq:calJ3omega-result}
  {\cal J}_{3}(\omega)
  =
  \frac{1-2\eta}{2\sqrt{\eta(1-\eta)}} \omega_{0} (\omega_{0}+\omega)
  |\gamma| \left[
  \beta e^{-i\theta} + \beta^{*} e^{+i\theta}
  \right]
  .
\end{eqnarray}
Similarly, we can evaluate the Fourier transformation
${\cal J}_{4}(\omega)$ of ${\cal J}_{4}(\tau)$ defined by
Eq.~(\ref{eq:calJ4tau-def}) for the case $\omega>0$ as
\begin{eqnarray}
  \label{eq:calJ4omega-result}
  {\cal J}_{4}(\omega)
  =
  \frac{1-2\eta}{2\sqrt{\eta(1-\eta)}} \omega_{0} (\omega_{0}-\omega)
  |\gamma| \left[
  \beta e^{-i\theta} + \beta^{*} e^{+i\theta}
  \right]
  .
\end{eqnarray}


Finally, we evaluate the Fourier transformation ${\cal J}_{5}(\omega)$
of ${\cal J}_{5}(\tau)$ defined by Eq.~(\ref{eq:calJ5tau-def}).
The vacuum fluctuations $\Delta_{l_{i}}(\tau)$ from the local
oscillator is also given by Eq.~(\ref{eq:Deltalineq0-Deltabneq0}) and
(\ref{eq:commutation-relation-E+E-}) as in the case of
${\cal J}_{2}(\omega)$.
We also consider the monochromatic local oscillator case
(\ref{eq:gammat-monochromatic}).
Then, we obtain the Fourier transformation ${\cal J}_{5}(\omega)$ for
the case $\omega>0$ as
\begin{eqnarray}
  \label{eq:calJ5omega-result}
  {\cal J}_{5}(\omega)
  =
  \frac{(1-2\eta)^{2}}{\eta(1-\eta)} (\omega_{0})^{2} |\gamma|^{2}
  .
\end{eqnarray}


Through the evaluated ${\cal J}_{i}(\omega)$ ($\omega=1,2,3,4,5$)
given by Eqs.~(\ref{eq:calJ1omega-result}),
(\ref{eq:calJ2omega-result}), (\ref{eq:calJ3omega-result}),
(\ref{eq:calJ4omega-result}), and (\ref{eq:calJ5omega-result}), we
obtain the contribution from the vacuum fluctuations of the local
oscillator field to the noise spectral density as
\begin{eqnarray}
  S_{s_{Nn}}^{({\rm loc.vac.})}(\omega)
  &:=&
       \int_{-\infty}^{+\infty} d\tau e^{+i\omega\tau}
       C_{({\rm av})s_{Nn}}^{({\rm loc.vac.})}(\tau)
       \nonumber\\
  &=&
      \frac{1}{2}
      \int_{0}^{+\infty} \frac{d\omega_{1}}{2\pi}
      (\omega_{1})^{2}
      \left(
      \SF_{b_{n}}(\omega_{1}) - 1
      \right)
      \nonumber\\
  &&
      -
      \frac{1}{4}
      \int_{0}^{\omega} \frac{d\omega_{1}}{2\pi}
      \omega_{1}(\omega_{1}-\omega)
      \left(
      \SF_{b_{n}}(\omega_{1}) - 1
      \right)
      +
      (\omega_{0})^{2} |\beta|^{2}
      \nonumber\\
  &&
      +
      \frac{1-2\eta}{\sqrt{\eta(1-\eta)}} (\omega_{0})^{2}
      |\gamma| \left[
      \beta e^{-i\theta} + \beta^{*} e^{+i\theta}
      \right]
      \nonumber\\
  &&
      +
      \frac{(1-2\eta)^{2}}{\eta(1-\eta)} (\omega_{0})^{2} |\gamma|^{2}
      .
      \label{eq:sNn-noise-spectral-density-from-loc.vac.}
\end{eqnarray}
Then the total noise spectral density is given from
Eqs.~(\ref{eq:2pideltaSNsnormal+sig.vac.-expression-homodyne-btheta-vs-Kimble})
and (\ref{eq:sNn-noise-spectral-density-from-loc.vac.}) as
\begin{eqnarray}
  S_{s_{Nn}}(\omega)
  &:=&
       S_{s_{Nn}}^{({\rm normal}+{\rm sig.vac.})}(\omega)
       +
       S_{s_{Nn}}^{({\rm loc.vac.})}(\omega)
       \nonumber\\
  &\sim&
         \omega_{0}^{2} |\gamma|^{2}
         \bar{S}_{b_{\theta}}^{(s)}(\omega)
      \nonumber\\
  &&
         +
         \frac{1}{2}
         \int_{0}^{+\infty} \frac{d\omega_{1}}{2\pi}
         (\omega_{1})^{2}
         \left(
         \SF_{b_{n}}(\omega_{1}) - 1
         \right)
      \nonumber\\
  &&
         -
         \frac{1}{4}
         \int_{0}^{\omega} \frac{d\omega_{1}}{2\pi}
         \omega_{1}(\omega_{1}-\omega)
         \left(
         \SF_{b_{n}}(\omega_{1}) - 1
         \right)
     +
     (\omega_{0})^{2} |\beta|^{2}
      \nonumber\\
  &&
     +
     \frac{1-2\eta}{\sqrt{\eta(1-\eta)}} (\omega_{0})^{2}
     |\gamma| \left[
     \beta e^{-i\theta} + \beta^{*} e^{+i\theta}
     \right]
      \nonumber\\
  &&
     +
     \frac{(1-2\eta)^{2}}{\eta(1-\eta)} (\omega_{0})^{2} |\gamma|^{2}
     .
     \label{eq:sNn-noise-spectral-density-total-full}
\end{eqnarray}
Here, we consider the ideal case where the beam splitter is ideal,
i.e., $\eta=1/2$.
Furthermore, we consider the situation of the signal field
$\hat{E}_{b}(t)$ is in the complete dark port $\beta=0$ in which the
leakage of the classical carrier field from the main interferometer
completely is shut out.
In this case the derived total spectral density of the quantum noise
for the measurement of the operator $\hat{s}_{N}$ is 
\begin{eqnarray}
  S_{s_{Nn}}(\omega)
  &\sim&
  \omega_{0}^{2} |\gamma|^{2}
  \bar{S}_{b_{\theta}}^{(s)}(\omega)
      \nonumber\\
  &&
  +
  \frac{1}{2}
  \int_{0}^{+\infty} \frac{d\omega_{1}}{2\pi}
  \omega_{1}^{2}
  \left(
  \SF_{b_{n}}(\omega_{1}) - 1
  \right)
      \nonumber\\
  &&
  +
  \frac{1}{4}
  \int_{0}^{\omega} \frac{d\omega_{1}}{2\pi}
  \omega_{1} (\omega-\omega_{1})
  \left(
  \SF_{b_{n}}(\omega_{1}) - 1
  \right)
  .
  \label{eq:sNn-noise-spectral-density-total-full-ideal}
\end{eqnarray}


\subsection{Quantum Noise in Balanced Homodyne Detections by Power Counting Detectors}
\label{sec:Quantum_Noise_in_Balanced_Homodyne_Detections_By_PowerCountingDetectors}


Next, we evaluate the noise spectral density for the measurement of
operator $\hat{s}_{P}(t)$ given by
Eq.~(\ref{eq:hatsPt-field-expression}) in
Sec.~\ref{sec:Homodyne_Detections_by_power-counting}.
To carry out this evaluation, as in the case of Glauber's photon
number case in
Sec.~\ref{sec:Quantum_Noise_in_Balanced_Homodyne_Detections_By_PhotonNumberDetectors},
it is convenient to introduce the states $|\hat{s}_{P}(t)\rangle$ and
$\langle\hat{s}_{P}(t)|$ which are defined by
\begin{eqnarray}
  |\hat{s}_{P}(t)\rangle
  &:=&
       \hat{s}_{P}(t)|\Psi\rangle
       \nonumber\\
  &=&
      2 \left[
      \left(
      \sqrt{\frac{{\cal A}c}{2\pi\hbar}} \hat{E}_{l_{i}}^{(-)}(t) - \gamma^{*}(t)
      \right)
      \sqrt{\frac{{\cal A}c}{2\pi\hbar}} \hat{E}_{b}(t)
      +
      \left(
      \gamma(t) + \gamma^{*}(t)
      \right)
      \sqrt{\frac{{\cal A}c}{2\pi\hbar}} \hat{E}_{b}(t)
      \right.
      \nonumber\\
  && \quad\quad
     \left.
      +
      \frac{1-2\eta}{2\sqrt{\eta(1-\eta)}}
      \left(
      \sqrt{\frac{{\cal A}c}{2\pi\hbar}} \hat{E}_{l_{i}}^{(-)}(t) - \gamma^{*}(t)
      \right)
      \right.
      \nonumber\\
  && \quad\quad\quad\quad\quad\quad
     \left.
     \times
     \left(
     \sqrt{\frac{{\cal A}c}{2\pi\hbar}} \hat{E}_{l_{i}}^{(-)}(t)
     + \gamma^{*}(t)
     + 2 \gamma(t)
     \right)
     \right]
     |\Psi\rangle
     ,
     \label{eq:sPt-state-ket}
     \\
  \langle\hat{s}_{P}(t)|
  &:=&
      \langle\Psi|
      \hat{s}_{P}(t)
       \nonumber\\
  &=&
      2
      \langle\Psi|
      \left[
      \sqrt{\frac{{\cal A}c}{2\pi\hbar}}
      \hat{E}_{b}(t)
      \left(
      \sqrt{\frac{{\cal A}c}{2\pi\hbar}} \hat{E}_{l_{i}}^{(+)}(t) - \gamma(t)
      \right)
      +
      \left(
      \gamma(t) + \gamma^{*}(t)
      \right)
      \sqrt{\frac{{\cal A}c}{2\pi\hbar}}
      \hat{E}_{b}(t)
      \right.
      \nonumber\\
  && \quad\quad\quad
     \left.
      +
      \frac{1-2\eta}{2\sqrt{\eta(1-\eta)}}
      \left(
      \sqrt{\frac{{\cal A}c}{2\pi\hbar}} \hat{E}_{l_{i}}(t)
      +
      \gamma(t)
      +
      2 \gamma^{*}(t)
      \right)
      \right.
      \nonumber\\
  && \quad\quad\quad\quad\quad\quad
     \left.
     \times
      \left(
      \sqrt{\frac{{\cal A}c}{2\pi\hbar}} \hat{E}_{l_{i}}^{(+)}(t)
      -
      \gamma(t)
      \right)
     \right]
      .
      \label{eq:sPt-state-bra}
\end{eqnarray}
We also use the states $|\hat{s}_{P}(t+\tau)\rangle$ and
$\langle\hat{s}_{P}(t+\tau)|$, which are given by the replacement
$t\rightarrow t+\tau$ in Eqs.~(\ref{eq:sPt-state-ket}) and
(\ref{eq:sPt-state-bra}).


We evaluate the noise-spectral density $S_{s_{Pn}}(\omega)$ for the
measurement of the power-counting operator $\hat{s}_{P}(t)$ and we
define the noise operator
$\hat{s}_{Pn}:=\hat{s}_{P}-\langle\hat{s}_{P}\rangle$.
In this section, we take into account of all contribution of the
vacuum fluctuations from the signal field $\hat{E}_{b}$ and from the
local oscillator $\hat{E}_{l_{i}}$.
Of course, it is possible to evaluate these contribution of the vacuum
fluctuations, separately, as in the case of the number counting
detector in
Sec.~\ref{sec:Quantum_Noise_in_Balanced_Homodyne_Detections_By_PhotonNumberDetectors}.
In this case, we have to evaluate the expectation value
$\langle:\hat{s}_{P}(t):\rangle$ of the normal ordered operator
$:\hat{s}_{P}(t):$ instead of the operator $\hat{s}_{P}(t)$ itself,
because the subtraction of the vacuum fluctuations from the local
oscillator is included in the definition
(\ref{eq:hatsPt-field-expression}) of the power-counting operator
$\hat{s}_{P}(t)$.
If we consistently treat these contribution of the vacuum
fluctuations, we obtain the corresponding results to the case of the
number counting operator $\hat{s}_{N}$ in
Secs.~\ref{sec:Normal_ordered_noise_spectral_density},
\ref{sec:Vacuum_fluctuations_from_main},
and~\ref{sec:Vacuum_fluctuations_from_local_oscillator}, respectively.
However, in this paper, we evaluate the noise-spectral density
$S_{s_{Pn}}(\omega)$ for the measurement of the operator
$\hat{s}_{P}(t)$ taking into account of all contributions of vacuum
fluctuations.


From the states defined in Eqs.~(\ref{eq:sPt-state-ket}) and
(\ref{eq:sPt-state-bra}), we obtain
\begin{eqnarray}
  \label{eq:CsPttau-def}
  C_{s_{Pn}}(t,\tau)
  &=&
      \frac{1}{2} \left\langle
      \hat{s}_{P}(t) \hat{s}_{P}(t+\tau)
      +
      \hat{s}_{P}(t+\tau) \hat{s}_{P}(t)
      \right\rangle
      -
      \left\langle
      \hat{s}_{P}(t)
      \right\rangle
      \left\langle
      \hat{s}_{P}(t+\tau)
      \right\rangle
  \\
  &=&
      \frac{1}{2}
      \left(
      \gamma(t+\tau) + \gamma^{*}(t+\tau)
      \right)
      \left(
      \gamma(t) + \gamma^{*}(t)
      \right)
      \nonumber\\
  && \quad
     \times
     \left\langle
     \sqrt{\frac{{\cal A}c}{2\pi\hbar}} \hat{E}_{bn}(t+\tau)
     \sqrt{\frac{{\cal A}c}{2\pi\hbar}} \hat{E}_{bn}(t)
     +
     \sqrt{\frac{{\cal A}c}{2\pi\hbar}} \hat{E}_{bn}(t)
     \sqrt{\frac{{\cal A}c}{2\pi\hbar}} \hat{E}_{bn}(t+\tau)
     \right\rangle
     \nonumber\\
  &&
     +
     \frac{1}{2}
     \Delta_{l_{i}}(t)
     \left\langle
     \sqrt{\frac{{\cal A}c}{2\pi\hbar}} \hat{E}_{b}(t+\tau)
     \sqrt{\frac{{\cal A}c}{2\pi\hbar}} \hat{E}_{b}(t)
     \right\rangle
     \nonumber\\
  &&
     +
     \frac{1}{2}
     \Delta_{l_{i}}(-\tau)
     \left\langle
     \sqrt{\frac{{\cal A}c}{2\pi\hbar}} \hat{E}_{b}(t)
     \sqrt{\frac{{\cal A}c}{2\pi\hbar}} \hat{E}_{b}(t+\tau)
     \right\rangle
     \nonumber\\
  &&
     +
     \frac{1-2\eta}{2\sqrt{\eta(1-\eta)}}
     \left(
     \Delta_{l_{i}}(\tau)
     +
     \Delta_{l_{i}}(-\tau)
     \right)
     \left[
     \left(
     \gamma(t+\tau) + \gamma^{*}(t+\tau)
     \right)
     \left\langle
     \sqrt{\frac{{\cal A}c}{2\pi\hbar}} \hat{E}_{b}(t)
     \right\rangle
     \right.
     \nonumber\\
  && \quad\quad\quad\quad\quad\quad\quad\quad\quad\quad\quad\quad\quad\quad\quad\quad
     \left.
     +
     \left(
     \gamma(t) + \gamma^{*}(t)
     \right)
     \left\langle
     \sqrt{\frac{{\cal A}c}{2\pi\hbar}} \hat{E}_{b}(t+\tau)
     \right\rangle
     \right]
     \nonumber\\
  &&
     +
     \frac{(1-2\eta)^{2}}{4\eta(1-\eta)}
     \left(
     \Delta_{l_{i}}(\tau) \Delta_{l_{i}}(\tau)
     +
     \Delta_{l_{i}}(-\tau) \Delta_{l_{i}}(-\tau)
     \right)
     \nonumber\\
  &&
     +
     \frac{(1-2\eta)^{2}}{2\eta(1-\eta)}
     \left(
     \Delta_{l_{i}}(\tau)
     +
     \Delta_{l_{i}}(-\tau)
     \right)
     \left(
     \gamma^{*}(t)
     +
     \gamma(t)
     \right)
     \left(
     \gamma^{*}(t+\tau)
     +
     \gamma(t+\tau)
     \right)
     .
     \label{eq:CsPttau-explicit}
\end{eqnarray}
All terms which include $\Delta_{l_{i}}(\pm\tau)$ are contributions
from the vacuum fluctuations of the local oscillator.
The averaged noise correlation function $C_{({\rm av})s_{Pn}}(\tau)$ and the
noise spectral density $S_{s_{Pn}}(\omega)$ are given by
\begin{eqnarray}
  \label{eq:sPn-CsPttau-correlation-defs}
  C_{({\rm av})s_{Pn}}(\tau)
  &:=&
       \lim_{T\rightarrow+\infty} \frac{1}{T} \int_{-T/2}^{T/2} dt
       C_{s_{Pn}}(t,\tau)
       ,
  \\
  S_{s_{Pn}}(\omega)
  &:=&
       \int_{-\infty}^{+\infty} d\tau e^{+i\omega\tau} C_{({\rm av})sPn}(\tau)
       =:
       \sum_{i=1}^{5} {\cal K}_{i}
       .
       \label{eq:sPn-noise-spectral-density-defs}
\end{eqnarray}
Here, we evaluate the noise spectral density $S_{s_{Pn}}(\omega)$
defined by Eq.~(\ref{eq:sPn-noise-spectral-density-defs}) only in the
case of the monochromatic local oscillator case.
In this case, ${\cal K}_{i}$ $(i=1,2,3,4,5)$ should also be evaluated
in this monochromatic case and these are defined as
\begin{eqnarray}
  {\cal K}_{1}(\omega)
  &:=&
       \int_{-\infty}^{+\infty} d\tau e^{+i\omega\tau}
       \lim_{T\rightarrow+\infty} \frac{1}{T} \int_{-T/2}^{T/2} dt
       \frac{1}{2} \omega_{0} |\gamma|^{2}
       \left(
       e^{+i\theta} e^{-i\omega_{0}t}
       +
       e^{-i\theta} e^{+i\omega_{0}t}
       \right)
       \nonumber\\
  &&
     \times
     \left\langle
     \sqrt{\frac{{\cal A}c}{2\pi\hbar}} \hat{E}_{bn}(t+\tau)
     \sqrt{\frac{{\cal A}c}{2\pi\hbar}} \hat{E}_{bn}(t)
     +
     \sqrt{\frac{{\cal A}c}{2\pi\hbar}} \hat{E}_{bn}(t)
     \sqrt{\frac{{\cal A}c}{2\pi\hbar}} \hat{E}_{bn}(t+\tau)
     \right\rangle
     \nonumber\\
  &&
     \times
     \left(
     e^{+i\theta} e^{-i\omega_{0}(t+\tau)}
     +
     e^{-i\theta} e^{+i\omega_{0}(t+\tau)}
     \right)
     ,
     \label{eq:calK1omega-def}
\end{eqnarray}
\begin{eqnarray}
  \label{eq:calK2omega-def}
  {\cal K}_{2}(\omega)
  &:=&
       \frac{1}{2}
       \int_{-\infty}^{+\infty} d\tau e^{+i\omega\tau}
       \lim_{T\rightarrow+\infty} \frac{1}{T} \int_{-T/2}^{T/2} dt
       \nonumber\\
  && \quad
     \times
       \left[
       \Delta_{l_{i}}(\tau)
       \left\langle
       \sqrt{\frac{{\cal A}c}{2\pi\hbar}} \hat{E}_{b}(t+\tau)
       \sqrt{\frac{{\cal A}c}{2\pi\hbar}} \hat{E}_{b}(t)
       \right\rangle
       \right.
       \nonumber\\
  && \quad\quad\quad\quad
     \left.
     +
     \Delta_{l_{i}}(-\tau)
     \left\langle
     \sqrt{\frac{{\cal A}c}{2\pi\hbar}} \hat{E}_{b}(t)
     \sqrt{\frac{{\cal A}c}{2\pi\hbar}} \hat{E}_{b}(t+\tau)
     \right\rangle
     \right]
     ,
\end{eqnarray}
\begin{eqnarray}
  \label{eq:calK3omega-def}
  {\cal K}_{3}(\omega)
  &:=&
       \frac{1-2\eta}{2\sqrt{\eta(1-\eta)}}
       \int_{-\infty}^{+\infty} d\tau e^{+i\omega\tau}
       \lim_{T\rightarrow+\infty} \frac{1}{T} \int_{-T/2}^{T/2} dt
       \left(
       \Delta_{l_{i}}(\tau) + \Delta_{l_{i}}(-\tau)
       \right)
       \sqrt{\omega_{0}} |\gamma|
       \nonumber\\
  && \quad
     \times
     \left[
     \left(
     e^{+i\theta} e^{-i\omega_{0}(t+\tau)}
     +
     e^{-i\theta} e^{+i\omega_{0}(t+\tau)}
     \right)
     \left\langle
     \sqrt{\frac{{\cal A}c}{2\pi\hbar}} \hat{E}_{b}(t)
     \right\rangle
     \right.
     \nonumber\\
  && \quad\quad\quad\quad
     \left.
     +
     \left(
     e^{+i\theta} e^{-i\omega_{0}t}
     +
     e^{-i\theta} e^{+i\omega_{0}t}
     \right)
     \left\langle
     \sqrt{\frac{{\cal A}c}{2\pi\hbar}} \hat{E}_{b}(t+\tau)
     \right\rangle
     \right]
     ,
\end{eqnarray}
\begin{eqnarray}
  \label{eq:calK4omega-def}
  {\cal K}_{4}(\omega)
  &:=&
       \frac{(1-2\eta)^{2}}{2\eta(1-\eta)}
       \int_{-\infty}^{+\infty} d\tau e^{+i\omega\tau}
       \lim_{T\rightarrow+\infty} \frac{1}{T} \int_{-T/2}^{T/2} dt
       \omega_{0} |\gamma|^{2}
       \left(
       e^{-i\theta} e^{+i\omega_{0}t}
       +
       e^{+i\theta} e^{-i\omega_{0}t}
       \right)
       \nonumber\\
  && \quad
     \times
     \left(
     e^{+i\theta} e^{-i\omega_{0}(t+\tau)}
     +
     e^{-i\theta} e^{+i\omega_{0}(t+\tau)}
     \right)
     \left(
     \Delta_{l_{i}}(\tau)
     +
     \Delta_{l_{i}}(-\tau)
     \right)
     ,
\end{eqnarray}
\begin{eqnarray}
  \label{eq:calK5omega-def}
  {\cal K}_{5}(\omega)
  &:=&
       \frac{(1-2\eta)^{2}}{4\eta(1-\eta)}
       \int_{-\infty}^{+\infty} d\tau e^{+i\omega\tau}
       \lim_{T\rightarrow+\infty} \frac{1}{T} \int_{-T/2}^{T/2} dt
       \nonumber\\
  && \quad
     \times
       \left(
       \Delta_{l_{i}}(\tau) \Delta_{l_{i}}(\tau)
       +
       \Delta_{l_{i}}(-\tau) \Delta_{l_{i}}(-\tau)
       \right)
       .
\end{eqnarray}


First, we evaluate ${\cal K}_{1}(\omega)$ defined by
Eq.~(\ref{eq:calK1omega-def}).
We use $\hat{E}_{bn}(t)=\hat{E}_{bn}^{(+)}(t)+\hat{E}_{bn}^{(-)}(t)$
and Eqs.~(\ref{eq:signal-electric-field-noise-positive}).
Furthermore, the situation $\omega_{0}\gg\omega>0$ and the same reason
that we used in the derivation of
Eq.~(\ref{eq:calI1omega-general-delta0-expression-2}) from
Eq.~(\ref{eq:normal-ordered-noise-spectral-density-monochro-calI1})
in the ${\cal I}_{1}(\omega)$ case leads
\begin{eqnarray}
  2 \pi \delta(\omega-\omega') {\cal K}_{1}(\omega)
  &\sim&
         \omega_{0}^{2} |\gamma|^{2}
         \left[
         e^{+2i\theta}
         \left\langle
         \hat{b}_{n}^{\dagger}(\omega_{0}-\omega)
         \hat{b}_{n}^{\dagger}(\omega_{0}+\omega')
         \right\rangle
         \right.
         \nonumber\\
  && \quad\quad\quad\quad
     \left.
     +
     \frac{1}{2}
     \left\langle
     \hat{b}_{n}^{\dagger}(\omega_{0}-\omega)
     \hat{b}_{n}(\omega_{0}-\omega')
     \right\rangle
     \right.
     \nonumber\\
  && \quad\quad\quad\quad
     \left.
     +
     \frac{1}{2}
     \left\langle
     \hat{b}_{n}(\omega_{0}-\omega')
     \hat{b}_{n}^{\dagger}(\omega_{0}-\omega)
     \right\rangle
     \right.
     \nonumber\\
  && \quad\quad\quad\quad
     \left.
     +
     \frac{1}{2}
     \left\langle
     \hat{b}_{n}^{\dagger}(\omega_{0}+\omega)
     \hat{b}_{n}(\omega_{0}+\omega')
     \right\rangle
     \right.
     \nonumber\\
  && \quad\quad\quad\quad
     \left.
     +
     \frac{1}{2}
     \left\langle
     \hat{b}_{n}(\omega_{0}+\omega')
     \hat{b}_{n}^{\dagger}(\omega_{0}+\omega)
     \right\rangle
     \right.
     \nonumber\\
  && \quad\quad\quad\quad
     \left.
     +
     e^{-2i\theta}
     \left\langle
     \hat{b}_{n}(\omega_{0}+\omega)
     \hat{b}_{n}(\omega_{0}-\omega')
     \right\rangle
     \right]
     .
     \label{eq:calK1omega-general-delta-expression}
\end{eqnarray}
At this moment, the frequency $\omega_{0}$ is just the central
frequency of the field $\hat{E}_{l_{i}}(t)$ from the local oscillator
and have nothing to do with the central frequency of the signal field
$\hat{E}_{b}(t)$.
Therefore, Eq.~(\ref{eq:calK1omega-general-delta-expression}) is also
valid even if the central frequency of the field $\hat{E}_{l_{i}}(t)$
does not coincide with the central frequency of the field
$\hat{E}_{b}(t)$, which is the ``{\it heterodyne detection}.''
On the other hand, if the central frequency $\omega_{0}$ of the
optical field from the local oscillator coincides with the central
frequency from the main interferometer, which is the ``{\it homodyne
  detection}'', we can use the sideband picture
$\hat{b}_{\pm}(\omega):=\hat{b}(\omega_{0}\pm\omega)$ and the above
$2\pi\delta(\omega-\omega') {\cal K}_{1}(\omega)$ is given by
\begin{eqnarray}
  \label{eq:calK1omega-general-delta-expression-homodyne}
  2 \pi \delta(\omega-\omega') {\cal K}_{1}(\omega)
  &\sim&
         \omega_{0}^{2} |\gamma|^{2}
         \left[
         e^{+2i\theta}
         \left\langle
         \hat{b}_{n-}^{\dagger}(\omega)
         \hat{b}_{n+}^{\dagger}(\omega')
         \right\rangle
         +
         \frac{1}{2}
         \left\langle
         \hat{b}_{n-}^{\dagger}(\omega)
         \hat{b}_{n-}(\omega')
         \right\rangle
         \right.
         \nonumber\\
  && \quad\quad\quad
     \left.
     +
     \frac{1}{2}
     \left\langle
     \hat{b}_{n-}(\omega')
     \hat{b}_{n-}^{\dagger}(\omega)
     \right\rangle
     +
     \frac{1}{2}
     \left\langle
     \hat{b}_{n+}^{\dagger}(\omega)
     \hat{b}_{n+}(\omega')
     \right\rangle
     \right.
     \nonumber\\
  && \quad\quad\quad
     \left.
     +
     \frac{1}{2}
     \left\langle
     \hat{b}_{n+}(\omega')
     \hat{b}_{n+}^{\dagger}(\omega)
     \right\rangle
     +
     e^{-2i\theta}
     \left\langle
     \hat{b}_{n+}(\omega)
     \hat{b}_{n-}(\omega')
     \right\rangle
     \right]
     .
\end{eqnarray}
Furthermore, using the definitions (\ref{eq:hatb1-hatb2-def}) of the
amplitude- and phase-quadrature and their noise operators, the
definition (\ref{eq:DBHD_20180805_1.1}) of the
$\hat{b}_{\theta}(\omega)$ and its noise operator
$\hat{b}_{\theta n}$, we obtain
\begin{eqnarray}
  \label{eq:calK1omega-general-delta-expression-homodyne-bthetan}
  2 \pi \delta(\omega-\omega') {\cal K}_{1}(\omega)
  &\sim&
         \omega_{0}^{2} |\gamma|^{2}
         \left\langle
         \hat{b}_{\theta n}(\omega)
         \hat{b}_{\theta n}^{\dagger}(\omega')
         +
         \hat{b}_{\theta n}^{\dagger}(\omega')
         \hat{b}_{\theta n}(\omega)
         \right\rangle
         .
\end{eqnarray}
Since we consider the situation $\omega,\omega'>0$, the right-hand
side of Eq.~(\ref{eq:calK1omega-general-delta-expression-homodyne})
is proportional to the single-sideband noise-spectral density
$\bar{S}_{b_{\theta}}^{(s)}(\omega)$ introduced in Ref.~\cite{H.J.Kimble-Y.Levin-A.B.Matsko-K.S.Thorne-S.P.Vyatchanin-2001}.
Then, we obtain
\begin{eqnarray}
  \label{eq:calK1omega-general-delta-expression-noise-spectral-density}
  {\cal K}_{1}(\omega)
  \sim
  \omega_{0}^{2} |\gamma|^{2}
  \bar{S}_{b_{\theta}}^{(s)}(\omega)
  .
\end{eqnarray}


Second, we evaluate ${\cal K}_{2}(\omega)$ defined by
Eq.~(\ref{eq:calK2omega-def}).
Here, we use the separation
$\hat{E}_{b}(t)=\hat{E}_{bn}(t)+\left\langle\hat{E}_{b}(t)\right\rangle$.
Then, ${\cal K}_{2}(\omega)$ is separated into two parts as
\begin{eqnarray}
  \label{eq:calK2omega-calK2-1omega+calK2-2omega}
  {\cal K}_{2}(\omega)
  =
  {\cal K}_{2-1}(\omega)
  +
  {\cal K}_{2-2}(\omega)
  ,
\end{eqnarray}
where
\begin{eqnarray}
  \label{eq:calK2-1omega-def}
  {\cal K}_{2-1}(\omega)
  &:=&
       \frac{1}{2}
       \int_{-\infty}^{+\infty} d\tau e^{+i\omega\tau}
       \lim_{T\rightarrow+\infty} \frac{1}{T} \int_{-T/2}^{T/2} dt
       \nonumber\\
  && \quad
     \times
     \left[
     \Delta_{l_{i}}(\tau)
     \left\langle
     \sqrt{\frac{{\cal A}c}{2\pi\hbar}} \hat{E}_{bn}(t+\tau)
     \sqrt{\frac{{\cal A}c}{2\pi\hbar}} \hat{E}_{bn}(t)
     \right\rangle
     \right.
     \nonumber\\
  && \quad\quad\quad\quad
     \left.
     +
     \Delta_{l_{i}}(-\tau)
     \left\langle
     \sqrt{\frac{{\cal A}c}{2\pi\hbar}} \hat{E}_{bn}(t)
     \sqrt{\frac{{\cal A}c}{2\pi\hbar}} \hat{E}_{bn}(t+\tau)
     \right\rangle
     \right]
     ,
\end{eqnarray}
and
\begin{eqnarray}
  \label{eq:calK2-2omega-def}
  {\cal K}_{2-2}(\omega)
  &:=&
       \frac{1}{2}
       \int_{-\infty}^{+\infty} d\tau e^{+i\omega\tau}
       \lim_{T\rightarrow+\infty} \frac{1}{T} \int_{-T/2}^{T/2} dt
       \nonumber\\
  && \quad
     \times
     \left[
     \Delta_{l_{i}}(\tau)
     \left\langle
     \sqrt{\frac{{\cal A}c}{2\pi\hbar}} \hat{E}_{b}(t+\tau)
     \right\rangle
     \left\langle
     \sqrt{\frac{{\cal A}c}{2\pi\hbar}} \hat{E}_{b}(t)
     \right\rangle
     \right.
     \nonumber\\
  && \quad\quad\quad\quad
     \left.
     +
     \Delta_{l_{i}}(-\tau)
     \left\langle
     \sqrt{\frac{{\cal A}c}{2\pi\hbar}} \hat{E}_{b}(t)
     \right\rangle
     \left\langle
     \sqrt{\frac{{\cal A}c}{2\pi\hbar}} \hat{E}_{b}(t+\tau)
     \right\rangle
     \right]
     .
\end{eqnarray}
We evaluate ${\cal K}_{2-1}(\omega)$ and ${\cal K}_{2-2}(\omega)$,
separately.


Here, we evaluate ${\cal K}_{2-1}(\omega)$.
To evaluate this, we use the Fourier decomposition
(\ref{eq:signal-electric-field-noise-positive}) of $\hat{E}_{bn}(t)$.
All terms in ${\cal K}_{2-1}(\omega)$ includes three integration by
$\int_{0}^{\infty}\frac{d\omega_{1}}{2\pi}$,
$\int_{0}^{\infty}\frac{d\omega_{2}}{2\pi}$, and
$\int_{0}^{\infty}\frac{d\omega_{3}}{2\pi}$ due to the Fourier
transformation of $\hat{E}_{bn}(t+\tau)$, $\hat{E}_{bn}(t)$, and
vacuum fluctuations $\Delta_{l_{i}}$, respectively.
We also note that ${\cal K}_{2-1}(\omega)$ includes the term which
have the factor
\begin{eqnarray}
  \label{eq:average-omega1+omega2}
  \lim_{T\rightarrow+\infty} \frac{1}{T} \int_{-T/2}^{T/2} dt
  e^{-i(\omega_{1}+\omega_{2})t}.
\end{eqnarray}
Since the integrations range over $\omega_{1}$ and $\omega_{2}$ is
$[0,\infty)$, the term which includes the factor
(\ref{eq:average-omega1+omega2}) vanishes.
This is due to the fact that the support of the factor
(\ref{eq:average-omega1+omega2}) is only on the single point
$\omega_{1}=-\omega_{2}$.
However, this point $\omega_{1}=-\omega_{2}$ is out of range of the
integration over $\omega_{1}$ and $\omega_{2}$.
Therefore, all terms which have the factor
(\ref{eq:average-omega1+omega2}) vanish.


On the other hand, the terms including the factor
\begin{eqnarray}
  \label{eq:average-omega1-omega2}
  \lim_{T\rightarrow+\infty} \frac{1}{T} \int_{-T/2}^{T/2} dt
  e^{-i(\omega_{1}-\omega_{2})t}
\end{eqnarray}
may give finite contributions.
From these discrimination of the terms appear in
${\cal K}_{2-1}(\omega)$, we obtain
\begin{eqnarray}
  {\cal K}_{2-1}(\omega)
  &=&
      \frac{1}{2}
      \int_{0}^{\infty} \frac{d\omega_{1}}{2\pi}
      \int_{0}^{\infty} \frac{d\omega_{2}}{2\pi}
      \sqrt{\omega_{1}\omega_{2}}
      \left[
      \Theta(\omega_{1}+\omega)
      (\omega_{1}+\omega)
      +
      \Theta(\omega_{2}-\omega)
      (\omega_{2}-\omega)
      \right]
      \nonumber\\
  && \quad\quad\quad\quad\quad
     \times
     \left\langle \hat{b}_{n}^{\dagger}(\omega_{1})\hat{b}_{n}(\omega_{2}) \right\rangle
     \lim_{T\rightarrow+\infty} \frac{1}{T} \int_{-T/2}^{T/2} dt e^{+i(\omega_{1}-\omega_{2})t}
     \nonumber\\
  &&
     +
     \frac{1}{2}
     \int_{0}^{\infty} \frac{d\omega_{1}}{2\pi}
     \int_{0}^{\infty} \frac{d\omega_{2}}{2\pi}
     \Theta(\omega-\omega_{1}) \sqrt{\omega_{1}\omega_{2}}
     (\omega-\omega_{1})
      \nonumber\\
  && \quad\quad\quad\quad\quad\quad
     \times
     \left\langle
     \hat{b}_{n}(\omega_{1})\hat{b}_{n}^{\dagger}(\omega_{2})
     \right\rangle
     \lim_{T\rightarrow+\infty} \frac{1}{T} \int_{-T/2}^{T/2} dt e^{+i(\omega_{2}-\omega_{1})t}
     ,
     \label{eq:calK2-1omega-tmp}
\end{eqnarray}
where we used the formula
\begin{eqnarray}
  \label{eq:integration-deltafunction-half-range-formula}
  \int_{0}^{+\infty} \frac{d\omega_{3}}{2\pi}
  2 \pi \delta(\omega_{3}-a) f(\omega_{3})
  =
  \Theta(a) f(a)
  .
\end{eqnarray}
Furthermore, the expectation values of the quadratures in
Eq.~(\ref{eq:calK2-1omega-tmp}) are given by the noise-spectral
density $\SF_{b_{n}}(\omega)$ defined by
Eq.~(\ref{eq:calS-bn-omega-def}).
For example,
\begin{eqnarray}
  \label{eq:exphatbndaggerhatbn}
  \left\langle
  \hat{b}_{n}^{\dagger}(\omega_{1})
  \hat{b}_{n}(\omega_{2})
  \right\rangle
  &=&
      \frac{1}{2}
      \left\langle
      \hat{b}_{n}^{\dagger}(\omega_{1}) \hat{b}_{n}(\omega_{2})
      +
      \hat{b}_{n}(\omega_{2}) \hat{b}_{n}^{\dagger}(\omega_{1})
      +
      \left[
      \hat{b}_{n}^{\dagger}(\omega_{1}), \hat{b}_{n}(\omega_{2})
      \right]
      \right\rangle
      \nonumber\\
  &=&
      \frac{1}{2}
      \left\langle
      \hat{b}_{n}^{\dagger}(\omega_{1}) \hat{b}_{n}(\omega_{2})
      +
      \hat{b}_{n}(\omega_{2}) \hat{b}_{n}^{\dagger}(\omega_{1})
      \right\rangle
      -
      \frac{1}{2}
      2 \pi \delta(\omega_{1}-\omega_{2})
      \nonumber\\
  &=&
      \frac{1}{2}
      2 \pi \delta(\omega_{1}-\omega_{2})
      \left(
      \SF_{b_{n}}(\omega_{1})
      -
      1
      \right)
      .
\end{eqnarray}
Similarly, we obtain
\begin{eqnarray}
  \label{eq:exphatbndaggerhatbn-2}
  \left\langle
  \hat{b}_{n}(\omega_{1})
  \hat{b}_{n}^{\dagger}(\omega_{2})
  \right\rangle
  &=&
      \frac{1}{2}
      2 \pi \delta(\omega_{1}-\omega_{2})
      \left(
      \SF_{b_{n}}(\omega_{1})
      +
      1
      \right)
      .
\end{eqnarray}
Through Eqs.~(\ref{eq:exphatbndaggerhatbn}) and
(\ref{eq:exphatbndaggerhatbn-2}), ${\cal K}_{2-1}(\omega)$ in
Eq.~(\ref{eq:calK2-1omega-tmp}) is given by
\begin{eqnarray}
  {\cal K}_{2-1}(\omega)
  &=&
      \frac{1}{2} \int_{0}^{\infty} \frac{d\omega_{1}}{2\pi}
      \omega_{1}^{2}
      \left(
      \SF_{b_{n}}(\omega_{1})
      -
      1
      \right)
      +
      \frac{1}{2}
      \int_{0}^{\omega} \frac{d\omega_{1}}{2\pi}
      \omega_{1}
      (\omega-\omega_{1})
      \SF_{b_{n}}(\omega_{1})
      ,
      \label{eq:calK2-1omega-result}
\end{eqnarray}


Similar calculations to the case of ${\cal J}_{2}(\omega)$ in
Eq.~(\ref{eq:calJ2tau-def}) with the expectation value
(\ref{eq:output-quadrature-expectation-value-main}) yield
\begin{eqnarray}
  \label{eq:calK2-2omega-result}
  {\cal K}_{2-2}(\omega)
  =
  \omega_{0}^{2} |\beta|^{2}
  .
\end{eqnarray}


From Eqs.~(\ref{eq:calK2omega-calK2-1omega+calK2-2omega}),
(\ref{eq:calK2-1omega-result}), and (\ref{eq:calK2-2omega-result}), we
obtain the final result of ${\cal K}_{2}(\omega)$ as
\begin{eqnarray}
  \label{eq:calK2omega-result}
  {\cal K}_{2}(\omega)
  =
  \frac{1}{2} \int_{0}^{\infty} \frac{d\omega_{1}}{2\pi}
  \omega_{1}^{2}
  \left(
  \SF_{b_{n}}(\omega_{1})
  -
  1
  \right)
  +
  \frac{1}{2}
  \int_{0}^{\omega} \frac{d\omega_{1}}{2\pi}
  \omega_{1}
  (\omega-\omega_{1})
  \SF_{b_{n}}(\omega_{1})
  +
  \omega_{0}^{2} |\beta|^{2}
  .
\end{eqnarray}


Third, we evaluate ${\cal K}_{3}(\omega)$ defined by
Eq.~(\ref{eq:calK3omega-def}).
Here, we use the separation
$\hat{E}_{b}(t)=\hat{E}_{bn}(t)+\left\langle\hat{E}_{b}(t)\right\rangle$,
and the Fourier transformation
(\ref{eq:signal-electric-field-noise-positive}), the explicit
expression (\ref{eq:commutation-relation-E+E-}) of the vacuum
fluctuation $\Delta_{l_{i}}(\tau)$, and the properties of the averaged
function function (\ref{eq:lim1overTintdte-iat-def-main}) which are
summarized in
Appendix~\ref{sec:Properties_of_a_time-averaged_function}.
Furthermore, we use the situation $\omega_{0}\gg\omega>0$ and the
expression of the expectation value $\langle\hat{b}(\omega)\rangle$
given by Eq.~(\ref{eq:output-quadrature-expectation-value-main}).
Then, ${\cal K}_{3}(\omega)$ is given by
\begin{eqnarray}
  \label{eq:calK3omega-result}
  {\cal K}_{3}(\omega)
  &:=&
       \frac{1-2\eta}{\sqrt{\eta(1-\eta)}} \omega_{0}^{2} |\gamma|
       \left[
       e^{-i\theta} \beta + e^{+i\theta} \beta^{*}
       \right]
       .
\end{eqnarray}


Finally, we evaluate ${\cal K}_{4}(\omega)$ and ${\cal K}_{5}(\omega)$
defined by Eqs.~(\ref{eq:calJ4tau-def}) and (\ref{eq:calJ5tau-def}),
respectively.
In this evaluation, we use the explicit expression of the vacuum
fluctuation $\Delta_{l_{i}}(\tau)$ and the situation
$\omega_{0}\gg\omega>0$.
Then, we obtain the following results:
\begin{eqnarray}
  \label{eq:calK4omega-result}
  {\cal K}_{4}(\omega)
  &=&
      \frac{(1-2\eta)^{2}}{\eta(1-\eta)}
      \omega_{0}^{2} |\gamma|^{2}
      ;
  \\
  \label{eq:calK5omega-result}
  {\cal K}_{5}(\omega)
  &=&
      \frac{(1-2\eta)^{2}}{4\eta(1-\eta)}
      \int_{0}^{\omega} \frac{d\omega_{1}}{2\pi} \omega_{1}
      (\omega-\omega_{1})
      .
\end{eqnarray}


In summary, from Eq.~(\ref{eq:sPn-noise-spectral-density-defs}), and
Eqs.~(\ref{eq:calK1omega-general-delta-expression-noise-spectral-density}),
(\ref{eq:calK2omega-result}), (\ref{eq:calK3omega-result}),
(\ref{eq:calK4omega-result}), and (\ref{eq:calK5omega-result}),
we have obtained the noise spectral density
$S_{s_{Pn}}(\omega)$ in the situation $\omega_{0}\gg\omega>0$ as
\begin{eqnarray}
  S_{s_{Pn}}(\omega)
  &=&
      \omega_{0}^{2} |\gamma|^{2}
      \bar{S}_{b_{\theta}}^{(s)}(\omega)
      \nonumber\\
  &&
     +
     \frac{1}{2} \int_{0}^{\infty} \frac{d\omega_{1}}{2\pi}
     \omega_{1}^{2}
     \left(
     \SF_{b_{n}}(\omega_{1})
     -
     1
     \right)
     +
     \frac{1}{2}
     \int_{0}^{\omega} \frac{d\omega_{1}}{2\pi}
     \omega_{1}
     (\omega-\omega_{1})
     \SF_{b_{n}}(\omega_{1})
     +
     \omega_{0}^{2} |\beta|^{2}
     \nonumber\\
  &&
     +
     \frac{1-2\eta}{\sqrt{\eta(1-\eta)}} \omega_{0}^{2} |\gamma|
     \left[
     e^{-i\theta} \beta + e^{+i\theta} \beta^{*}
     \right]
     \nonumber\\
  &&
     +
     \frac{(1-2\eta)^{2}}{\eta(1-\eta)}
     \left(
     \omega_{0}^{2} |\gamma|^{2}
     +
     \int_{0}^{\omega} \frac{d\omega_{1}}{2\pi} \omega_{1}
     (\omega-\omega_{1})
     \right)
     .
     \label{eq:noise-spectral-power-counting-general-result}
\end{eqnarray}


In the case of the ideal beam splitter $\eta=1/2$ and the complete
dark port $\beta=0$ of the main interferometer, the noise spectral
density (\ref{eq:noise-spectral-power-counting-general-result}) yields
\begin{eqnarray}
  S_{s_{Pn}}(\omega)
  &=&
      \omega_{0}^{2} |\gamma|^{2}
      \bar{S}_{b_{\theta}}^{(s)}(\omega)
     \nonumber\\
  &&
     +
     \frac{1}{2}
     \int_{0}^{\infty} \frac{d\omega_{1}}{2\pi}
     \omega_{1}^{2}
     \left(
     \SF_{b_{n}}(\omega_{1}) - 1
     \right)
     +
     \frac{1}{2}
     \int_{0}^{\omega} \frac{d\omega_{1}}{2\pi}
     \omega_{1} (\omega-\omega_{1})
     \SF_{b_{n}}(\omega_{1})
     .
     \label{eq:noise-spectral-power-counting-ideal-result}
\end{eqnarray}
This noise spectral density is slightly different from the noise
spectral density
(\ref{eq:sNn-noise-spectral-density-total-full-ideal}) for the
number-counting detector.


\subsection{Local oscillator from the main interferometer}
\label{sec:Local_oscillator_from_the_main_interferometer}

\begin{figure}[ht]
  \centering
  \includegraphics[width=0.7\textwidth]{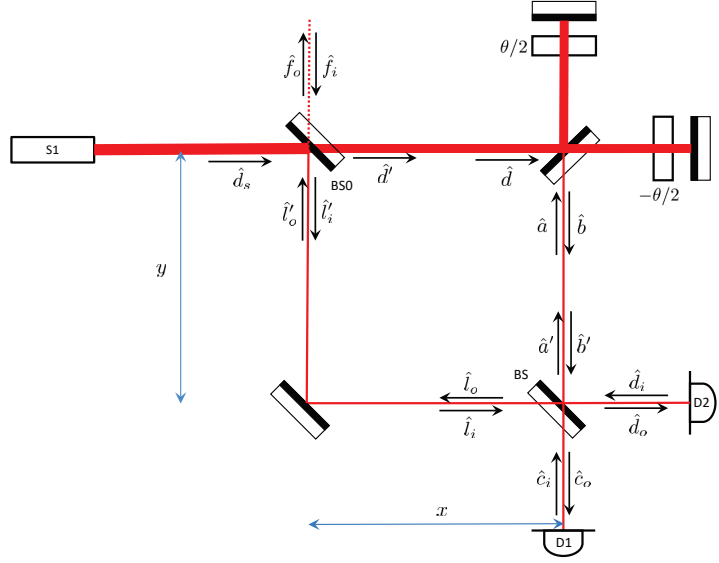}
  \caption{
    Introduction of the optical field of the local oscillator from the
    main interferometer.
  }
  \label{fig:MichelsonWithBHDByLocalOsciFromMain}
\end{figure}

Here, we consider the introduction of the optical field as the local
oscillator from the main interferometer as depicted in
Fig.~\ref{fig:MichelsonWithBHDByLocalOsciFromMain}.
We consider the optical field junction at the beam splitter BS0 with
the transmissivity $\zeta$.
In addition to the above notation, $\hat{E}_{d_{s}}(t)$ is the
incident field directly from the light source,
$\hat{E}_{d'}(t)$ is the output field from the beam splitter BS0 to
the main interferometer, $\hat{E}_{l'_{i}}(t)$ is the output field to
the local oscillator, and $\hat{E}_{f_{i}}(t)$ is the incident field
to BS0 as depicted in
Fig.~\ref{fig:MichelsonWithBHDByLocalOsciFromMain}.
We assume that the state of the optical field $\hat{E}_{f_{i}(t)}$ is
in the vacuum state $|0\rangle_{f_{i}}$, i.e.,
$\hat{f}_{i}|0\rangle_{f_{i}}=0$.
The optical field junction conditions at BS0 yield
\begin{eqnarray}
  \label{eq:BS0-junction-electric-field}
  \hat{E}_{d'}(t)
  =
  \sqrt{\zeta} \hat{E}_{d_{s}}(t)
  +
  \sqrt{1-\zeta} \hat{E}_{f_{i}}(t),
  \quad
  \hat{E}_{l'_{i}}(t)
  =
  - \sqrt{1-\zeta} \hat{E}_{d_{s}}(t)
  +
  \sqrt{\zeta} \hat{E}_{f_{i}}(t).
  .
\end{eqnarray}
These conditions are equivalent to
Eqs.~(\ref{eq:BS0-junction-condition-1}) and
(\ref{eq:BS0-junction-condition-2}).
The propagation of the optical field $\hat{E}_{d'}(t)$ to the main
interferometer yields $\hat{E}_{d}(t)=\hat{E}_{d'}(t-x/c)$ and the
propagation of the optical field $\hat{E}_{l'_{i}}(t)$ to BS of the
balanced homodyne detection yields
$\hat{E}_{l_{i}}(t)=\hat{E}(t-(x+y)/c)$.
These are equivalent to the conditions
(\ref{eq:hatd-hatdprimee-iomegax}) and
(\ref{eq:hatli-hatliprimee-iomegax+y}), respectively.


Here, we consider the situation where the optical field from the laser
source S1 in Fig.~\ref{fig:MichelsonWithBHDByLocalOsciFromMain} is a
coherent state with the complex amplitude $\gamma_{s}(t)$, i.e.,
\begin{eqnarray}
  \label{eq:ds-coherent-state}
  \hat{E}^{(+)}_{d_{s}}(t) |\Psi\rangle
  =
  \sqrt{\frac{2\pi\hbar}{{\cal A}c}} \gamma_{s}(t) |\Psi\rangle,
  \quad
  \gamma_{s}(t)
  =
  \int_{0}^{+\infty} \frac{d\omega}{2\pi} \sqrt{\omega}
  \gamma_{s}(\omega) e^{-i\omega t}
  .
\end{eqnarray}
Then, we obtain
\begin{eqnarray}
  \label{eq:lprimei+-dprime-coherent-state}
  \hat{E}_{l'_{i}}^{(+)}(t)|\Psi\rangle
  =
  \sqrt{\frac{2\pi\hbar}{{\cal A}c}} \sqrt{1-\zeta} \gamma_{s}(t)
  |\Psi\rangle,
  \quad
  \hat{E}_{d'}^{(+)}(t)|\Psi\rangle
  =
  \sqrt{\frac{2\pi\hbar}{{\cal A}c}} \sqrt{\zeta} \gamma_{s}(t) |\Psi\rangle.
\end{eqnarray}
From the propagation conditions $\hat{E}_{d}(t)=\hat{E}_{d'}(t-x/c)$ and
$\hat{E}_{l_{i}}(t)=\hat{E}(t-(x+y)/c)$, and the vacuum condition for
the optical field $\hat{E}_{f_{i}}(t)$, we obtain
\begin{eqnarray}
  \label{eq:lprimei+-dprime-coherent-state-propagated-1}
  \hat{E}_{l_{i}}^{(+)}(t)|\Psi\rangle
  &=&
      \sqrt{\frac{2\pi\hbar}{{\cal A}c}} \sqrt{1-\zeta} \gamma_{s}(t-(x+y)/c)
      |\Psi\rangle
      ,
      \\
  \label{eq:lprimei+-dprime-coherent-state-propagated-2}
  \hat{E}_{d}^{(+)}(t)|\Psi\rangle
  &=&
      \sqrt{\frac{2\pi\hbar}{{\cal A}c}} \sqrt{\zeta} \gamma_{s}(t-x/c)
      |\Psi\rangle
      .
\end{eqnarray}
From the definition of the coherent state
Eq.~(\ref{eq:coherent-state-def-a-time-domain}) and
(\ref{eq:alphat-alphaomega-relation}) and the above
Eqs.~(\ref{eq:lprimei+-dprime-coherent-state-propagated-1}) and
(\ref{eq:lprimei+-dprime-coherent-state-propagated-2}),
the complex amplitude $\gamma(t)$ for the coherent state
$|\gamma\rangle_{l_{i}}$ which is the eigenstate of the operator
$\hat{E}_{l_{i}}^{(+)}(t)$ is given by
\begin{eqnarray}
  \label{eq:complex-amplitude-replacement}
  \gamma(t) = \sqrt{1-\zeta} \gamma_{s}(t-(x+y)/c).
\end{eqnarray}
From the relation (\ref{eq:alphat-alphaomega-relation}) between
$\gamma(t)$ and $\gamma(\omega)$, we obtain
\begin{eqnarray}
  \label{eq:gammat-gammaomega-relation-loc.osc.-main}
  \gamma(\omega)
  =
  \sqrt{1-\zeta}
  \gamma_{s}(\omega)
  e^{+i\omega (x+y)/c}
  .
\end{eqnarray}
In the monochromatic local oscillator case, we obtain
\begin{eqnarray}
  \label{eq:complex-amplitude-replacement-freq.-domain}
  \gamma 2 \pi \delta(\omega-\omega_{0})
  =
  \sqrt{1-\zeta}
  \gamma_{s}
  e^{+i\omega_{0} (x+y)/c}
  2 \pi
  \delta(\omega-\omega_{0})
  ,
\end{eqnarray}
or equivalently
\begin{eqnarray}
  \label{eq:complex-amplitude-replacement-freq.-domain-2}
  |\gamma| e^{+i\theta}
  =
  \sqrt{1-\zeta}
  |\gamma_{s}| e^{+i(\theta_{s}+\omega_{0} (x+y)/c)}
  .
\end{eqnarray}
Then, we obtain
\begin{eqnarray}
  \label{eq:gamma-abs-phase-relation}
  |\gamma| = \sqrt{1-\zeta} |\gamma_{s}|, \quad
  \theta = \theta_{s}+\omega_{0} (x+y)/c
\end{eqnarray}


On the other hand, the commutation relation is unchanged.
Therefore, in the expression of the noise-spectral densities, we
should just replace the eigenvalue of the coherent state from the local
oscillator as Eqs.~(\ref{eq:gamma-abs-phase-relation}) with unchanged
vacuum fluctuations.


\section{Summary and Discussion}
\label{sec:Summary}


In summary, we re-examined the estimation of quantum noise in the
balanced homodyne detection.
We consider the both cases in which the direct observable is Glauber's
photon-number operator (\ref{eq:multi-mode-photon-number-def}) and the
power operator (\ref{eq:multi-mode-power-def}) of the optical field,
respectively.
In our estimation, we did not use the two-photon formulation which is
widely used in the gravitational-wave community.
We concentrate on the stationary noise of the system through the
time-average procedure.
We also carefully treat vacuum fluctuations in our noise estimation.
Furthermore, we introduce the imperfection of the beamsplitter of the
balanced homodyne detection and the leakage of the classical carrier
field from the main interferometer as the noise sources.


In spite of the introduction of the imperfections as the noise
sources, the balanced homodyne detection of the both models of
Glauber's photon-number operator and the power operator yields the
expectation value of the operator $\hat{b}_{\theta}(\omega)$ as
Eqs.~(\ref{eq:hatsnomega-expvalue-delta-gamma-sideband-picture-result})
and
(\ref{eq:hatsPt-exp-valu-Fourier-monochro-local-omega0gtromegagrt0}).
In this sense, the balanced homodyne detections enable us to measure
the operator $\hat{b}_{\theta}(\omega)$ as their expectation values.


In the noise estimation, we have derived the deviations from the
Kimble's noise spectral density
(\ref{eq:Kimble-noise-spectral-density-single-side}) which are beyond
the two-photon formulation in both models of Glauber's photon number
operator and the power operator.
As expected, the imperfection of the beamsplitter and the leakage of
the classical carrier field, which are introduced as the imperfections
of the interferometer configuration, contribute to the noise spectral
density.
These imperfections appear due to the vacuum fluctuations from the
local oscillator as shown in
Eqs.~(\ref{eq:sNn-noise-spectral-density-total-full}) and
(\ref{eq:noise-spectral-power-counting-general-result}).
As the result of the coupling with the vacuum fluctuations from the
local oscillator, the leakage of the classical carrier field and its
coupling with the imperfection of the beamsplitter leads the white
noise in both cases of Glauber's photon number operator and the
power operator.
Even if the leakage of the classical carrier field from the main
interferometer is absent, the white noise appears due to the coupling
with the vacuum fluctuation from the local oscillator the imperfection
of the beamsplitter of the homodyne detection.
In addition to these white noises, the coupling between the vacuum
fluctuations and the imperfection of the beamsplitter leads to the
frequency dependent noise in the power-counting detector model as
shown in Eq.~(\ref{eq:noise-spectral-power-counting-general-result}),
while such terms do not appear in the model of Glauber's photon-number
counting mode.
Thus, the difference between the photodetector models appears the
frequency dependence of the noise spectral densities, in principle.


Even in the ideal model where there is no leakage of the classical
carrier field from the main interferometer ($\beta=0$) and the beam
splitter of the homodyne detection is ideal ($\eta=1/2$), the noise
spectral densities
(\ref{eq:sNn-noise-spectral-density-total-full-ideal}) and
(\ref{eq:noise-spectral-power-counting-ideal-result}) have the terms
due to the coupling between the vacuum fluctuations from the local
oscillator and the low frequency fluctuations from the main
interferometer.
These ideal noise spectral densities
(\ref{eq:sNn-noise-spectral-density-total-full-ideal}) and
(\ref{eq:noise-spectral-power-counting-ideal-result}) are proportional
to the Kimble noise spectral density when the amplitude $|\gamma|$ of
the coherent state from the local oscillator is sufficiently large.
In this sense, our result supports the noise spectral densities in the
conventional two-photon formulation.
On the other hand, when the amplitude of the coherent state from the
local oscillator is small, these noise spectral densities
(\ref{eq:sNn-noise-spectral-density-total-full-ideal}) and
(\ref{eq:noise-spectral-power-counting-ideal-result}) yield the
deviations from Kimble's noise spectral density.


We evaluate the order of magnitude of these deviations.
As noted in Sec.~\ref{sec:Preliminary}, the integration range
$[0,+\infty]$ is replaced as the minimum and the maximum of the
measurement time scales $[\omega_{\min},\omega_{\max}]$ in the second
term in Eq.~(\ref{eq:sNn-noise-spectral-density-total-full-ideal}) and
the second term in Eq.~(\ref{eq:noise-spectral-power-counting-ideal-result}).
If the contribution of the noise spectral density
$\SF_{b_{n}}(\omega_{1})$ is the only vacuum fluctuations of the
operator $\hat{b}_{n}$, $\SF_{b_{n}}(\omega_{1})$ becomes unity.
In this case, the deviation from the Kimble's noise spectral density
in Eq.~(\ref{eq:sNn-noise-spectral-density-total-full-ideal}) for
the photon-number counting case vanishes.
On the other hand, only the remaining term of the deviation from the
Kimble's noise spectral density in
Eq.~(\ref{eq:noise-spectral-power-counting-ideal-result}) for the
power-counting case is the last term, which yields $\frac{1}{2}
\int_{0}^{\omega} \frac{d\omega_{1}}{2\pi}
\omega_{1}(\omega-\omega_{1}) = \frac{1}{6} \omega^{3}$.
When the output-frequency $0<\omega<10^{5}$Hz, we compare the first
term in Eq.~(\ref{eq:noise-spectral-power-counting-ideal-result}) with
the above integration, the condition that the deviation is dominate the
Kimble's noise spectral density is given by
\begin{eqnarray}
  \label{eq:deiation-order-estimate-1}
  \frac{I_{0}}{\hbar\omega_{0}}
  < 4 \times 10^{-15}\ \mbox{Hz}
  \left(\frac{1}{\bar{S}_{b_{\theta}}^{(s)}(\omega)}\right)
  \left(\frac{10^{15}\mbox{Hz}}{\omega_{0}}\right)^{2}
  \left(\frac{\omega}{10^{5}\mbox{Hz}}\right)^{3},
\end{eqnarray}
where we used $\gamma=\sqrt{I_{0}/\hbar\omega_{0}}$ and $I_{0}$ is the
power of the laser from the local oscillator.
The inequality (\ref{eq:deiation-order-estimate-1}) indicates that the
deviation from Kimble's noise spectral density is extremely small in
realistic situations.
We may regard that the noise spectral densities
(\ref{eq:sNn-noise-spectral-density-total-full-ideal}) and
(\ref{eq:noise-spectral-power-counting-ideal-result}) are regarded
identical and coincide with the Kimble's noise spectral density,
when the contribution to the noise spectral density $\SF_{b_{n}}$ is
the only vacuum fluctuations of the operator $\hat{b}_{n}$.


In the case where $\hat{b}_{n}$ is modulated in the frequency range
$[\omega_{\min},\omega_{\max}]$, the noise spectral density
$\SF_{b_{n}}(\omega)$ may not be unity.
Here, we choose $\SF_{b_{n}}(\omega)\sim O(10)$.
In this case, the dominant term in the deviations from Kimble's noise
spectral density is the second terms in
Eqs.~(\ref{eq:sNn-noise-spectral-density-total-full-ideal}) and
(\ref{eq:noise-spectral-power-counting-ideal-result}), respectively.
Comparing with the term of Kimble's noise spectral density in these
equations, the conditions that the deviations from Kimble's noise
spectral density dominates the Kimble noise spectral density is given
by
\begin{eqnarray}
  \label{eq:deiation-order-estimate-2}
  \frac{I_{0}}{\hbar\omega_{0}}
  < 10^{-14}
  \left(\frac{10^{15}\mbox{Hz}}{\omega_{0}}\right)^{2}
  \left(\frac{\omega_{\max}}{10^{5}\mbox{Hz}}\right)^{3}
  \left(\frac{1}{\bar{S}_{b_{\theta}}^{(s)}(\omega)}\right)
  \left(\frac{\SF_{b_{n}}(\omega)}{10}\right).
\end{eqnarray}
The inequality (\ref{eq:deiation-order-estimate-2}) also indicates
that the deviation from Kimble's noise spectral density is extremely
small in realistic situations.
Again, we may regard that the noise spectral densities
(\ref{eq:sNn-noise-spectral-density-total-full-ideal}) and
(\ref{eq:noise-spectral-power-counting-ideal-result}) are regarded
identical and coincide with Kimble's noise spectral density,
even when the noise spectral density $\SF_{b_{n}}$ is modulated in the
frequency range $[\omega_{\min},\omega_{\max}]$.


Even in the difference between the last term in
Eqs.~(\ref{eq:sNn-noise-spectral-density-total-full}) and
(\ref{eq:noise-spectral-power-counting-general-result}), we can
evaluate the order estimate of this difference and conclude that this
term is quite small compared with the Kimble noise spectral term.


Thus, we conclude that the noise spectral densities of the two
ideal detector models of Glauber's photon number counting and the
power counting are physically same and we cannot distinguish these
models within the analyses of this paper.
Our derived noise-spectral densities are based on the ideal
premise that the direct observable of the photodetector is Glauber's
photon number operator (\ref{eq:multi-mode-photon-number-def}) or the
power operator (\ref{eq:multi-mode-power-def}) of the optical field.
Therefore, our derived noise spectral densities will characterize
these ideal photodetectors model.
Since these ideal models are not fully supported from theoretical
point of view as mentioned in Sec.~\ref{sec:Introduction}, it will be
better to keep in our mind the possibility that the deviations from
the Kimble noise-spectral density might also appear due to the
deviations of the physical properties of actual photodetectors from
our ideal photodetector models.


On the other hand, in the case where the directly measured operator of the
photo-detector is the number operator
(\ref{eq:each-mode-photon-number-def}) of each frequency modes as in
Refs.~\cite{K.Nakamura-M.-K.Fujimoto-double-balanced-letter,K.Nakamura-M.-K.Fujimoto-double-balanced-full},
we cannot measure the operator $\hat{b}_{\theta}(\omega)$ by the
balanced homodyne detection.
The arguments in
Refs.~\cite{K.Nakamura-M.-K.Fujimoto-double-balanced-letter,K.Nakamura-M.-K.Fujimoto-double-balanced-full}
together with those in this paper indicate that the choice of the
directly measured operator at the photodetector affects the result not
only of the noise properties but also of the expectation value of the
output signal itself, in general.
Therefore, we conclude that the specification of the directly measured
operator is crucial in the development and applications of the
mathematical quantum measurement theory.


\section*{Acknowledgements}


The author deeply acknowledged to Prof. Masa-Katsu Fujimoto for his
valuable comments, discussion, and encouragements during the
author is carrying out this work.
The author also acknowledged to Prof. Takayuki Tomaru and
Prof. Tomotada Akutsu for their continuous encouragements.


\appendix


\section*{Appendix}


\section{Commutation relation $\left[\hat{E}_{b}(t),\hat{E}_{l_{i}}(t')\right]$}
\label{sec:Commutation_relation_Eb_Eli}

\begin{figure}[ht]
  \centering
  \includegraphics[width=0.7\textwidth]{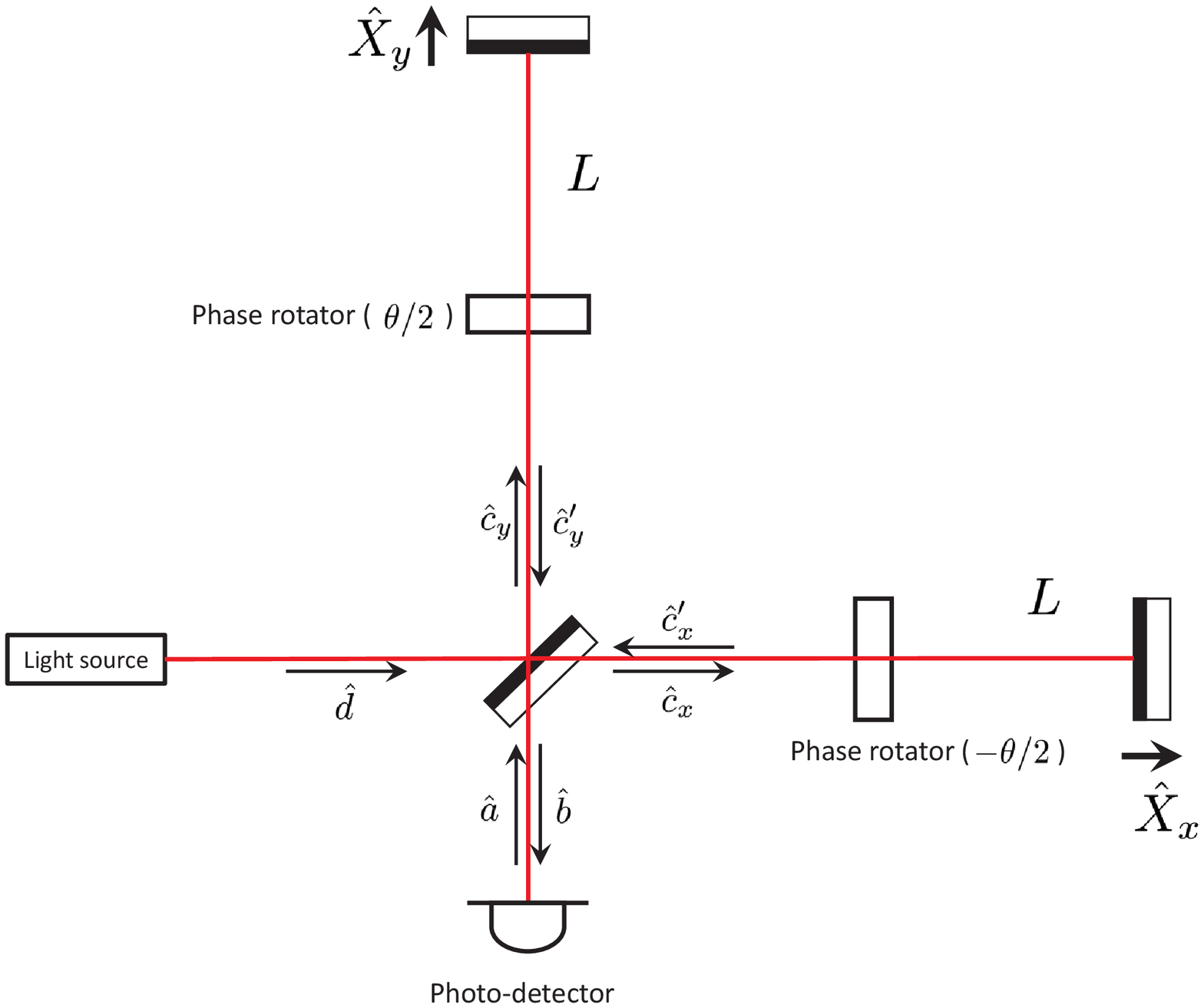}
  \caption{
    Configuration of the Michelson gravitational-wave detector.
    The notations of the quadratures $\hat{a}$, $\hat{b}$,
    $\hat{d}$ are given in this figure.
  }
  \label{fig:Michelson-gravitational-wave-detector}
\end{figure}

In this appendix, we evaluate the commutation relation
$\left[\hat{E}_{b}(t),\hat{E}_{l_{i}}(t')\right]$.
To evaluate this commutation relation, we have to consider the main
interferometer.
For example, we consider the Michelson gravitational-wave detector
with the phase offset $\phi$ as depicted in
Fig.~\ref{fig:Michelson-gravitational-wave-detector}.


The input-output relation of the Michelson gravitational-wave detector
is given by
\begin{eqnarray}
  \hat{b}_{\pm}
  &=&
      \sin\left(\frac{\phi}{2}\right) \left[
      i + \kappa \cos\left(\frac{\phi}{2}\right)
      \right]
      \sqrt{\frac{I_{0}}{\hbar\omega_{0}}}
      2 \pi \delta(\omega_{0}\pm\omega)
      \nonumber\\
  &&
     +
     e^{+2 i (\omega-\omega_{0})\tau} \left[
     i \sin\left(\frac{\phi}{2}\right) \hat{d}_{\pm}
     +
     \cos\left(\frac{\phi}{2}\right) \hat{a}_{\pm}
     \right]
      \nonumber\\
  &&
     +
     e^{\pm 2 i \omega\tau} \frac{\kappa}{2} \left[
     \sin\phi \left(\hat{d}_{\mp}^{\dagger}+\hat{d}_{\pm}\right)
     +
     i
     \cos\phi \left(\hat{a}_{\mp}^{\dagger}+\hat{a}_{\pm}\right)
     \right]
     \nonumber\\
  &&
     -
     i e^{+ i (\omega-\omega_{0})\tau}\sqrt{\kappa}
     \cos\left(\frac{\phi}{2}\right)
     \frac{h(\pm(\omega-\omega_{0}))}{h_{SQL}}
     .
     \label{eq:Michelson-input-output-relation-with-phi}
\end{eqnarray}
where the subscription $\pm$ or $\mp$ of the quadratures indicates the
upper- and the lower-sideband quadrature as in
Sec.~\ref{sec:simple_Homodyne_Detection}, $\kappa$, $h_{SQL}$, and
$\tau$ are given by
\begin{eqnarray}
  \label{eq:kappa-hSQL-tau-Michelson-input-output-relation-with-phi}
  \kappa
  :=
  \frac{8\omega_{0}I_{0}}{mc^{2}(\omega-\omega_{0})^{2}}
  , \quad
  h_{SQL}
  :=
  \sqrt{\frac{8\hbar}{m(\omega-\omega_{0})^{2}}} L^{2}
  , \quad
  \omega_{0} \tau = \omega_{0} \frac{L}{c} = 2\pi n, \quad n\in\ZF.
\end{eqnarray}
Here, the first line in
Eq.~(\ref{eq:Michelson-input-output-relation-with-phi}) is the leakage
of the classical carrier field due to the offset $\phi$.
The first term in the second line in
Eq.~(\ref{eq:Michelson-input-output-relation-with-phi}) is the shot
noise of the optical field, the second term in the second line in
Eq.~(\ref{eq:Michelson-input-output-relation-with-phi}) is the
radiation pressure noise.
These two terms are regarded as quantum noise.
The last line in
Eq.~(\ref{eq:Michelson-input-output-relation-with-phi}) includes the
gravitational-wave signal.


\begin{figure}[ht]
  \centering
  \includegraphics[width=0.7\textwidth]{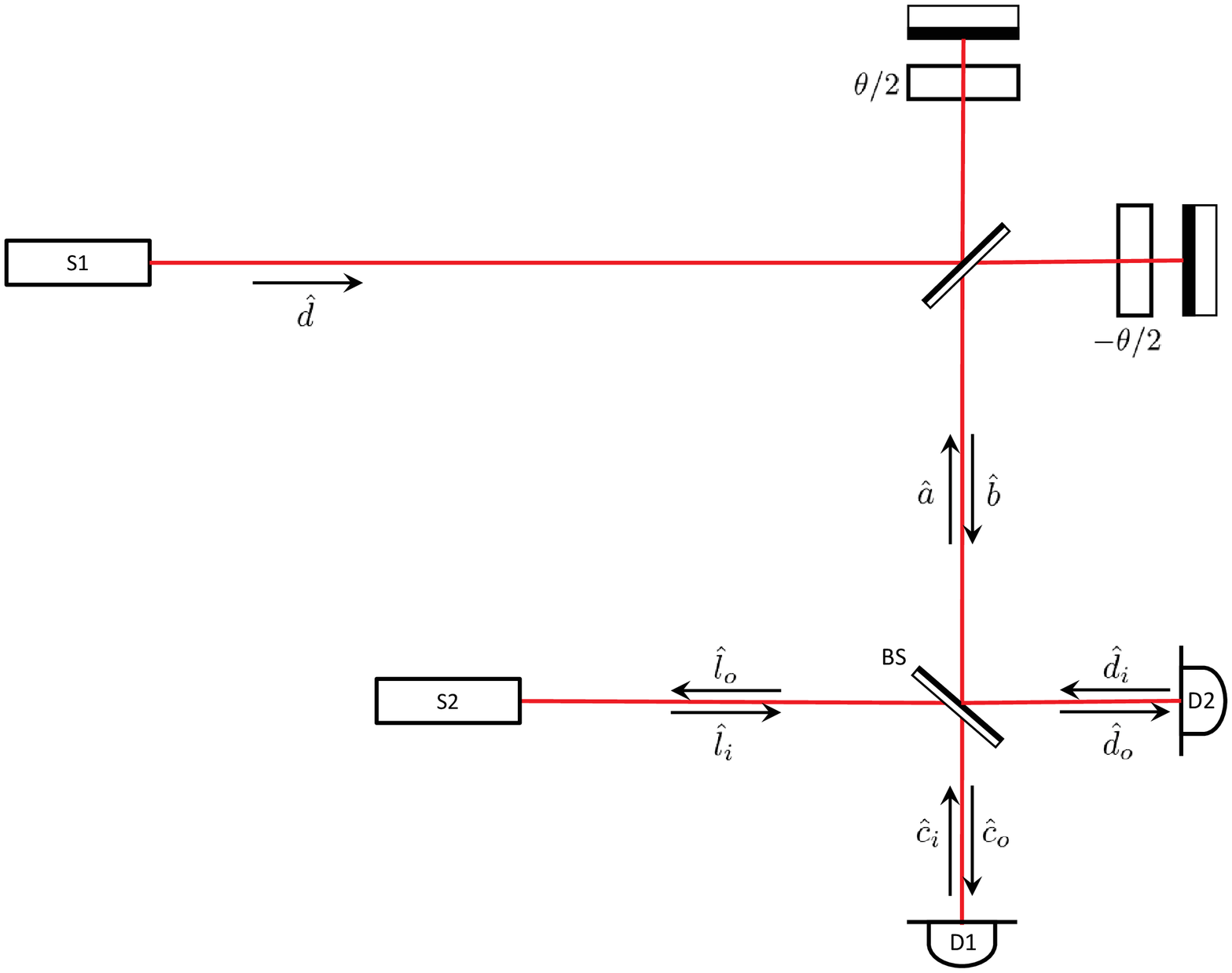}
  \caption{
    Introduction of the independent optical field as the local oscillator.
  }
  \label{fig:MichelsonWithBHDByIndepLocalOsci}
\end{figure}


If the electric field $\hat{E}_{l_{i}}(t)$ have nothing to do with the
electric field $\hat{E}_{b}(t)$ as depicted in
Fig.~\ref{fig:MichelsonWithBHDByIndepLocalOsci}, we may regard that
$\left[\hat{E}_{b}(t),\hat{E}_{l_{i}}(t')\right]=0$.
However, if we introduce the optical field of the local oscillator
from the main interferometer as depicted in
Fig.~\ref{fig:MichelsonWithBHDByLocalOsciFromMain}, the optical field
$\hat{E}_{l_{i}}(t)$ is not independent of the optical field
$\hat{E}_{d}(t)$ through the input-output relation.
Actually, the input-output relation
(\ref{eq:Michelson-input-output-relation-with-phi}) does include the
quadrature $\hat{d}$ unless the offset $\phi$ vanishes, i.e., the
complete dark port in which the classical carrier field does not leak
from the main interferometer.
Therefore, we concentrate on the interferometer setup depicted in
Fig.~\ref{fig:MichelsonWithBHDByLocalOsciFromMain}.
Furthermore, to check the commutation relation
$\left[\hat{E}_{b}(t),\hat{E}_{l_{i}}(t)\right]=0$, we may concentrate
on the commutation relation of the quadrature $\hat{l}_{i}(\omega)$
from the local oscillator and the quadrature $\hat{d}(\omega)$ in the
input-output relation.


We consider the optical field junction at the beam splitter BS2 in
Fig.~\ref{fig:MichelsonWithBHDByLocalOsciFromMain} with the
transmissivity $\zeta$.
The optical field junction condition at BS0 is given by
\begin{eqnarray}
  \label{eq:BS0-junction-condition-1}
  \hat{d}'(\omega)
  &=&
      \sqrt{\zeta} \hat{d}_{s}(\omega)
      +
      \sqrt{1-\zeta} \hat{f}_{i}(\omega)
      ,
  \\
  \label{eq:BS0-junction-condition-2}
  \hat{l}_{i}'
  &=&
      \sqrt{\zeta} \hat{f}_{i}(\omega)
      -
      \sqrt{1-\zeta} \hat{d}_{s}(\omega)
      ,
\end{eqnarray}
where $\hat{d}_{s}(\omega)$ is the quadrature for the incident field
from the light source; $\hat{d}'(\omega)$ is the quadrature for the
output field to the main interferometer; $\hat{l}'_{i}(\omega)$ the
quadrature for the output field to the local oscillator; and
$\hat{f}_{i}(\omega)$ is the quadrature for the incident vacuum field
to BS0, i.e.,
\begin{eqnarray}
  \label{eq:hatfi-vacuum-condition}
  \hat{f}_{i}(\omega)|0\rangle_{f_{i}} = 0.
\end{eqnarray}
The propagation of the field $\hat{E}_{d'}$ associated with the
quadrature $\hat{d}'(\omega)$ to the main interferometer yields
\begin{eqnarray}
  \label{eq:hatd-hatdprimee-iomegax}
  \hat{d}(\omega) = \hat{d}'(\omega) e^{-i\omega x/c}.
\end{eqnarray}
Furthermore, the propagation of the field $\hat{E}_{l_{i}'}$ to BS of
the balanced homodyne detection yields
\begin{eqnarray}
  \label{eq:hatli-hatliprimee-iomegax+y}
  \hat{l}_{i}(\omega)
  =
  \hat{l}_{i}'(\omega) e^{-i\omega(x+y)/c}
  .
\end{eqnarray}
Through the above setup, we obtain the quadratures $\hat{d}$ and
$\hat{l}_{i}$ as
\begin{eqnarray}
  \label{eq:hatd-hatds-hatfi-relation}
  \hat{d}(\omega)
  &=&
      e^{-i\omega x} \left[
      \sqrt{\zeta} \hat{d}_{s}(\omega)
      +
      \sqrt{1-\zeta} \hat{f}_{i}(\omega)
      \right]
      ,
  \\
  \label{eq:hatli-hatds-hatfi-relation}
  \hat{l}_{i}(\omega)
  &=&
      e^{-i\omega(x+y)} \left[
      \sqrt{\zeta} \hat{f}_{i}(\omega)
      -
      \sqrt{1-\zeta} \hat{d}_{s}(\omega)
      \right]
      .
\end{eqnarray}


Then, we can evaluate the commutation relations:
\begin{eqnarray}
  \label{eq:hatd-hatli-commutation}
  &&
     \left[
     \hat{d}(\omega'), \hat{l}_{i}(\omega)
     \right]
     =
     0
     ,
     \quad
     \left[
     \hat{d}^{\dagger}(\omega'), \hat{l}_{i}^{\dagger}(\omega)
     \right]
     =
     0
     ,
  \\
  &&
     \left[
     \hat{d}^{\dagger}(\omega'), \hat{l}_{i}(\omega)
     \right]
     =
     e^{+i\omega'x/c} e^{-i\omega(x+y/c)} \sqrt{\zeta(1-\zeta)}
     \nonumber\\
  && \quad\quad\quad\quad\quad\quad\quad\quad\quad
     \times
     \left\{
     -
     \left[
     \hat{d}_{s}^{\dagger}(\omega'), \hat{d}_{s}(\omega)
     \right]
     +
     \left[
     \hat{f}^{\dagger}_{i}(\omega'), \hat{f}_{i}(\omega)
     \right]
     \right\}
     \nonumber\\
  && \quad\quad\quad\quad\quad\quad
     =
     0
     ,
     \label{eq:hatddagger-hatli-commutation}
  \\
  &&
     \left[
     \hat{d}(\omega'), \hat{l}_{i}^{\dagger}(\omega)
     \right]
     =
     e^{-i\omega'x/c} e^{+i\omega(x+y)/c} \sqrt{\zeta(1-\zeta)}
     \nonumber\\
  && \quad\quad\quad\quad\quad\quad\quad\quad\quad
     \times
     \left\{
     -
     \left[
     \hat{d}_{s}(\omega'), \hat{d}_{s}^{\dagger}(\omega)
     \right]
     +
     \left[
     \hat{f}_{i}(\omega'), \hat{f}_{i}^{\dagger}(\omega)
     \right]
     \right\}
     \nonumber\\
  && \quad\quad\quad\quad\quad\quad
     =
     0
     .
     \label{eq:hatd-hatlidagger-commutation}
\end{eqnarray}
From these commutation relations, even in the situation depicted in
Fig.~\ref{fig:MichelsonWithBHDByLocalOsciFromMain}, we reach to the
conclusion
\begin{eqnarray}
  \label{eq:hatEb-hatEli-commutation-vanishes}
  \left[
  \hat{E}_{b}(t), \hat{E}_{l_{i}}
  \right]
  =
  0
  .
\end{eqnarray}


\section{Properties of a time-averaged function}
\label{sec:Properties_of_a_time-averaged_function}


To evaluate the integration of the noise spectral densities, we have
to consider the integration of a function
\begin{eqnarray}
  \label{eq:lim1overTintdte-iat-def}
  f(a) := \lim_{T\rightarrow+\infty} \frac{1}{T} \int_{-T/2}^{T/2} dt
  e^{-iat}
  ,
  \quad a\in\RF
  .
\end{eqnarray}
Here, we summarize some properties of this function $f(a)$ which are
used in the estimation of the noise spectral densities.


The first trivial property is the value of $f(a=0)$:
\begin{eqnarray}
  \label{eq:lim1overTintdte-iat-a=0}
  f(a=0) = \lim_{T\rightarrow+\infty} \frac{1}{T} \int_{-T/2}^{T/2} dt 1 = 1.
\end{eqnarray}
On the other hand, when $a\neq 0$, we can estimate this function as
\begin{eqnarray}
  |f(a)|
  &=&
      \left|
      \lim_{T\rightarrow+\infty} \frac{1}{T} \frac{1}{-ia}
      \left(e^{-iaT/2} - e^{+iaT/2}\right)
      \right|
      \nonumber\\
  &\leq&
         \lim_{T\rightarrow+\infty} \frac{1}{T} \frac{1}{|a|} \left(
         \left|e^{-iaT/2}\right| + \left|e^{+iaT/2}\right|
         \right)
         =
         \lim_{T\rightarrow+\infty} \frac{1}{T} \frac{2}{|a|}
         \rightarrow
         0
         .
         \label{eq:lim1overTintdte-iat-aneq0-absolute-value}
\end{eqnarray}
Thus, we have shown that
\begin{eqnarray}
  \label{eq:lim1overTintdte-iat-aneq0}
  f(a) =
  \left\{
  \begin{array}{lcccl}
    1 & \quad & \mbox{for} & \quad & a=0; \\
    0 & \quad & \mbox{for} & \quad & a\neq 0.
  \end{array}
                                     \right.
\end{eqnarray}


Furthermore, this function $f(a)$ is bounded, and its support is
measure zero.
Then, we have
\begin{eqnarray}
  \label{eq:intdagafa=0}
  \int_{-\infty}^{+\infty} g(a) f(a) da = 0
\end{eqnarray}
for a finite function $g(a)$.
On the other hand, when $g(a)$ is a $\delta$-function, we obtain
\begin{eqnarray}
  \label{eq:intdadeltaafa=1}
  \int_{-\infty}^{+\infty} \delta(a) f(a) da = 1, \quad
  \int_{-\infty}^{+\infty} \delta(b\neq a) f(a) db = 0.
\end{eqnarray}
These properties of integrations characterize the stationary of the
modes in noises, i.e., which modes survive in the stationary
situation where the noise spectral density depends only on $\tau=t-t'$.


We use these properties when we evaluate the noise spectral
densities.


\section{Evaluation of ${\cal I}_{1}(\omega)$ in
  Eq.~(\ref{eq:normal-ordered-noise-spectral-density-monochro-calI1})
  through the Michelson example}
\label{sec:calI1omega-in-Michelson}


In this appendix, we consider an example of the input-output relation
(\ref{eq:Michelson-input-output-relation-with-phi}) in the Michelson
interferometer.
More generally, the input-output relation
(\ref{eq:Michelson-input-output-relation-with-phi}) is also written as
\begin{eqnarray}
  \hat{b}_{n}(\omega)
  &:=&
       \hat{b}(\omega)
       -
       \langle\hat{b}(\omega)\rangle
       \nonumber\\
  &=&
      \int_{0}^{\omega} \frac{d\omega'}{2\pi}
      \left[
      {\cal A}(\omega,\omega') \hat{a}(\omega')
      +
      {\cal B}(\omega,\omega') \hat{a}^{\dagger}(\omega')
      \right.
      \nonumber\\
  && \quad\quad\quad\quad
     \left.
     +
     {\cal C}(\omega,\omega') \hat{d}(\omega')
     +
     {\cal D}(\omega,\omega') \hat{d}^{\dagger}(\omega')
     \right]
     \label{eq:output-noise-quadrature-general}
     ,
\end{eqnarray}
where the expectation value $\langle\hat{b}(\omega)\rangle$ is given
by Eq.~(\ref{eq:output-quadrature-expectation-value-main}).
Here, the input-output relation
(\ref{eq:Michelson-input-output-relation-with-phi}) is realized as
\begin{eqnarray}
  \label{eq:calA-KN-MKF-Annals-Phys}
  {\cal A}(\omega_{0}\pm\omega,\omega')
  &=&
      \left[
      +  e^{\pm 2i\omega\tau} \cos\left(\frac{\theta}{2}\right)
      + i \frac{\kappa(\omega) e^{\pm 2 i \omega\tau}}{2} \cos\theta
      \right]
      2\pi \delta(\omega'-(\omega_{0}\pm\omega))
      ,
  \\
  \label{eq:calB-KN-MKF-Annals-Phys}
  {\cal B}(\omega_{0}\pm\omega,\omega')
  &=&
      i \frac{\kappa(\omega) e^{\pm 2 i \omega\tau}}{2} \cos\theta
      2 \pi \delta(\omega'-(\omega_{0}\mp\omega))
      ,
  \\
  \label{eq:calC-KN-MKF-Annals-Phys}
  {\cal C}(\omega_{0}\pm\omega,\omega')
  &=&
      \left[
      + i e^{\pm 2i\omega\tau} \sin\left(\frac{\theta}{2}\right)
      + \frac{\kappa(\omega) e^{\pm 2 i \omega\tau}}{2} \sin\theta
      \right] 2\pi \delta(\omega'-(\omega_{0}\pm\omega))
      ,
  \\
  \label{eq:calD-KN-MKF-Annals-Phys}
  {\cal D}(\omega_{0}\pm\omega,\omega')
  &=&
      \frac{\kappa(\omega) e^{\pm 2 i \omega\tau}}{2} \sin\theta
      2 \pi \delta(\omega'-(\omega_{0}\mp\omega)
      .
\end{eqnarray}
and
\begin{eqnarray}
  \label{eq:alpha-KN-MKF-Annals-Phys}
  \alpha(\omega)
  &=&
      -
      i e^{+ i (\omega-\omega_{0})\tau}\sqrt{\kappa}
      \cos\left(\frac{\phi}{2}\right)
      \frac{h(\pm(\omega-\omega_{0}))}{h_{SQL}}
      ,
  \\
  \label{eq:beta-KN-MKF-Annals-Phys}
  \beta
  &=&
      \sin\left(\frac{\phi}{2}\right) \left[
      i + \kappa(\omega_{0}) \cos\left(\frac{\phi}{2}\right)
      \right]
      \sqrt{\frac{I_{0}}{\hbar\omega_{0}}}
      .
\end{eqnarray}
Although $\kappa(\omega_{0})$ in Eq.~(\ref{eq:beta-KN-MKF-Annals-Phys})
diverge in the Michelson interferometer, we do not take this
divergence seriously because this divergence disappears in the
Fabri-P\'erot
interferometer~\cite{H.J.Kimble-Y.Levin-A.B.Matsko-K.S.Thorne-S.P.Vyatchanin-2001}.


In this example,
$\hat{b}_{n}(\omega_{0}+\omega)\hat{b}_{n}(\omega_{2})$ in the term
${\cal I}_{1}$ in
Eq.~(\ref{eq:normal-ordered-noise-spectral-density-monochro-calI1})
is given by
\begin{eqnarray}
  &&
     \langle
     \hat{b}_{n}(\omega_{0}+\omega)
     \hat{b}_{n}(\omega_{2})
     \rangle
     \nonumber\\
  &=&
      \int_{0}^{+\infty} \frac{d\omega_{3}}{2\pi}
      \left[
      {\cal A}(\omega_{0}+\omega,\omega_{3})
      {\cal B}(\omega_{2},\omega_{3})
      +
      {\cal C}(\omega_{0}+\omega,\omega_{3})
      {\cal D}(\omega_{2},\omega_{3})
      \right]
      \label{eq:langlehatbnomega0+omegahatbnomega2rangle-general}
  \\
  &=&
      \frac{\kappa(|\omega-\omega_{0}|)}{2} \left[
      + i \cos\left(\frac{\theta}{2}\right)
      - \frac{\kappa(|\omega-\omega_{0}|)}{2} \cos(2\theta)
      \right]
      \nonumber\\
  && \quad\quad\quad\quad\quad\quad
     \times
     \Theta(\omega+\omega_{0})
     2 \pi \delta(\omega_{2}-(\omega_{0}-\omega))
     .
     \label{eq:langlehatbnomega0+omegahatbnomega2rangle-Michelson}
\end{eqnarray}
The appearance of the $\delta$-function $2 \pi
\delta(\omega_{2}-(\omega_{0}-\omega))$ in
Eq.~(\ref{eq:langlehatbnomega0+omegahatbnomega2rangle-Michelson}) is
important.
Due to this delta function, we obtain the finite result in the above
example as
\begin{eqnarray}
  {\cal I}_{1}(\omega)
  &=&
      \omega_{0} |\gamma|^{2} e^{-2i\theta}
      \Theta(\omega_{0}+\omega)
      \sqrt{(\omega_{0}+\omega)(\omega_{0}-\omega)}
      \frac{\kappa(|\omega-\omega_{0}|)}{2}
      \nonumber\\
  && \quad\quad
     \times
     \left[
     i \cos\left(\frac{\theta}{2}\right)
     -
     \frac{\kappa(|\omega-\omega_{0}|)}{2}
     \cos(2\theta)
     \right]
     .
     \label{eq:calI1omega-Michelson-explicit-result}
\end{eqnarray}


Similar calculations yields the finite ${\cal I}_{2}(\omega)$,
${\cal I}_{3}(\omega)$, ${\cal I}_{4}(\omega)$ through
Eq.~(\ref{eq:normal-ordered-noise-spectral-density-monochro-calI2}),
(\ref{eq:normal-ordered-noise-spectral-density-monochro-calI3}),
and
(\ref{eq:normal-ordered-noise-spectral-density-monochro-calI4}),
respectively.



\end{document}